\documentclass[a4paper,11pt]{article}
\pdfoutput=1 

\usepackage{jheppub} 
\usepackage[]{float} 
\usepackage[T1]{fontenc} 
\usepackage{graphicx}
\usepackage{dcolumn}
\usepackage{bm}
\usepackage{hyperref}
\usepackage{multirow}
\usepackage{wasysym}

\title{\boldmath Genuine NMSSM deviations in the 125 GeV Higgs boson decays}

\author{C. Beskidt}
\author{and W. de Boer}
\affiliation{Dept. of Phys., Karlsruhe Institute for Technology (KIT), Karlsruhe, Germany}

\emailAdd{Conny.Beskidt@kit.edu}
\emailAdd{Wim.de.Boer@kit.edu}

\abstract{In Supersymmetry deviations from the SM signal strengths of the 125 GeV Higgs boson can occur, because of possible SUSY contributions in diagrams with loops, which leads to different deviations in signal strengths for processes with and without loop diagrams. In the Next-to Minimal Supersymmetric Standard Model (NMSSM) additional deviations may occur, because of the mixing with the additional singlet-like Higgs boson and/or additional decays into pairs of light particles, like neutralinos, pseudo-scalar Higgs bosons or singlet-like Higgs bosons. In this paper we study these ``genuine" NMSSM deviations in detail to check not only their possible size, but also look for correlations or anti-correlations with respect to other channels in the hope to find specific patterns in the deviations as for the SUSY contributions, which occur predominantly in processes including loops in the diagrams. The whole NMSSM parameter space is sampled in a deterministic way. The novel scanning method is discussed in detail in the Appendix. We found three different regions with ``genuine" NMSSM deviations, which are largely independent of the production mode. What was surprising: some regions show negativ   correlations between final states with fermions and bosons meaning if the signal strengths for fermions decrease the signal strengths for bosonic final states increase, while other regions show positive correlations. The sign of the correlations is a strong function of the mass difference between the observed 125 GeV Higgs boson and the Higgs singlet, so looking for correlations could give useful hints about the existence and mass of the singlet Higgs boson. The features of these ``genuine" NMSSM effects have been investigated in representative benchmark points for each of the three regions where a single effect is dominant. These benchmark points have been detailed in the Appendix.}

\newlength{\dslashwidth}

\newcommand{\beq}{\begin{equation}} 
\newcommand{\eeq}{\end{equation}}
\newcommand{\beqa}{\begin{eqnarray}} 
\newcommand{\eeqa}{\end{eqnarray}}
\newcommand{\newc}{\newcommand}
\newcommand{\bq}{\begin{equation}}
\newcommand{\eq}{\end{equation}}
\newcommand{\ba}{\begin{array}}
\newcommand{\ea}{\end{array}}
\newcommand{\bqa}{\begin{eqnarray}}
\newcommand{\eqa}{\end{eqnarray}}

\newcommand{\lnf}{{\ifmmode \Lambda^{(N_f)} \else $\Lambda^{(N_f)}$\fi}}
\newcommand{\ms}{{\ifmmode \overline{MS} \else $\overline{MS}$\fi}}
\newcommand{\dr}{{\ifmmode \overline{DR} \else $\overline{DR}$\fi}}
\newcommand{\lms}{{\ifmmode \Lambda^{(5)}_{\overline{MS}} \else $\Lambda^{(5)}_{\overline{MS}}$\fi}}
\newcommand{\lam}{{\ifmmode \Lambda \else $\Lambda$\fi}}
\newcommand{\mev}{{\ifmmode {\rm MeV} \else ${\rm MeV}$\fi}}
\newcommand{\gev}{{\ifmmode {\rm GeV} \else ${\rm GeV}$\fi}}
\newcommand{\gevc}{{\ifmmode {\rm GeV/c^2} \else ${\rm GeV/c^2}$\fi}}
\newcommand{\tev}{{\ifmmode {\rm TeV} \else ${\rm TeV}$\fi}}
\newcommand{\tevc}{{\ifmmode {\rm TeV/c^2} \else ${\rm TeV/c^2}$\fi}}
\newcommand{\lp}{{\ifmmode L^+  \else $L^+$\fi}}
\newcommand{\lm}{{\ifmmode L^-  \else $L^-$\fi}}
\newcommand{\mlp}{{\ifmmode M(L^-) \else $M(L^-)$\fi}}
\newcommand{\mlz}{{\ifmmode M(L^0) \else $M(L^0)$\fi}}
\newcommand{\lz}{{\ifmmode L^0 \else $L^0$\fi}}
\newcommand{\ev}{{\ifmmode GeV/c^2 \else $GeV/c^2$\fi}}
\newcommand{\tri}{{\ifmmode \triangleup \else $\triangleup$\fi}}
\newcommand{\unl}{{\ifmmode U_{lL^0} \else $U_{lL^0}$\fi}}\newcommand{\gL}{{\ifmmode g_L \else $g_{L}$\fi}}
\newcommand{\gR}{{\ifmmode g_R  \else $g_{R}$\fi}}
\newcommand{\gumu}{{\ifmmode \gamma^{\mu} \else $\gamma^{\mu}$\fi}}
\newcommand{\gunu}{{\ifmmode \gamma^{\nu} \else $\gamma^{\nu}$\fi}}
\newcommand{\gdmu}{{\ifmmode \gamma_{\mu} \else $\gamma_{\mu}$\fi}}
\newcommand{\gdnu}{{\ifmmode \gamma_{\nu} \else $\gamma_{\nu}$\fi}}
\newcommand{\stw}{{\ifmmode\sin^2\theta_W \else $\sin^{2}\theta_{W}$ \fi}}
\newcommand{\sws}{{\ifmmode \;\sin^2\theta_W  \else $\;\sin^{2}\theta_{W}$ \fi}}
\newcommand{\cws}{{\ifmmode \;\cos^2\theta_W  \else $\;\cos^{2}\theta_{W}$ \fi}}
\newcommand{\sw}{{\ifmmode \;\sin\theta_W  \else $\sin\theta_{W}$ \fi}}
\newcommand{\cw}{{\ifmmode \;\cos\theta_W  \else $\;\cos\theta_{W}$ \fi}}
\newcommand{\tw}{{\ifmmode \;\tan\theta_W  \else $\;\tan\theta_{W}$ \fi}}
\newcommand{\qq}{{\ifmmode q\overline{q} \else $q\overline{q}$\fi}}
\newcommand{\lR}{{\ifmmode l_R \else $l_R$\fi}}
\newcommand{\lL}{{\ifmmode l_L \else $l_L$\fi}}
\newcommand{\nt}{{\ifmmode \nu_{\tau} \else $\nu_{\tau}$\fi}}
\newcommand{\nuR}{{\ifmmode \nu_R  \else $\nu_R$\fi}}
\newcommand{\nuL}{{\ifmmode \nu_L  \else $\nu_L$\fi}}
\newcommand{\qR}{{\ifmmode g_R  \else $q_R$\fi}}
\newcommand{\qL}{{\ifmmode q_L  \else $q_L$\fi}}
\newcommand{\qRp}{{\ifmmode q_R'  \else $q_{R}$'\fi}}
\newcommand{\qLp}{{\ifmmode q_L'  \else $q_{L}$'\fi}}
\newcommand{\est}{{\ifmmode e^{\bf \ast} \else $e^{\bf \ast}$\fi}}
\newcommand{\lst}{{\ifmmode l^{\bf \ast} \else $l^{\bf \ast}$\fi}}
\newcommand{\must}{{\ifmmode \mu^{\bf \ast} \else $\mu^{\bf \ast}$\fi}}
\newcommand{\taust}{{\ifmmode \tau^{\bf \ast} \else $\tau^{\bf \ast}$ \fi}}
\newcommand{\pperp}{{\ifmmode p_t  \else $p_t$\fi}}
\newcommand{\et}{{\ifmmode E_t  \else $E_t$\fi}}
\newcommand{\xt}{{\ifmmode x_t  \else $x_t$\fi}}
\newcommand{\smumu}{{\ifmmode \sigma_{\mu\mu}  \else $\sigma_{\mu\mu}$ \fi}}
\newcommand{\eg}{{\ifmmode e\gamma  \else $e\gamma$\fi}}
\newcommand{\epem}{{\ifmmode e^+e^-  \else $e^+e^-$\fi}}
\newcommand{\lplm}{{\ifmmode L^+L^-  \else $L^+L^-$\fi}}
\newcommand{\pp}{{\ifmmode p\overline p  \else $p\overline p$\fi}}
\newcommand{\llz}{{\ifmmode L^0\overline{L}^0 \else $L^0\overline{L}^0$\fi}}
\newcommand{\epemt}{{\ifmmode e^+e^- \to  \else $e^+e^- \to$\fi}}
\newcommand{\eb}{{\ifmmode E_{beam}  \else $E_{beam}$\fi}}
\newcommand{\ip}{{\ifmmode pb^{-1}  \else $pb^{-1}$\fi}}
\newcommand{\upm}{{\ifmmode ^{\pm}  \else $^{\pm}$\fi}}
\newcommand{\de}{{\ifmmode ^{\circ}  \else $^{\circ}$ \fi}}
\newcommand{\appr}{{\ifmmode \sim \else $\sim$ \fi}}
\newcommand{\corresp}{{\ifmmode \stackrel{\wedge}{=} \else $\stackrel{\wedge}{=}$ \fi}}
\newcommand{\sqrts}{{\ifmmode \sqrt{s} \else $\sqrt{s}$\fi}}
\newcommand{\zz}{{\ifmmode Z^0  \else $Z^0$\fi}}
\newcommand{\mz}{{\ifmmode M_{Z}  \else $M_{Z}$\fi}}
\newcommand{\mzs}{{\ifmmode M_{Z}^2  \else $M_{Z}^2$\fi}}
\newcommand{\mw}{{\ifmmode M_{W}  \else $M_{W}$\fi}}
\newcommand{\mws}{{\ifmmode M_{W}^2  \else $M_{W}^2$\fi}}
\newcommand{\mh}{{\ifmmode M_{Higgs}  \else $M_{Higgs}$\fi}}
\newcommand{\msusy}{{\ifmmode M_{SUSY}  \else $M_{SUSY}$\fi}}
\newcommand{\msusys}{{\ifmmode M_{SUSY}^2  \else $M_{SUSY}^2$\fi}}
\newcommand{\su}{{\ifmmode SU(3)_C\otimes\- SU(2)_L\otimes\- U(1)_Y \else $SU(3)_C\otimes\A0SU(2)_L\otimes U(1)_Y$\fi}}
\newcommand{\suthree}{{\ifmmode SU(3)_C  \else $SU(3)_C$\fi}}
\newcommand{\sutwo}{{\ifmmode  SU(2)_L\otimes U(1)_Y \else $SU(2)_L\otimes U(1)_Y$\fi}}
\newcommand{\taup}{{\ifmmode \tau_{proton} \else $\tau_{proton}$\fi}}
\newcommand{\as}{{\ifmmode \alpha_{s}  \else $\alpha_{s}$\fi}}
\newcommand{\mgut}{{\ifmmode M_{GUT}  \else $M_{GUT}$\fi}}
\newcommand{\mguts}{{\ifmmode M_{GUT}^2  \else $M_{GUT}^2$\fi}}
\newcommand{\mzero}{{\ifmmode m_0        \else $m_0$\fi}}
\newcommand{\mhalf}{{\ifmmode m_{1/2}    \else $m_{1/2}$\fi}}
\newcommand{\sq}{{\ifmmode \tilde{q}    \else $\tilde{q}$\fi}}
\newcommand{\gl}{{\ifmmode \tilde{g}    \else $\tilde{g}$\fi}}
\newcommand{\mb}{{\ifmmode m_{b}    \else $m_{b}$\fi}}
\newcommand{\mt}{{\ifmmode m_{t}    \else $m_{t}$\fi}}
\newcommand{\mts}{{\ifmmode m_{t}^2    \else $m_{t}^2$\fi}}

\newcommand{\mtau}{{\ifmmode m_{\tau}  \else $m_{\tau}$\fi}}
\newcommand{\dpp}{{\ifmmode \delta_{pert} \else $\delta_{pert}$\fi}}
\newcommand{\dnp}{{\ifmmode\delta_{non-pert}\else$\delta_{non-pert}$\fi}}
\newcommand{\dew}{{\ifmmode \delta_{\rm EW}\else $\delta_{\rm EW}$\fi}}
\newcommand{\rt}{{\ifmmode R_{\tau}  \else $R_{\tau} $\fi}}
\newcommand{\rz}{{\ifmmode R_{Z}  \else $R_{Z} $\fi}}

\newcommand{\swb}{{\ifmmode \sin^2\theta_{\overline{MS}} \else $\sin^2\theta_{\overline{MS}}$\fi}}
\newcommand{\cwb}{{\ifmmode \cos^2\theta_{\overline{MS}} \else $\cos^2\theta_{\overline{MS}}$\fi}}

\newc\AIPCP[3] {{\em AIP Conf. Proc.} {\bf #1} (#2) #3}
\newc\AJ[3] {{\em Astrophys. J.} {\bf #1} (#2) #3}
\newc\AMS[3] {{\em Ann. Math. Statist.} {\bf #1} (#2) #3}
\newc\AP[3] {{\em Ann. Phys.} {\bf #1} (#2) #3}
\newc\APJ[3] {{\em Astropart. J.} {\bf #1} (#2) #3}
\newc\APP[3] {{\em Astropart. Phys.} {\bf #1} (#2) #3}
\newc\APS[3] {{\em Astrophys. J. Suppl.} {\bf #1} (#2) #3}
\newc\ARNPS[3] {{\em Ann. Rev. Nucl. Part. Sci.} {\bf C#1} (#2) #3}
\newc\BA[3] {{\em Bayesian Anal.} {\bf C#1} (#2) #3}
\newc\CPC[3] {{\em Comput. Phys. Commun.} {\bf C#1} (#2) #3}
\newc\CP[3] {{\em Contemp. Phys.} {\bf #1} (#2) #3}
\newc\EPJ[3] {{\em Euro. Phys. Journ.} {\bf C#1} (#2) #3}
\newc\JCAP[3] {{\em JCAP} {\bf #1} (#2) #3}
\newc\JHEP[3] {{\em JHEP} {\bf #1} (#2) #3}
\newc\JPG[3] {{\em J. Phys.} {\bf G #1} (#2) #3}
\newc\IJMP[3] {{\em Int. J. Mod. Phys.} {\bf A #1} (#2) #3}
\newc\MNRAS[3] {{\em Mon. Not. Roy. Astron. Soc.} {\bf #1} (#2) #3}
\newc\MPL[3] {{\em Mod. Phys. Lett.} {\bf A #1} (#2) #3}
\newc\NAR[3] {{\em New Astron. Rev.} {\bf #1} (#2) #3}
\newc\NCA[3] {{\em Nuovo Cimento} {\bf #1} (#2) #3}
\newc\NIM[3] {{\em Nucl. Instrum. Methods} {\bf #1} (#2) #3}
\newc\NIMA[3] {{\em Nucl. Instrum. Methods} {\bf A #1} (#2) #3}
\newc\NAT[3] {{\em Nature} {\bf #1} (#2) #3}
\newc\NPB[3] {{\em Nucl. Phys.} {\bf B #1} (#2) #3}
\newc\NPA[3] {{\em Nucl. Phys.} {\bf A #1} (#2) #3}
\newc\NPPS[3] {{\em Nucl. Phys. Proc. Suppl.} {\bf #1} (#2) #3}
\newc\PLB[3] {{\em Phys. Lett.} {\bf B #1} (#2) #3}
\newc\PR[3] {{\em Phys. Rep.} {\bf #1} (#2) #3}
\newc\PRL[3] {{\em Phys. Rev. Lett.} {\bf #1} (#2) #3}
\newc\PRD[3] {{\em Phys. Rev.} {\bf D #1} (#2) #3}
\newc\PRC[3] {{\em Phys. Rev.} {\bf C #1} (#2) #3}
\newc\PTP[3] {{\em Prog. Theor. Phys.} {\bf #1} (#2) #3}
\newc\RMP[3] {{\em Rev. Mod. Phys.} {\bf #1} (#2) #3 }
\newc\RPP[3] {{\em Rept. Prog. Phys.} {\bf #1} (#2) #3 }
\newc\SC[3] {{\em Science} {\bf #1} (#2) #3 }
\newc\ZPC[3] {{\em Z. Phys.} {\bf C #1} (#2) #3}
\newc\Err[3] {{\em Erratum-ibid.} {\bf #1} (#2) #3 }

\begin{document} 

\maketitle
\flushbottom

\section{Introduction}
\label{Introduction}

After the discovery of the 125 GeV Higgs boson \cite{Aad:2012tfa,Chatrchyan:2012xdj} deviations from SM expectations were found, e.g. in the $\gamma\gamma$ decay, which led to speculations about new physics \cite{Carena:2012mw,Ellwanger:2011aa,Arvanitaki:2011ck,Gunion:2012zd,Basso:2012tr,Mahmoudi:2012eh,
Baer:2012up,Vasquez:2012hn,Espinosa:2012ir,Choi:2019yrv,Baum:2019uzg}. However, with higher statistics the measurements became consistent with SM predictions, although the errors are  still too large  to be sensitive to deviations induced by Supersymmetry (SUSY) \cite{Tanabashi:2018oca}.
Deviations may occur, e.g. because of SUSY particles contributing in processes with loop diagrams, like the Higgs production via  gluon fusion, where both, top quarks and their supersymmetric partners, the stop quarks, can contribute.   Or the loop-induced decays of Higgs bosons into photons may have contributions from SUSY particles.

Experimentally one can only measure cross sections times branching ratios, so if deviations of the signal strengths of the observed 125 GeV Higgs boson are observed, one does not know a priori if the deviations occur in the cross sections or branching ratios. One can disentangle these effects by looking for correlations of the deviations, e.g. by the correlation in signal strengths for processes with and without loop diagrams. 
Deviations in addition to the loop-induced ones are expected in an extension of the Minimal Supersymmetric Standard Model (MSSM), namely the Next-to-Minimal Supersymmetric Standard Model (NMSSM). Reviews on SUSY can be found in Refs. \cite{Haber:1984rc,Kane:1993td,deBoer:1994dg,Martin:1997ns,Kazakov:2010qn,Kazakov:2015ipa}.
 for the MSSM and in Ref. \cite{Ellwanger:2009dp} for the NMSSM.

One may ask  why one should  study the NMSSM, if there is not yet a sign of the MSSM. 
There are good motivations for the NMSSM in comparison with the MSSM. They differ by the introduction of a singlet Higgs boson in the NMSSM in addition to the two doublets of the MSSM. 
The main theoretical motivation for the singlet is the solution to the $\mu$ problem: the $\mu$ parameter with the dimension of a mass in the Lagrangian could take any value up to the GUT scale, but radiative electroweak symmetry breaking requires $\mu$ to have a value of the order of the electroweak scale. Assuming $\mu$ to correspond to the vev of a singlet would  bring this parameter naturally in the range of the electroweak scale, thus solving the $\mu$ problem, see e.g. Refs. \cite{Kim:1983dt,Miller:2003ay,Ellwanger:2009dp}. Experimentally, the singlet has the advantage that the tree level value of the  Higgs mass is increased  by the mixing with the singlet, so there is no  need for large loop corrections from multi-TeV stop quarks to raise the Higgs mass from below the electroweak scale  - expected in the MSSM - to 125 GeV \cite{Hall:2011aa,Arvanitaki:2011ck,Gunion:2012zd,King:2012is,Kang:2012sy,Cao:2012fz,Ellwanger:2012ke,Beskidt:2013gia}. And last, but not least, the superpartner of the singlet provides an electroweak scale dark matter candidate consistent with all experimental data \cite{Hugonie:2007vd,Kozaczuk:2013spa,Ellwanger:2014dfa,Beskidt:2014oea,Cao:2016nix,
Xiang:2016ndq,Beskidt:2017xsd,Ellwanger:2018zxt}.

We focus on the semi-constrained NMSSM \cite{Djouadi:2008yj,Ellwanger:2009dp,Kowalska:2012gs}, a well motivated subspace of the general NMSSM assuming unification at the GUT scale. In this case   all radiative corrections are integrated up to the GUT scale using the renormalization group equations (RGEs). Especially, radiative electroweak symmetry breaking and the important fixed point solutions for the trilinear couplings are taken into account, thus avoiding trilinear coupling values at the SUSY scale not allowed by  solutions of the RGEs.
As the name fixed point solution indicates, the values of the trilinear couplings at the SUSY scale are largely independent of GUT scale values for given Higgs and SUSY masses.

For the lightest Higgs boson in the MSSM one expects SM-like couplings in the so-called decoupling regime of the Higgs bosons, which is obtained  if the heavy Higgs masses are well beyond the electroweak mass scale  \cite{Djouadi:2005gj}. This decoupling regime is presumably reached, given the non-observation of any signs of SUSY or additional Higgs bosons at the LHC. In the NMSSM one expects,  instead of a single light Higgs boson, two light Higgs bosons, where one is singlet-like  and the second one is expected to be again SM-like in what is called the alignment limit, which is the equivalent of the decoupling limit in the MSSM \cite{Carena:2015moc}. In the alignment limit both off-diagonal matrix elements for the mixing of the SM-like Higgs boson with the two other scalar Higgs bosons are  zero after including the one-loop corrections from the third generation.

Given that a singlet Higgs boson has by definition no quantum numbers of the SM it can couple to other particles only by mixing with the other Higgs bosons.   The mixing can lead to additional decay modes of the observed Higgs boson, if kinematically allowed, see e.g. Refs. \cite{King:2014xwa, Carena:2015moc,Beskidt:2016egy,Ellwanger:2017skc,Baum:2019uzg,Baum:2019pqc}. 
 Apart from new decay modes the NMSSM provides  additional possibilities for deviations from the SM expectation because of the mixing with the singlet.  We call all these NMSSM specific deviations  ``genuine" NMSSM deviations in contrast to the deviations from SUSY particles in loop diagrams, which are present for both, the MSSM and NMSSM.
  Observing deviations only in loop-induced processes  would be a clear sign of non-SM contributions in the loop diagrams, which can happen in both, the production and the decays. But the ``genuine" NMSSM deviations  are largely independent of the production mode. Therefore, it is important to  search for patterns of deviations, which  we call correlated deviations.  

We shortly introduce the NMSSM Higgs sector in Sect. \ref{theory}. The analysis method is described in Sect. \ref{analysis}. The results are presented in Sects. \ref{loops}, \ref{genuine}, \ref{single} and \ref{muscan}. The paper  provides a survey of expected correlations in deviations from SM expectations, both from SUSY particles in loop diagrams and from ``genuine" NMSSM effects. The latter show a completely different pattern of deviations in comparison with the ones from loop diagrams, namely correlated deviations between final states with fermions and bosons.  The origins of these correlations are  identified and the regions of  parameter space with correlated or anti-correlated deviations are determined using an efficient sampling technique.
It does not sample the   7D NMSSM parameter space with its spiky likelihood distribution, but samples  the smoother and lower dimensionality of the 3D Higgs mass space, which can be done in a deterministic way, as discussed  in detail in Appendix \ref{method}. Here the questions of coverage and uniqueness are addressed as well as the advantages in speed with respect to numerous investigations of the NMSSM phenomenology, using mostly a stochastic sampling of the parameter space  \cite{Ellwanger:2012ke,Gunion:2012zd,Cao:2013gba,Badziak:2013bda,Barbieri:2013nka,Guchait:2015owa,King:2014xwa,King:2012is,Potter:2015wsa,Bomark:2015hia,
Dermisek:2008uu,Bernon:2014nxa,King:2014xwa,Cao:2016uwt,Muhlleitner:2017dkd,Das:2016eob,Baum:2017gbj,Bandyopadhyay:2015tva}.
 Searching for  correlations in signal strengths deviating from SM expectations could help to disentangle the origins of possible deviations in  future data. At present all observed signal strengths of the 125 GeV Higgs boson are consistent with the SM expectations, but the experimental errors are still above 50\% for the channels of interest, as summarized in Appendix \ref{error}.

\section{The Higgs sector in the semi-constrained NMSSM}
\label{theory}

 Within the NMSSM the Higgs fields consist of the two Higgs doublets  ($H_u, H_d$), which appear in the MSSM as well, but together with
an additional complex Higgs singlet $S$ \cite{Ellwanger:2009dp}
The neutral components from the two Higgs doublets and singlet mix to form three physical CP-even scalar bosons and two physical CP-odd pseudo-scalar bosons. The mass eigenstates of the neutral Higgs bosons are determined by the diagonalization of the mass matrix, so the scalar Higgs bosons $H_i$, where the index $i$ increases with increasing mass, are mixtures of the CP-even weak eigenstates $H_d, H_u$ and $S$: 
\begin{eqnarray}\label{eq0}
H_i=S_{id}  H_d  + S_{iu}  H_u  + S_{is}  S, 
\end{eqnarray}
where $S_{ij}$ with $i=1,2,3$ and $j=d,u,s$ are the elements of the Higgs mixing matrix, which can be found in Appendix \ref{higgsmixing}.
The Higgs mixing matrix elements enter the Higgs couplings to quarks and leptons of the third generation:
\begin{align}
H_i t_L t_R^c &:  -\frac{h_t}{\sqrt{2}}S_{iu} & h_t &= \frac{m_t}{v \sin\beta},\notag\\
H_i b_L b_R^c &: \frac{h_b}{\sqrt{2}}S_{id} & h_b &= \frac{m_b}{v \cos\beta},\label{coupling}\\
H_i \tau_L \tau_R^c &: \frac{h_\tau}{\sqrt{2}}S_{id} & h_\tau &= \frac{m_\tau}{v \cos\beta},\notag
\end{align}
where $h_t$, $h_b$ and $h_\tau$ are the corresponding Yukawa couplings, $\tan\beta$ corresponds to the ratio of the vev*s of the Higgs doublets, i.e. $\tan\beta=v_u/v_d$. 
The relations include the quark and lepton masses $m_t$, $m_b$ and $m_\tau$ of the third generation and $v^2=v_u^2+v_d^2$. The couplings to fermions of the first and second generation are analogous to Eq. \ref{coupling} with different quark and lepton masses. The couplings are crucial for the corresponding branching ratios and cross sections for each Higgs boson. 
While the couplings to fermions are proportional to either the mixing element from the up- or down-type state, namely $S_{iu}$ or $S_{id}$ as can be seen from Eq. \ref{coupling}, the couplings to gauge bosons consist of a linear combination of both mixing elements:

\begin{align}
H_i Z_\mu Z_\nu &:  g_{\mu\nu}\frac{g_1^2+g_2^2}{\sqrt{2}} \left ( v_d S_{id} + v_u S_{iu} \right ),\notag\\
H_i W_\mu^+ W_\nu^- &: g_{\mu\nu}\frac{g_2^2}{\sqrt{2}}\left ( v_d S_{id} + v_u S_{iu} \right ),\label{coupling2}
\end{align} 
Thus the couplings to fermions and bosons are  correlated, leading to correlated signal strengths as well.

\begin{figure}
\begin{center}
\hspace{-1cm}
\includegraphics[width=0.5\textwidth]{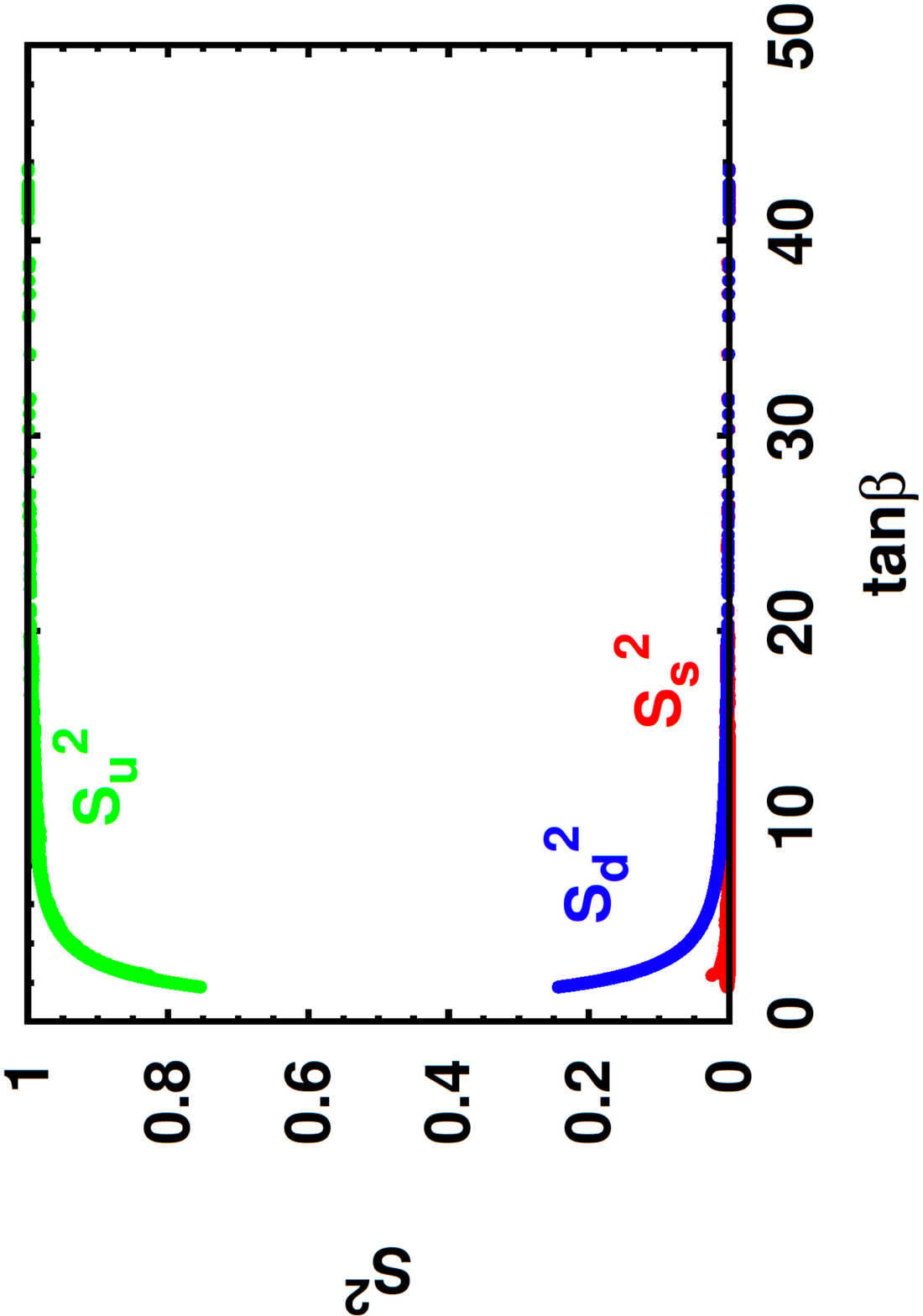}
\caption[]{ 
The mixing matrix elements squared for the SM-like Higgs boson as function of $\tan\beta$. They correspond to the matrix elements in Eq. \ref{eq0} for $i=2$. The  contributions from $S_d^2$, $S_u^2$ and $S_s^2$ have been indicated. The singlet component $S_s^2$ is small for most regions of parameter space and is here shown by the thin (red) line at the bottom of the figure. The points result from the fit for each mass combination in the 3D Higgs mass space, as described in Sect. \ref{analysis}.    
}
\label{fig1}
\end{center}
\end{figure}

The reduced couplings, meaning couplings in units of the SM couplings, only include  the Higgs mixing matrix elements and $\tan\beta$:
\begin{align}
c_u&=\frac{S_{iu}}{\sin\beta} & c_d&=\frac{S_{id}}{\cos\beta} &c_{W/Z}&= \cos\beta \cdot S_{id} + \sin\beta \cdot S_{iu}, \label{coupling3}
\end{align} 
where $c_{W/Z}$,  $c_u$ and  $c_d$ denote the couplings to vector  bosons, up-type and down-type fermions, respectively.

The loop diagrams  needed for the reduced couplings to gluons $c_{gluon}$ and photons $c_\gamma$ are parametrized as effective couplings  within NMSSMTools \cite{Das:2011dg}.  
If the reduced couplings are SM-like, the Higgs mixing matrix elements  are adjusted such that the reduced couplings are one as function of $\tan\beta$, which is possible  if one chooses $S_{iu}\approx \sin\beta$ and $S_{id}\approx \cos\beta$, as is obvious from Eq. \ref{coupling3}.  In this case the couplings to gauge bosons take automatically SM-like values: $c_{W/Z}=\cos\beta\cdot S_{id}+\sin\beta \cdot S_{iu} \approx \cos^2\beta + \sin^2\beta \approx 1$. 
The components of the Higgs mixing matrix elements squared of the 125 GeV SM-like Higgs boson are shown in Fig. \ref{fig1} as function of $\tan\beta$. The points result from the fit for each mass combination in the 3D Higgs mass space, as described in Sect. \ref{analysis} and indicating that that multiple values of $\tan\beta$ can describe the signal strengths and mass of the observed 125 GeV Higgs boson. This is not surprising, since the reduced coupling strengths to fermions depend only on the {\it ratio} of the matrix element and either $\sin\beta$ or $\cos\beta$, as shown by Eq. \ref{coupling3}.
 One observes  that $S_{u}$ is the dominant component in the linear combination of Eq. \ref{eq0} for the SM-like Higgs boson for $\tan\beta>5$. In contrast, for the heavy Higgs $H_3$ the dominant component is the down-type component, as can be seen from the term $S_{3d}>0.97$  for the benchmark point in Table \ref{t3} of Appendix \ref{output}, which will be discussed later in detail. The square of the singlet component is hardly visible and represented by the thin (red) line at the bottom of the figure.

The cross section times BR relative to the SM values is also known as the signal strength $\mu$ and defined as: 
\begin{align}
\mu_j^i=\frac{\sigma_i \times BR_j}{(\sigma_i \times BR_j)_{SM}}=c_i^2 \cdot \frac{BR_j}{(BR_j)_{SM}}.\label{coupling4} 
\end{align}

So the signal strength is obtained by the production cross section $\sigma_i$ for mode $i$ times  the corresponding  BR for decay $j$, each normalized to  the SM expectation. Normalized cross sections  are called reduced cross sections, which are given by the square of the reduced couplings $c_i$. In the following we focus on the main Higgs boson production modes with the following  reduced couplings $c_i$: the effective reduced gluon coupling $c_{gluon}$ for gluon fusion (ggf), $c_{W/Z}$ for vector boson fusion (VBF) and Higgs Strahlung (VH) and $c_u$ for top fusion (tth). We consider two \textit{fermion} final states (b-quarks and $\tau$-leptons) and two \textit{boson} final states ($W/Z$ and $\gamma\gamma$) for  different production modes.  VBF and VH share the same reduced couplings, so they can be combined to VBF/VH. This leads  to 8 signal strengths in total, namely four for \textit{fermionic} final states and four for \textit{bosonic} final states:   
\begin{align}
\mu_{fermions}&:\mu_{\tau\tau}^{VBF/VH}&, &\mu_{\tau\tau}^{ggf}&, &\mu_{bb}^{VBF/VH}&, &\mu_{bb}^{tth},\notag\\
\mu_{bosons}&:\mu_{ZZ/WW}^{VBF/VH}&, &\mu_{ZZ/WW}^{ggf}&, &\mu_{\gamma\gamma}^{VBF/VH}&, &\mu_{\gamma\gamma}^{ggf}.\label{coupling5}
\end{align}

In addition, these eight signal strengths can be divided into  signal strengths for processes with and without loop diagrams: 
\begin{align}
\mu_{no-loop}&:\mu_{\tau\tau}^{VBF/VH}&, &\mu_{ZZ/WW}^{VBF/VH}&, &\mu_{bb}^{VBF/VH}&, &\mu_{bb}^{tth},\notag\\
\mu_{loop}&:\mu_{\tau\tau}^{ggf}&, &\mu_{ZZ/WW}^{ggf}&, &\mu_{\gamma\gamma}^{VBF/VH}&, &\mu_{\gamma\gamma}^{ggf}.\label{coupling6}
\end{align}

\begin{figure}
\begin{center}
\hspace{-1cm}
\includegraphics[width=0.6\textwidth]{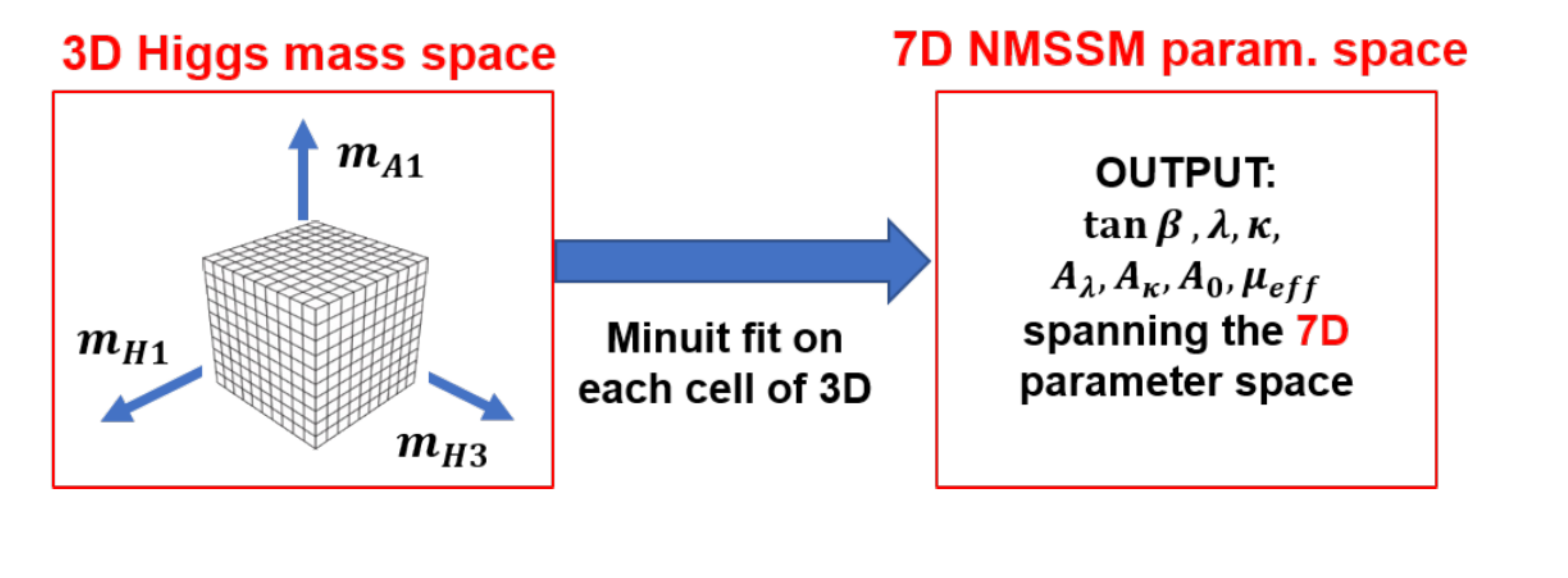}
\caption[]{ Sketch of the sampling technique to determine the allowed NMSSM parameter space. The sampling is done by performing
 a Minuit fit on each cell of the 3D Higgs mass space (left box) to determine the corresponding seven free NMSSM parameters (right box). The corresponding constraints entering the fit are listed in Table \ref{t1-diff} and detailed in Appendix \ref{method}. The relation between the NMSSM parameters and the Higgs masses is encoded in the public software package NMSSMTools \cite{Das:2011dg}. The second-lightest Higgs boson is chosen to be the 125 GeV Higgs boson, but we repeat the fit in case $m_{H_1}$=125 GeV. Then the $m_{H_1}$  becomes an $m_{H_2}$ axis in the grid on the left. The 3D Higgs mass space can be restricted to the experimentally accessible Higgs masses and the low dimensionality allows a fit to each cell of the grid. Avoiding stochastic sampling guarantees complete coverage of the Higgs mass space. In addition, each point on the grid can be fitted in parallel, so the  complete sampling can be done in the order of hours on a  cluster with parallel processors. 
}
\label{fig2}
\end{center}
\end{figure}
\begin{figure}
\begin{center}
\small{\boldmath$m_0=m_{1/2}$}  \small{\boldmath$=0.7$ \textbf{TeV}} \\
\includegraphics[width=0.32\textwidth]{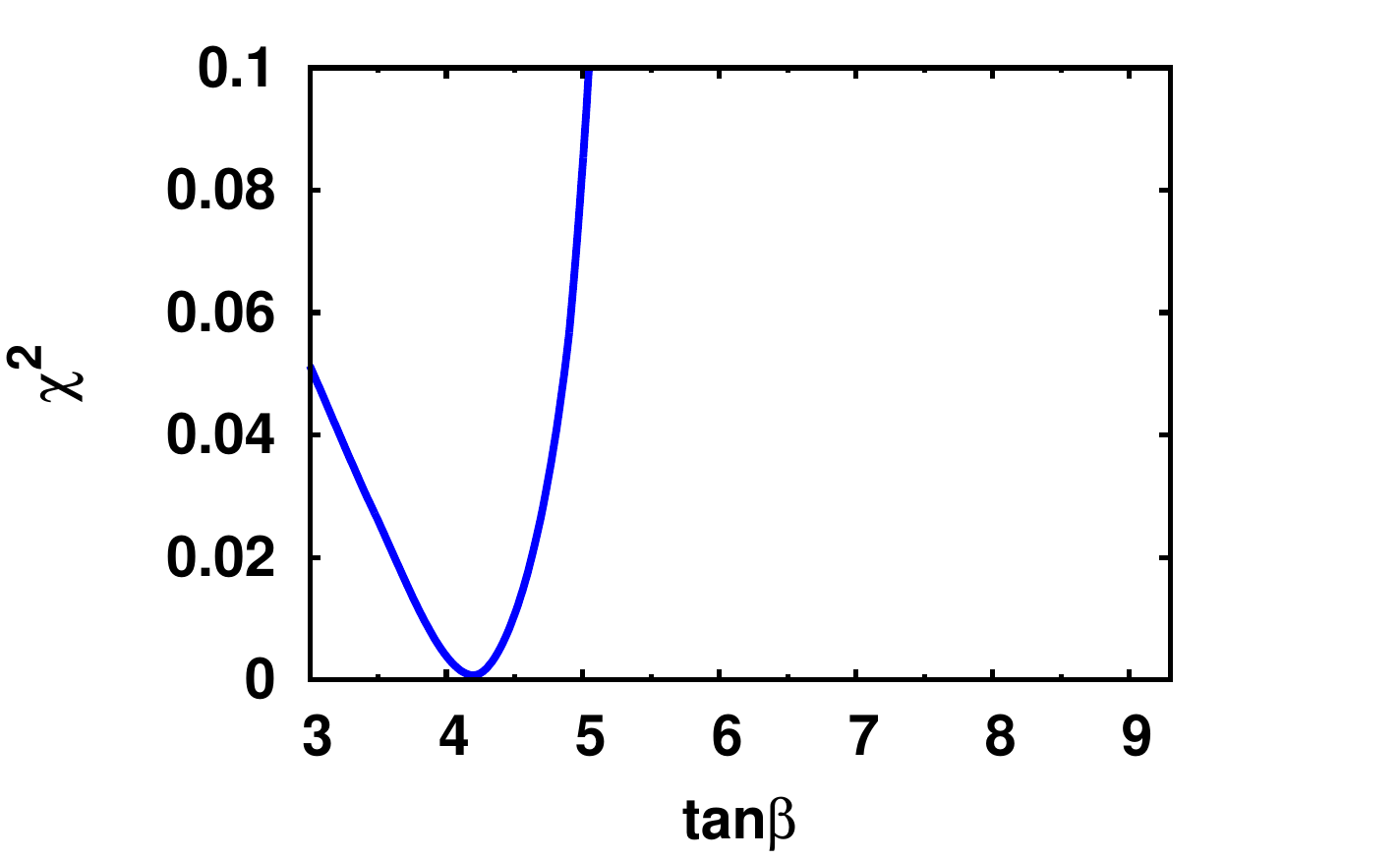} 
\includegraphics[width=0.32\textwidth]{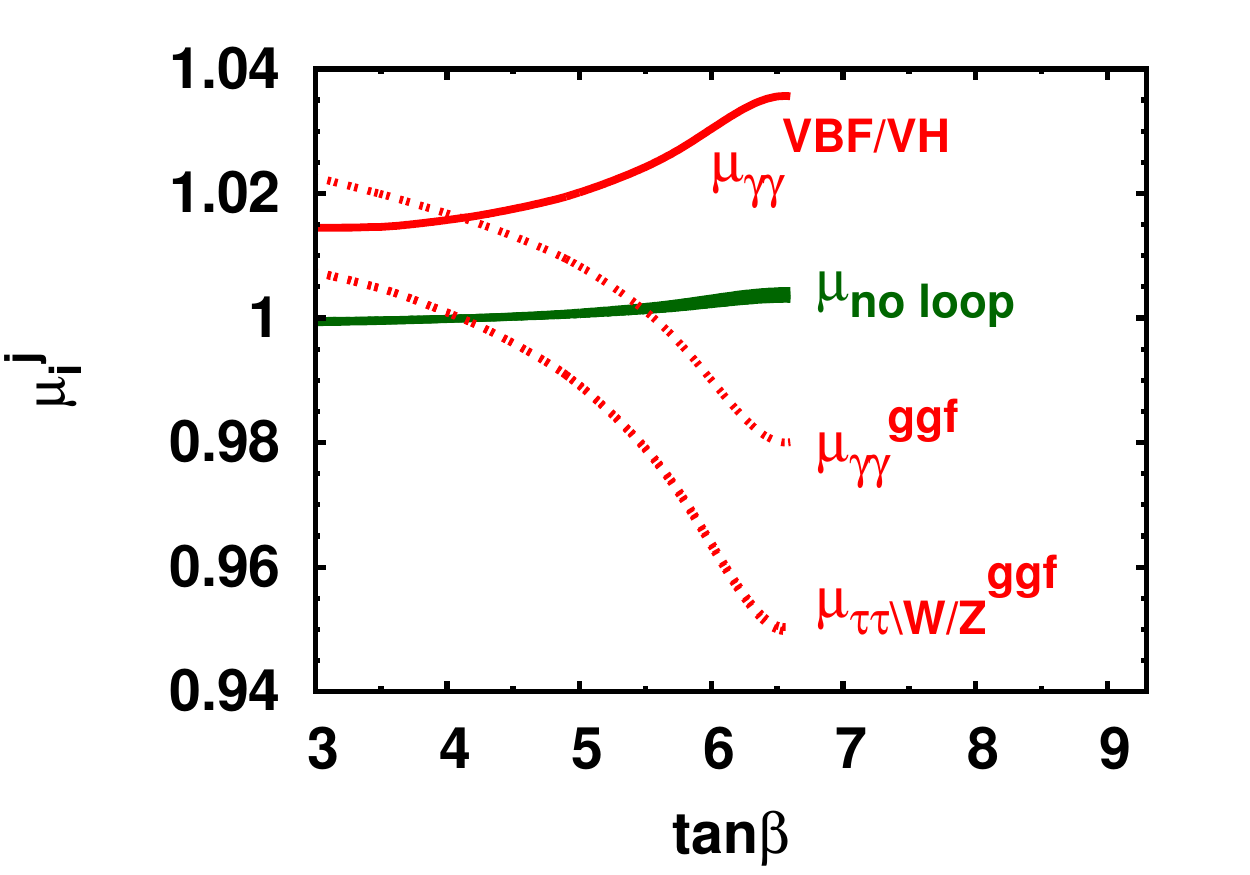}  
\includegraphics[width=0.32\textwidth]{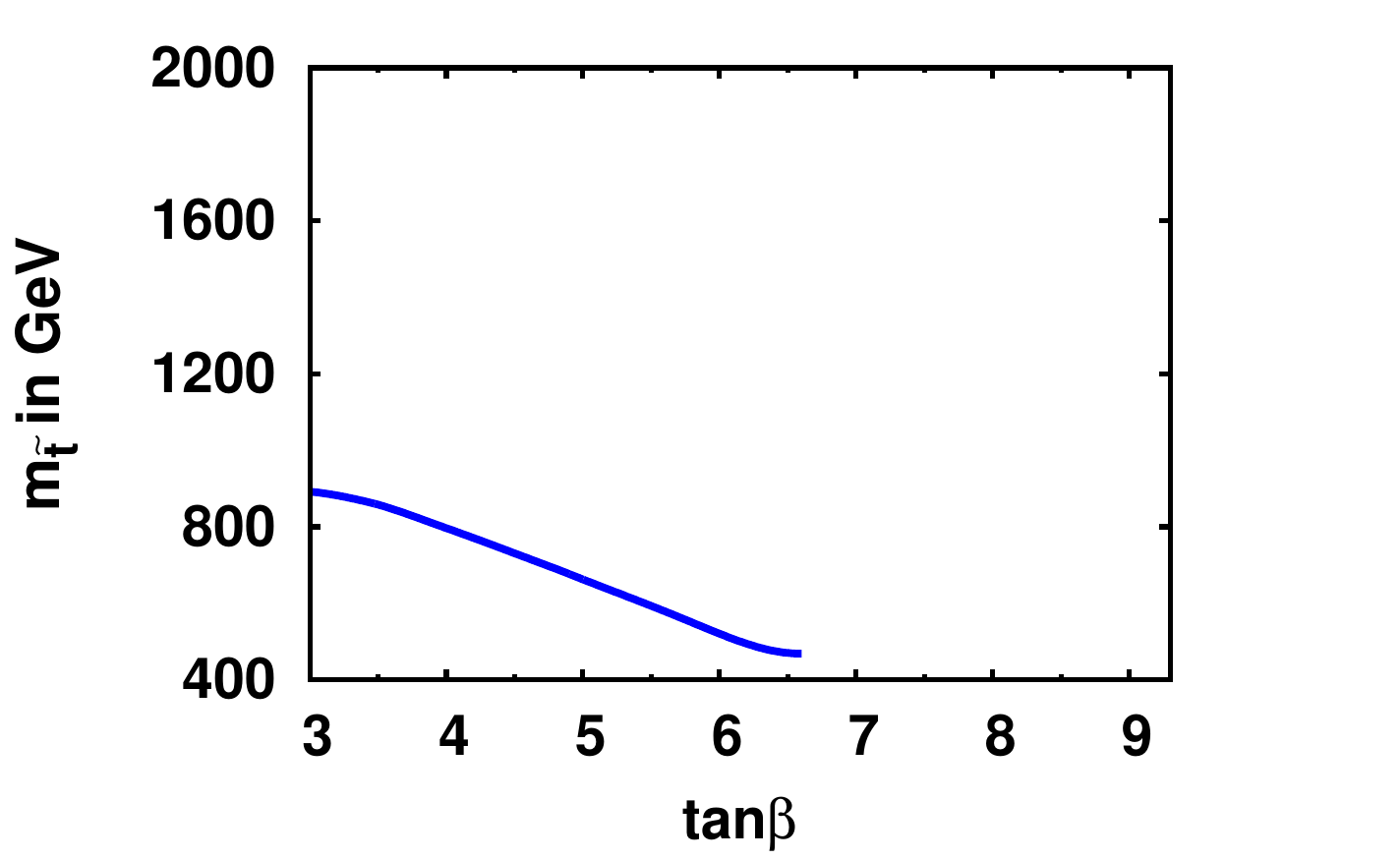} \\
\small{\boldmath$m_0=m_{1/2}$}  \small{\boldmath$=1.0$ \textbf{TeV}} \\
\includegraphics[width=0.32\textwidth]{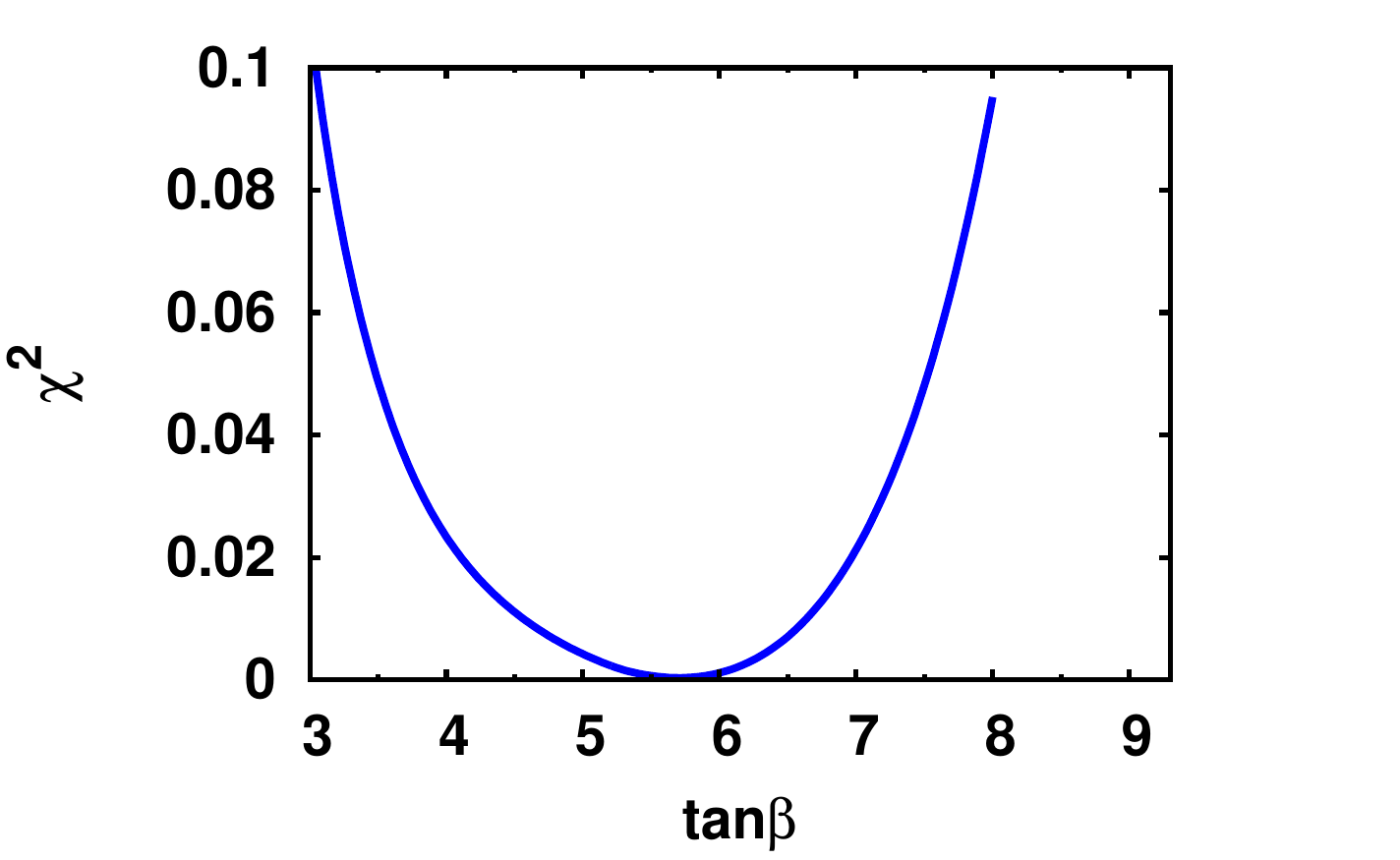} 
\includegraphics[width=0.32\textwidth]{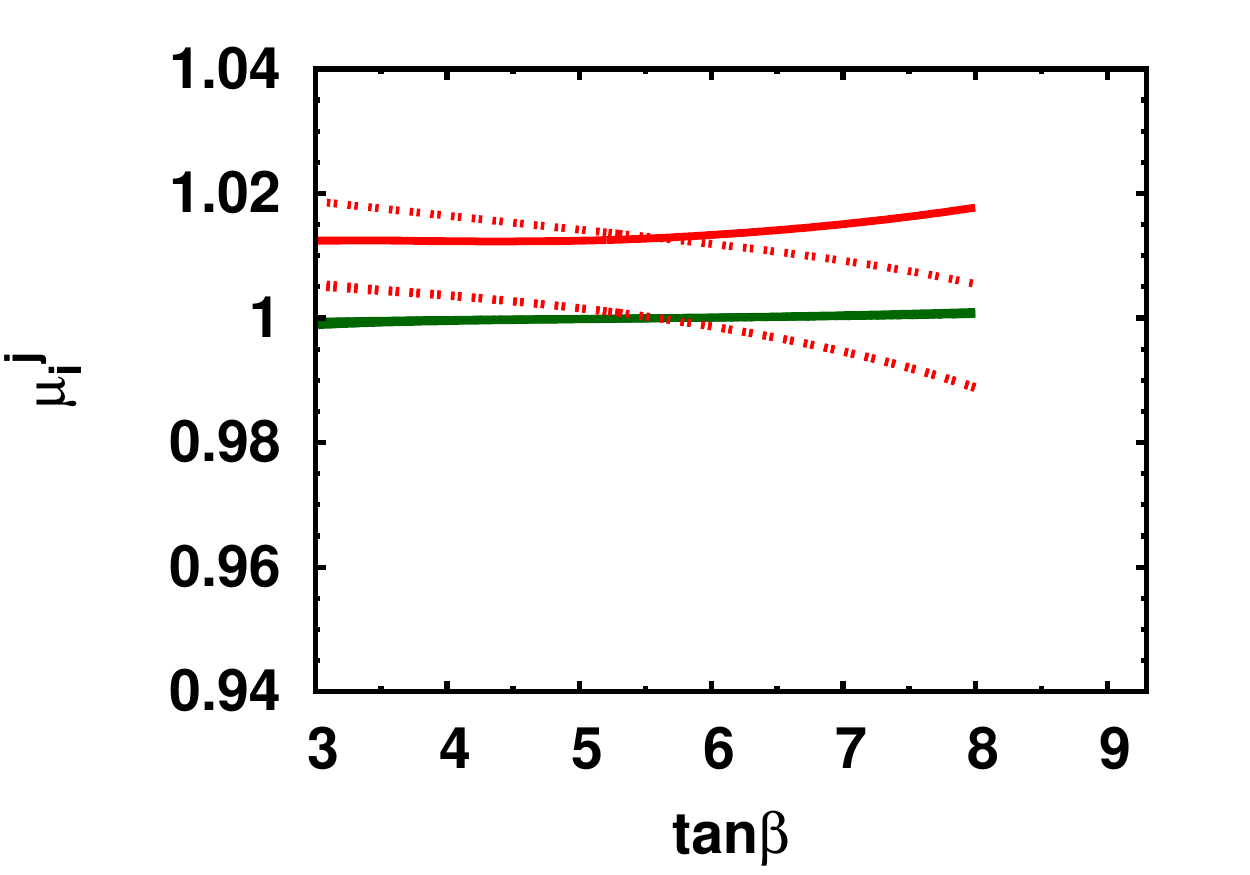}  
\includegraphics[width=0.32\textwidth]{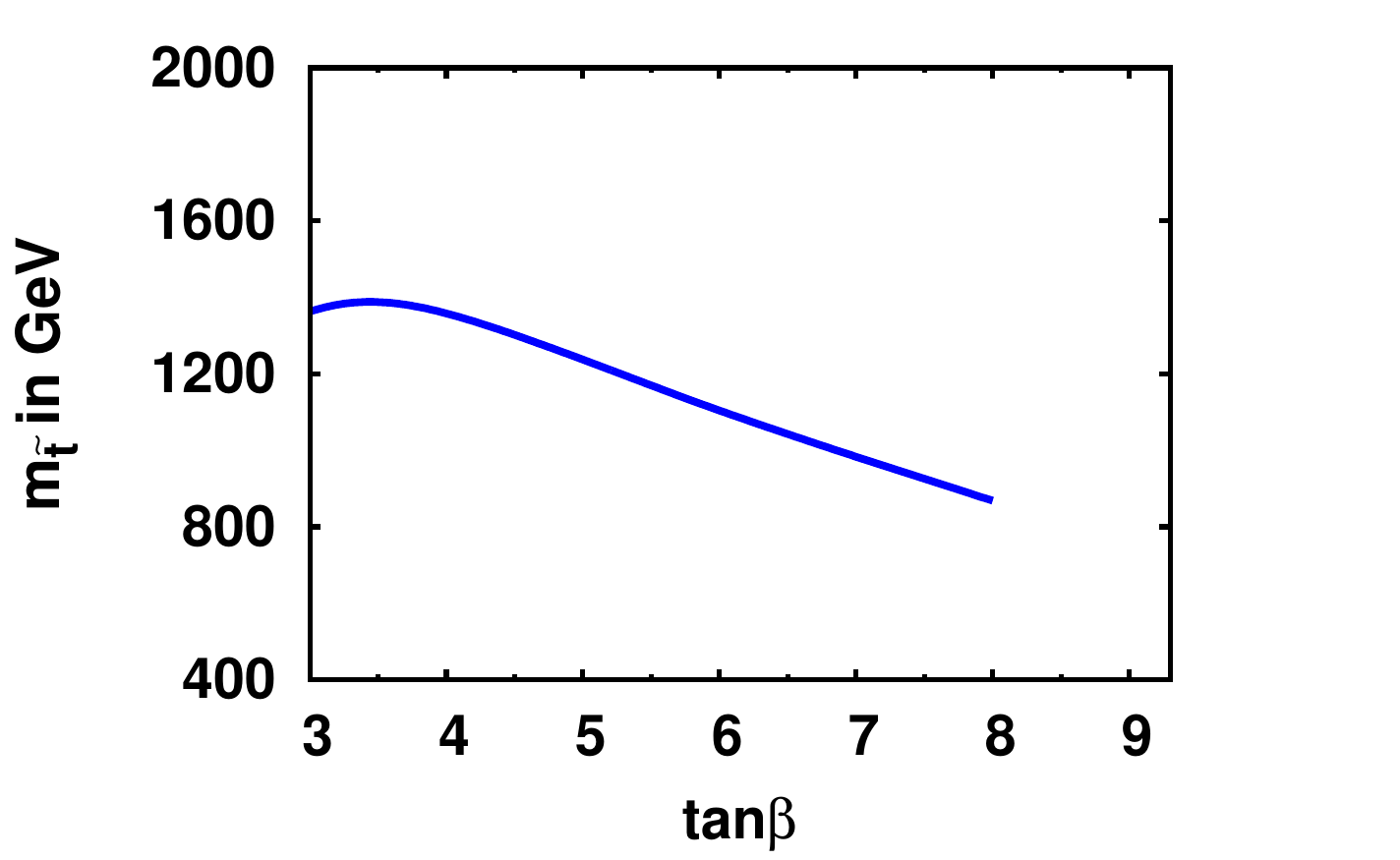} \\
\small{\boldmath$m_0=m_{1/2}$}  \small{\boldmath$=1.3$ \textbf{TeV}} \\
\includegraphics[width=0.32\textwidth]{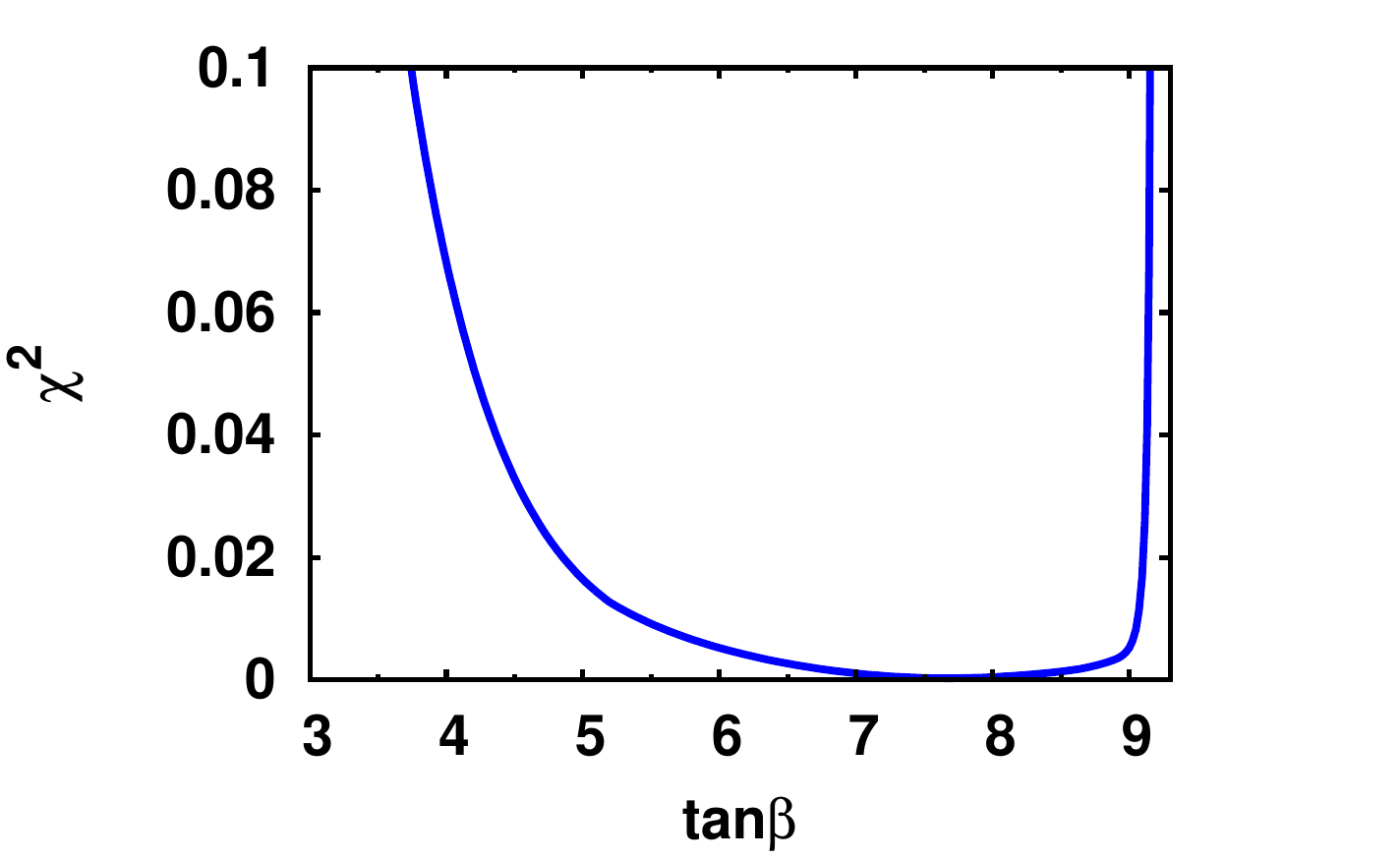} 
\includegraphics[width=0.32\textwidth]{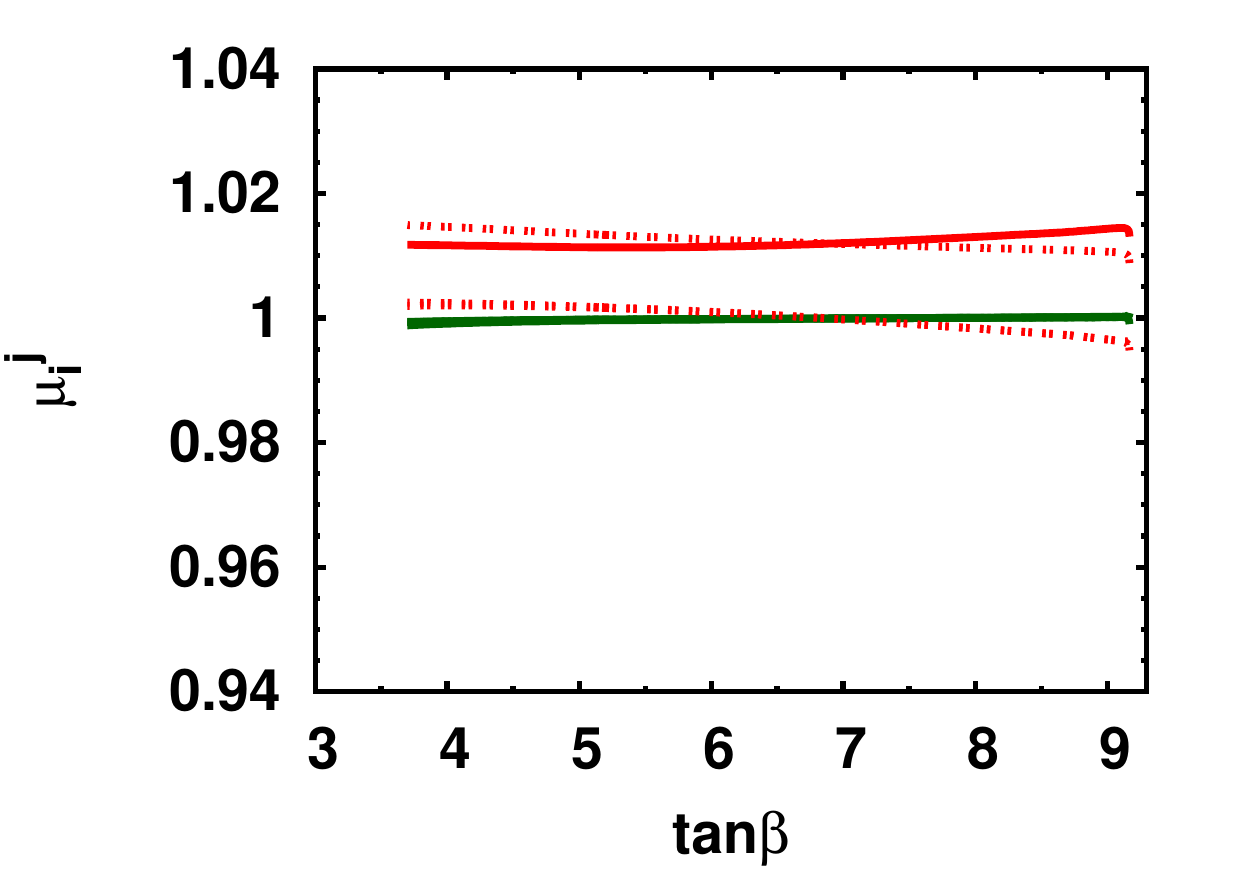}  
\includegraphics[width=0.32\textwidth]{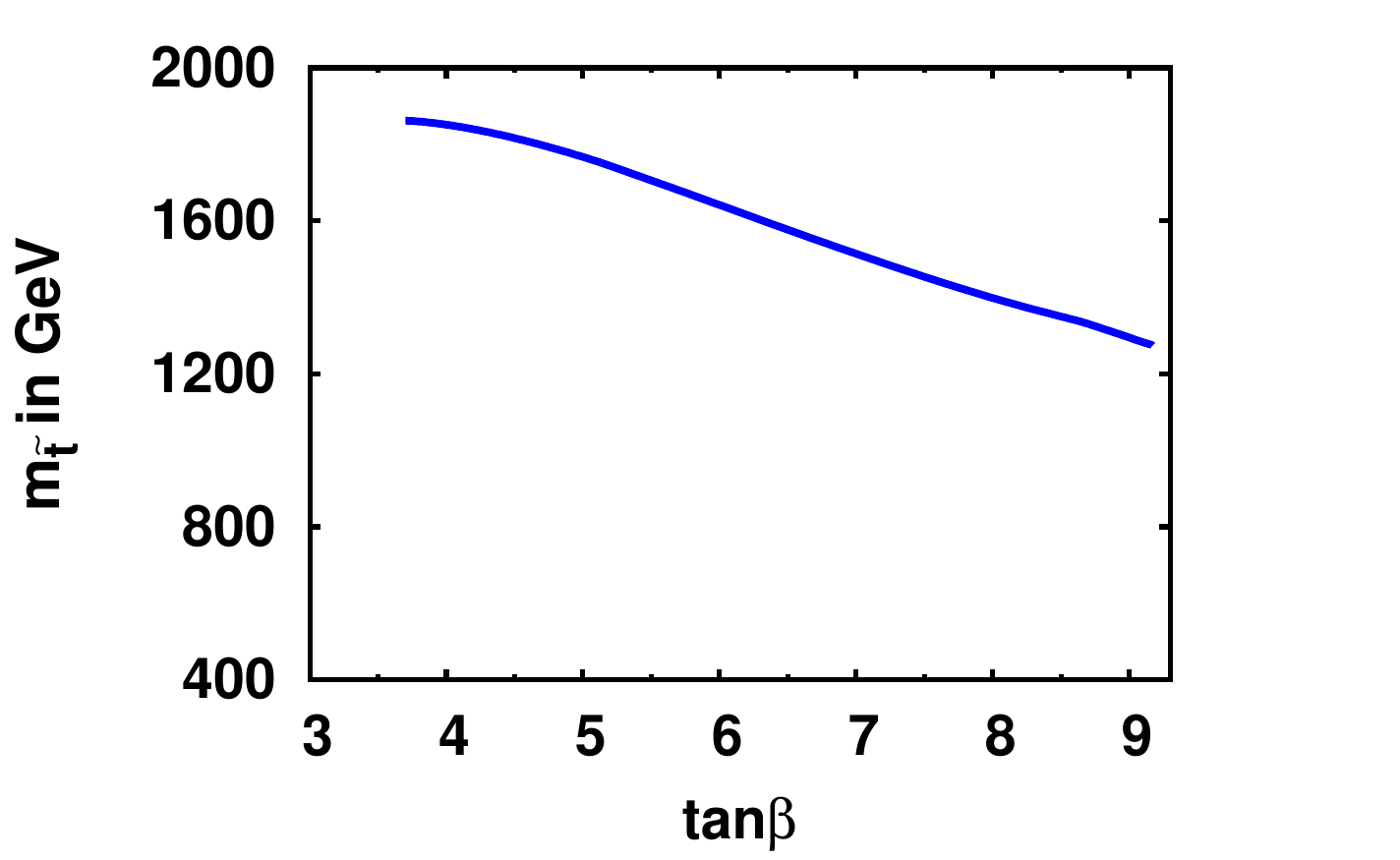} \\
\end{center}
\caption[]{
The left side shows the $\chi^2$ distribution as function of $\tan\beta$ for $m_0=m_{1/2}=$ 0.7/1.0/1.3 TeV from top to bottom, respectively. The fits are for an exemplary  Higgs mass combination  of $m_{H1}=90$ GeV, $m_{H3}=1000$ GeV, $m_{A1}=200$ GeV. The signal strengths including gammas and/or gluons deviate from the SM prediction, since the corresponding reduced couplings, and hence the signal strengths, are sensitive to SUSY contributions. These contributions vary with the mass of the stop quark, which varies strongly as function of  $\tan\beta$ and the SUSY mass scales, as can be seen from the right panels for $m_0=m_{1/2}=$ 0.7/1.0/1.3 TeV, respectively.
}
\label{fig3}
\end{figure}
\begin{figure}[!h]
\begin{center}
\includegraphics[width=0.45\textwidth]{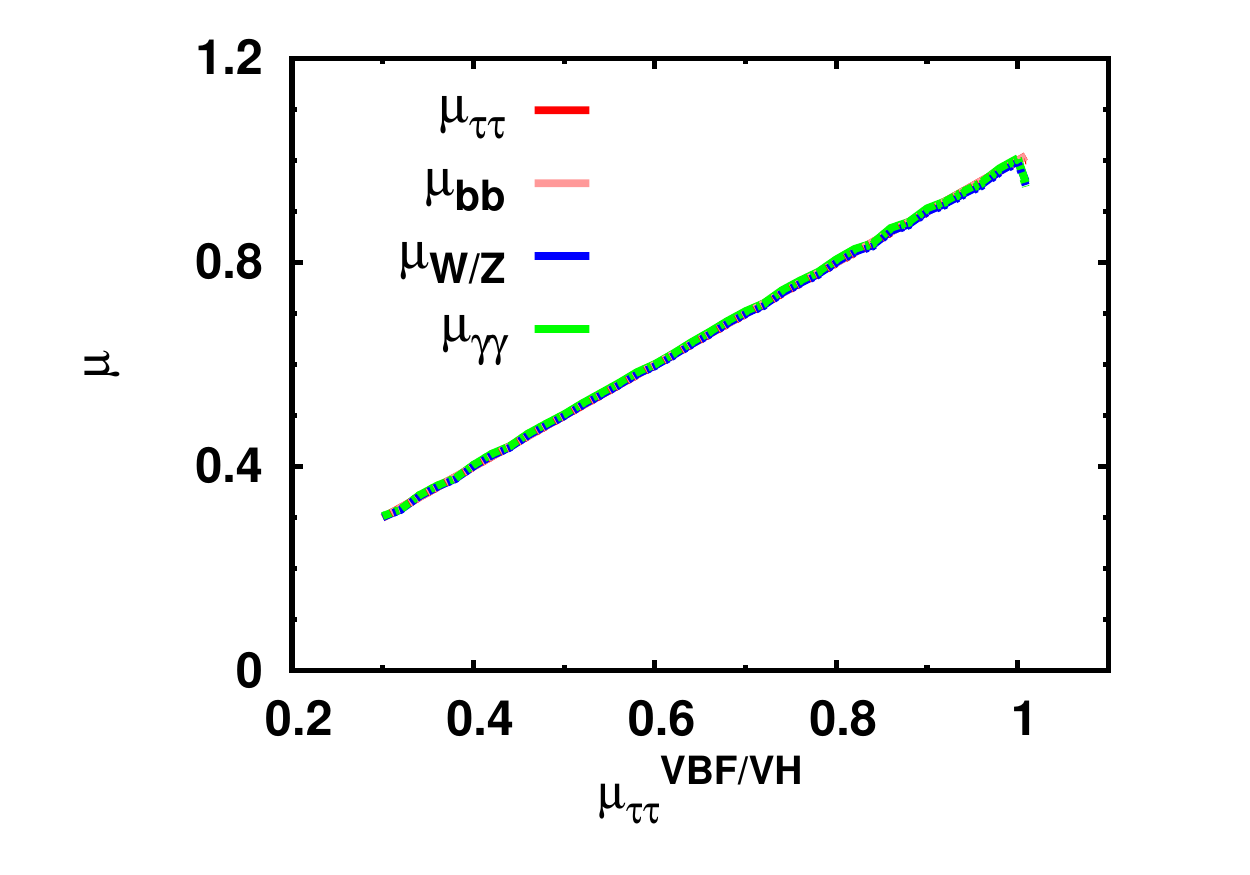}  
 \includegraphics[width=0.45\textwidth]{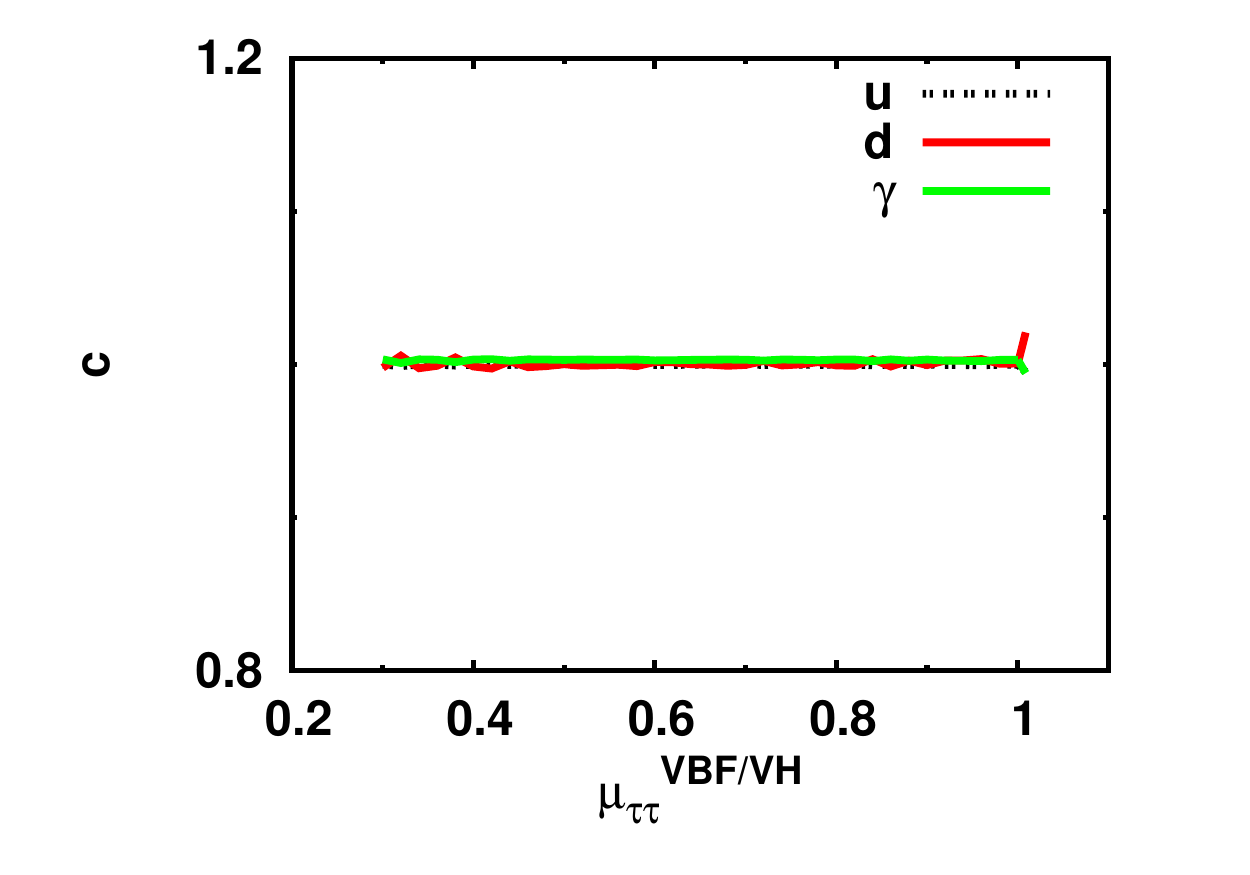} \\
(a) \hspace{0.45\textwidth} (b) \\
\includegraphics[width=0.45\textwidth]{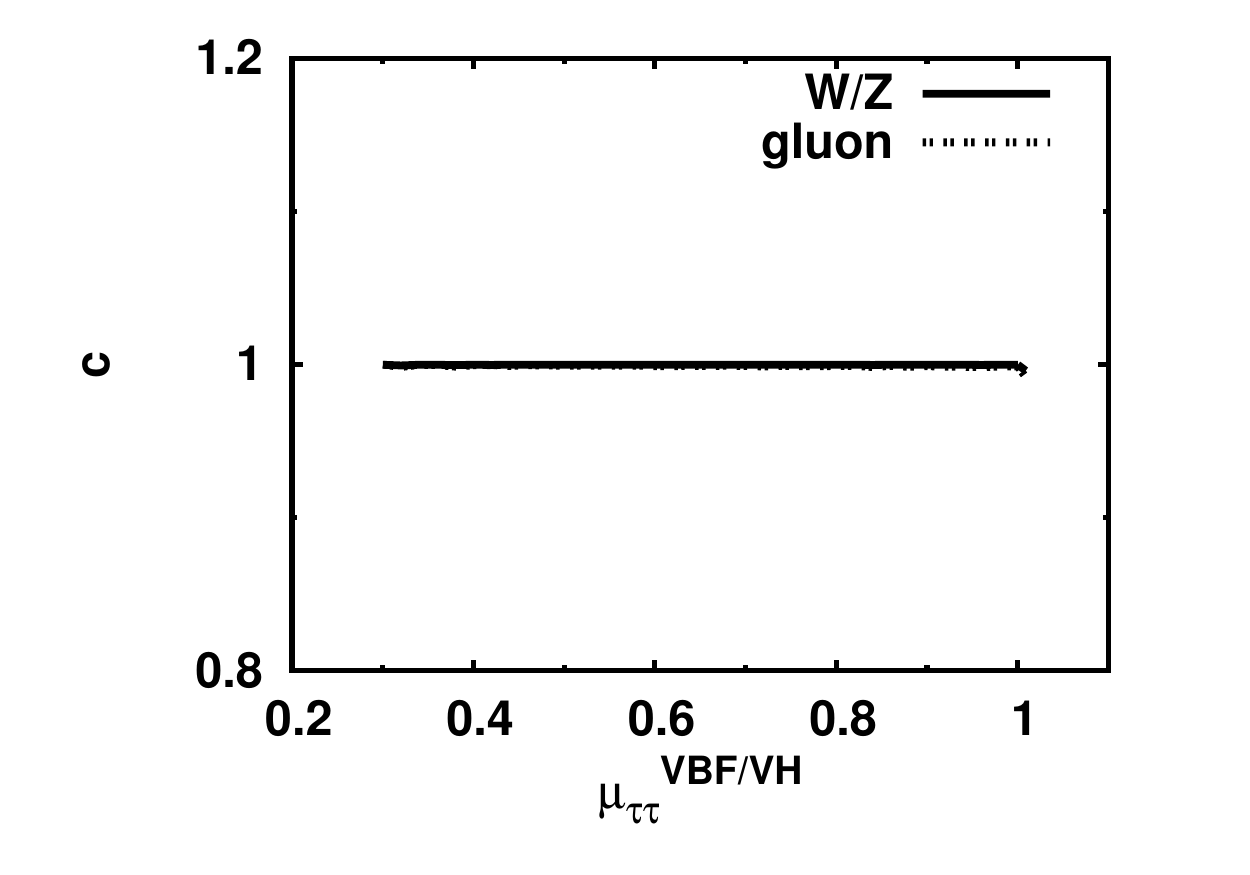}  
 \includegraphics[width=0.45\textwidth]{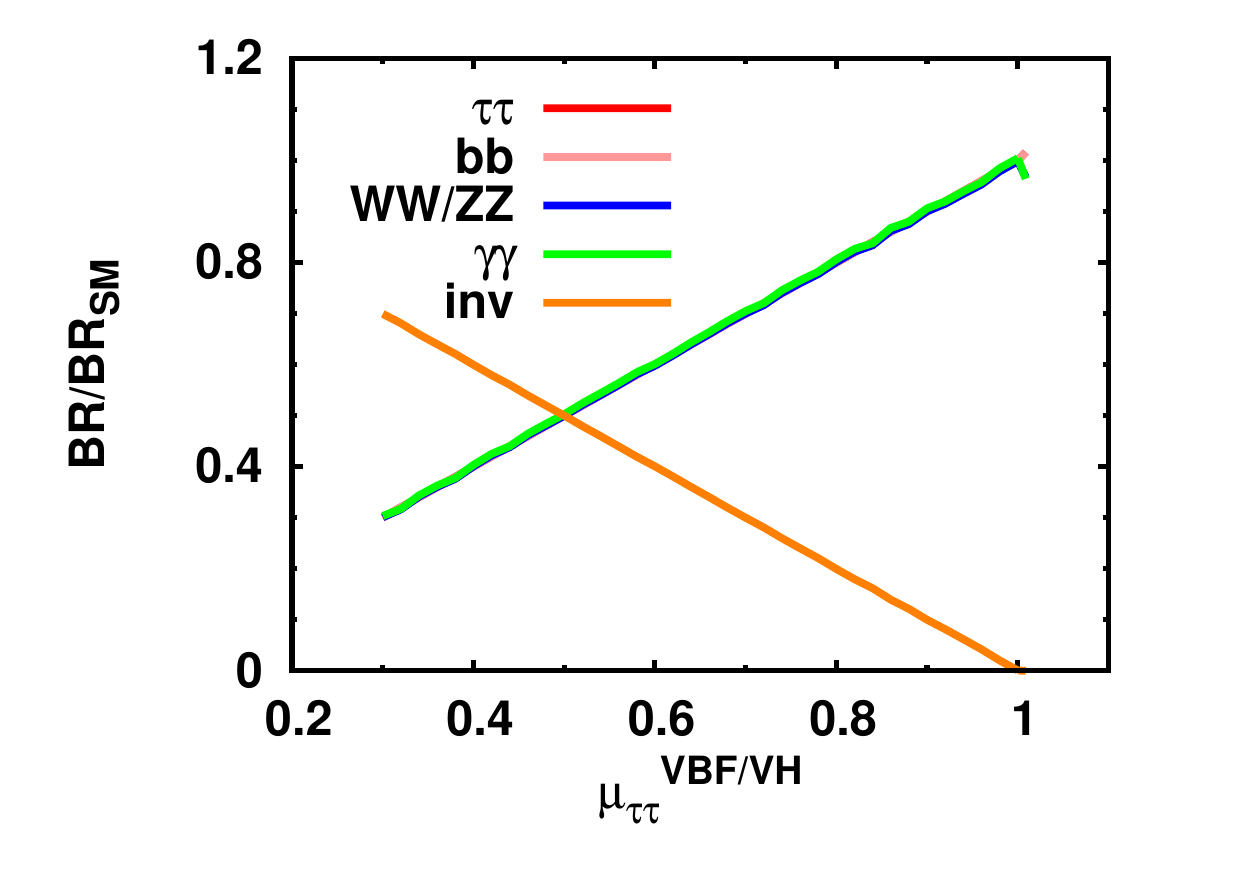} \\
(c) \hspace{0.45\textwidth} (d) \\
\caption[]{The figure  demonstrates  deviations of SM-like couplings in case  decays into neutralinos are allowed (CASE Ia as discussed in Sect. \ref{I}).  All signal strengths in the VBF/VH production mode deviate in the same way as   $\mu_{\tau\tau}^{VBF/VH}$, which was required to deviate from the SM expectation by  fitting it to a value $\mu_{theo}<1$. This is shown by the overlapping lines in the top left panel (a).   The couplings $c$ in panels (b) and (c) are defined in Eq. \ref{coupling3} for the  lines denoted by u, d, W/Z, while the lines denoted by gluon and $\gamma$ are the effective couplings of the 125 GeV boson to gluons and photons. The fact that the coupling to gluons in (d) overlaps with the coupling to W/Z implies that the influence from stop quarks is small for the chosen value of $\mzero=\mhalf$=1000 GeV. Deviations in signal strengths can either come from deviations of the couplings or from the BRs. Here it is clearly caused by the BRs, since all reduced couplings  in panels (b) and (c)   correspond to the SM expectation of one.   This in contrast to the BRs into invisible final states (in this case neutralinos), shown by the orange line with a negative slope in panel (d),  so this BR increases with decreasing values of $\mu_{\tau\tau}^{VBF/VH}$.  The invisible decays increase the total width, which reduces the BRs for all other channels in a correlated way (overlapping lines with positive slope in panel (d)). This leads to the same (=correlated) change in all signal strengths (overlapping lines with positive slope in panel (a)) given the constant couplings in panels (b) and (c). These results are independent of the production modes, as demonstrated by the overlap of the solid and dashed lines in (a) representing the signal strengths for the VBF/VH and ggf production mode, respectively. 
}
\label{fig4}
\end{center}
\end{figure} 
\section{Analysis method}
\label{analysis}

As mentioned in the Introduction we concentrate on the semi-constrained NMSSM\cite{Djouadi:2008yj,Ellwanger:2009dp}, which has in total nine free parameters: 

\begin{equation}
 \mzero,~ \mhalf,~A_0,~ \tan\beta,  ~ \lambda, ~\kappa,  ~A_\lambda, ~A_\kappa, ~\mu_{eff}.
\label{parameter}
\end{equation}

\begin{table}
\centering
\caption{ \label{t1-diff}
Differences of the fitting procedures searching for loop-induced  deviations (middle column) and ``genuine" NMSSM deviations (last column). In the first case the signal strengths are required to be one ($\mu_{no-loop}=\mu_{theo}=1$), while in the second case deviations from one for a selected channel are imposed ($\mu_{sel}=\mu_{theo}$), if $\mu_{theo}\neq 1$. From the fitted parameters one can calculate other signal strengths and check for correlations, like the correlations between $\mu_{no-loop}$ and $\mu_{loop}$ in the first case and the correlation between $\mu_{sel}$ and all other signal strengths in the second case.  The listed input parameters are for the case that $m_{H_2}=$125 GeV, but the fits have been repeated for the case $m_{H_1}=$125 GeV. The $\chi^2$ constraints are detailed in Eq. \ref{eq5} in  Appendix \ref{constraints}.
}
\begin{tabular}{l|p{4.7cm}|p{4.7cm}}
	\hline\noalign{\smallskip}
	Procedure & SM-like 125 GeV Higgs boson& non-SM-like \\
	\noalign{\smallskip}\hline\noalign{\smallskip}
	Input & $m_{A_1}, m_{H_1}, m_{H_3}$, $m_{H_2}=$125 GeV, all $\mu_{no-loop}=\mu_{theo}=1$ & $m_{A_1}, m_{H_1}, m_{H_3}$, $m_{H_2}=$125 GeV, $\mu_{sel}=\mu_{theo}\neq 1$\\
	Output &  $\tan\beta, A_0, A_\kappa, A_\lambda, \mu_{eff}, \lambda, \kappa$ & $\tan\beta, A_0, A_\kappa, A_\lambda, \mu_{eff}, \lambda, \kappa$ \\
	$\chi^2$ constraints & $\chi^2_{H_1}$, $\chi^2_{H_3}$, $\chi^2_{A_1}$, $\chi^2_{H_{2}}$, $\chi^2_{LEP}$, $\chi^2_{LHC}$ , $\chi^2_{\mu_{no-loop}}$& $\chi^2_{H_1}$, $\chi^2_{H_3}$, $\chi^2_{A_1}$, $\chi^2_{H_{2}}$, $\chi^2_{LEP}$, $\chi^2_{LHC}$, {$\chi^2_{\mu_{sel}}$} \\
	\noalign{\smallskip}\hline
\end{tabular}
\end{table}
The latter six parameters in Eq. \ref{parameter} enter the Higgs mixing matrix at tree level, see Appendix \ref{higgsmixing}, and thus form the 6D parameter space of the NMSSM Higgs sector. 
The coupling $\lambda$ represents the coupling between the Higgs singlet and doublets, while $\kappa$ determines  the self-coupling of the singlet. $A_{\lambda}$ and $A_{\kappa}$ are the corresponding trilinear soft breaking terms. $\mu_{eff}$ represents an effective Higgs mixing parameter, which is related to the vev of the singlet $s$ via the coupling $\lambda$, i.e. $\mu_{eff}=\lambda\cdot s$. 
In addition, we have the GUT scale parameters of the constrained MSSM (CMSSM) $\mzero$ and $\mhalf$ denoting the common mass scales of the spin 0 and spin {1/2} SUSY particles at the GUT scale. The trilinear coupling $A_0$ at the GUT scale is  correlated with $A_{\lambda}$ and $A_{\kappa}$, so fixing it would restrict the range of $A_{\lambda}$ and $A_{\kappa}$ severely. Therefore, $A_0$ is considered a free parameter in the fit, which leads in total to 7 free parameters and thus a 7D NMSSM parameter space.
For each set of the six NMSSM parameters the six Higgs boson masses and  couplings are completely determined. The masses of the heavy Higgs bosons  $A_2$ and $H^\pm$ can be derived from the mass of H$_3$ and the NMSSM parameters. They are approximately equal in the alignment limit, which is reached if they have masses well above the electroweak scale. Furthermore, one of the masses has to be 125 GeV. Then only 3 Higgs masses are free, e.g. $m_{A_1}$, $m_{H_1}$ and $m_{H_3}$. Each set of parameters in the 7D NMSSM parameter space determines a mass combination in the 3D Higgs mass space and their Higgs mixing matrix elements. Alternatively, one can try to reconstruct from a given mass combination in the 3D Higgs mass space with certain signal strengths the allowed region in the 7D NMSSM parameter space.  Here there is no unique solution, but one finds  regions with  equivalent solutions for masses and signal strengths.  This can be inferred already from Fig. \ref{fig1}, since here each entry corresponds to the same mass and signal strengths of the 125 GeV Higgs boson.  Given the miniscule differences between these almost degenerate solutions, we call them quasi-degenerate.  This is discussed  in  Appendix \ref{method} together with relevant questions for such  mapping of the 3D to the 7D space concerning the coverage. The sampling method has been used before in Refs\cite{Beskidt:2016egy,Beskidt:2017dil,Beskidt:2017xsd,Beskidt:2019mos}
 
The mapping of the 3D Higgs mass space to the corresponding 7D NMSSM parameter space can be obtained from a Minuit fit \cite{James:1975dr} - as sketched in Fig. \ref{fig2} - with the parameters of the 7D space as free parameters  to be fitted to the input, given by  the masses of the 3D space and one or more signal strengths. The relations between masses and parameters are encoded in  the publicly available software package NMSSMTools 5.2.0 \cite{Das:2011dg}, which can be obtained from the web site \cite{NMSSMToolsweb}. We switched on the radiative corrections available in NMSSMTools, which is important, since the NMSSM radiative corrections to the Higgs boson can lower the mass by several GeV, see e.g. Ref. \cite{Degrassi:2009yq}.

The inputs and  free parameters  have been summarized in Table \ref{t1-diff}.  The middle column describes the inputs and parameters for a fit to check if the SM expectations can be fulfilled, at least by the 125 GeV mass in combination with SM-like signal strengths for all four processes without loop diagrams, defined in Eq. \ref{coupling6}. This is what the input $\mu_{no-loop}=\mu_{theo}=1$ means, which is expected in the alignment limit \cite{Carena:2015moc}. The corresponding $\chi^2$ terms are defined in Appendix \ref{method}. One can study the effect of SUSY particles on the signal strengths $\mu_{loop}$ from the fitted   parameters found for $\mu_{no-loop}=\mu_{theo}=1$. The results are discussed in Sect. \ref{loops}. 

The constraint $\mu_{no-loop}=\mu_{theo}=1$ is not suitable, if one wants to study  deviations from SM-like signal strengths  for the no-loop processes, as expected for the ``genuine" NMSSM effects. For this we require  $\mu_{no-loop}=\mu_{theo} \neq 1$. However, requiring {\it all}  $\mu_{no-loop}$ to differ from one by the same amount turns out to be a too strong constraint, since we found regions with anti-correlated signal strengths for the no-loop processes. These can only be found by requiring a single $\mu_{no-loop}$ - called $\mu_{sel}$ -  to deviate from one and see what happens to all other signal strengths calculated from the fitted parameters.  The fits are performed as function of $\mu_{theo}$. The inputs and constraints for this case have been summarized in the last column of Table \ref{t1-diff}. For the signal strength deviating from the SM expectation ($\mu_{sel}$) we usually select $\mu_{\tau\tau}^{VBF/VH}$, but other choices from the four no-loop signal strengths in Eq. \ref{coupling6} can be taken as well, which leads to similar results. In comparison with the  fit in the middle column of Table \ref{t1-diff} one has three constraints less, since only one no-loop signal strength is required to equal $\mu_{theo}$ instead of all four. The reduction of the constraints still leads to converging fits, because the signal strengths are highly correlated, as will be discussed in Sect. \ref{single}, which means if one constrains  a single signal strength the others are determined as well.

\section{Loop-induced MSSM contributions}
\label{loops}

 As mentioned in the introduction, the SUSY particles contributing to loop diagrams can lead to correlations in the signal strengths.
 This is well known, but in order to compare these correlations with the correlations from ``genuine" NMSSM effects, we repeat here shortly the results from Ref. \cite{Beskidt:2019mos}, which serves  to illustrate the  analysis method too.
 
The left side of Fig. \ref{fig3}  shows the $\chi^2$ distribution as function of $\tan\beta$ for $m_0=m_{1/2}=$ 0.7/1.0/1.3 TeV from top to bottom, respectively for a given cell  in the left panel of Fig. \ref{fig2}. In this case    $m_{H1}=90$ GeV, $m_{H3}=1000$ GeV, $m_{A1}=200$ GeV was chosen. The main contribution to the $\chi^2$ is coming from the signal strengths, which are close to the SM expectation of one for a large range of $\tan\beta$ for processes without loop diagrams, as shown on the middle panels of Fig. \ref{fig3}. However, the signal strengths {\it including} loop contributions deviate from one because of the  SUSY contributions. The sign of the deviations depends on interferences in the diagrams. The deviations vary with $\tan\beta$, while  the stop masses vary with $\tan\beta$, as shown on the panels on the right for the different choices of  $m_0,m_{1/2}$. With increasing stop masses the signal strengths approach the SM expectation of one, as can be seen from  the middle column of Fig. \ref{fig3}. This is expected, since in the limit of infinite stop masses the loop corrections vanish.

    The shift in the minimum of the $\chi^2$ function to higher $\tan\beta$  in the left column of the middle and bottom row is  caused by the increase in the stop mass - shown in the last column -, which increases the stop  corrections to the 125 GeV Higgs boson mass. The broadening is caused by the larger stop mass and the larger value of $\tan\beta$, which both increase the MSSM tree level terms of the Higgs boson mass to a mass range, where it becomes less sensitive to $\tan\beta$. 

\section{``Genuine" NMSSM-induced contributions}
\label{genuine}

  Using the deterministic scan guaranteeing a complete coverage of the Higgs mass space we found three different regions with ``genuine" NMSSM deviations by checking in which regions  $\mu_{sel}\neq 1$  is allowed, as discussed in Sect. \ref{analysis}. The deviations are not restricted to processes including loops, as discussed above, but by correlations between final states with either  fermions or bosons, largely independent of the production mode. What was hard to understand initially: some regions had positive correlations between fermions and bosons, while others showed negative correlations. It turned out, that the correlations depended not only on the opening up of new decay channels in some regions, but in other regions  from the strength of the mixing between the 125 GeV Higgs boson and the singlet. The mixing is a strong function of the mass difference, so looking for correlations in possible deviations of the 125 GeV boson from SM expectations could give useful hints about the existence and mass of the singlet Higgs boson.   Before discussing the results of the scan over the whole parameter space  we present the  deviations in a representative mass point for each of the three regions, which we call CASE I, II and III. They have the following characteristics:

\begin{itemize} 
\item CASE I: \textit{Additional decays.} The decay of the 125 GeV Higgs boson with SM couplings into particles with a mass $m < 0.5 m_{Higgs}$ leads to modifications of the total width, which changes all BRs in a correlated way and thus leads to {\it correlated} deviations in the signal strengths.
\item CASE II: \textit{Small Higgs mixing between the 125 GeV  and the singlet  Higgs boson.}  A modification of the Higgs mixing matrix elements leads to anti-correlated deviations because the total width is not much affected by the small mixing. The BR to down-type \textit{fermions}  decreases,   which can only be compensated by an increase of the BRs to \textit{vector bosons} for an almost constant total width. This leads to {\it anti-correlated} signal strengths for \textit{fermionic} and \textit{bosonic}   final states.
\item CASE III: \textit{Strong Higgs mixing between the 125 GeV  and the singlet-like Higgs boson.}  A strong mixing and thus a large singlet component of the 125 GeV Higgs boson can be reached, if both masses are close to each other. This leads to correlated deviations of the signal strengths, because the singlet component of the 125 GeV Higgs boson does not couple to SM particles, so the reduced couplings decrease in a {\it correlated} way. 
\end{itemize}

\section{Examples of genuine NMSSM contributions}
\label{single}

Before discussing the results in the search for deviation from the scan over the whole Higgs mass space we first examplify  the three cases (CASE I, II, III) for a represenative point in the Higgs mass space, i.e. fixed Higgs masses, and vary $\mu_{theo}$ in a large range, since  the experimental errors on $\mu_{sel}\equiv \mu_{\tau\tau}^{VBF/VH}$ are typically larger than 50\%, as shown in Appendix \ref{error}.

\subsection{CASE I: Deviations by decays of the 125 GeV Higgs boson into non-SM light particles}\label{I}

For CASE I either $m_{\tilde{\chi}_1^0}$ and/or $m_{A_1}$ and/or $m_{H_1}$ have to be smaller than about 60 GeV to allow decays of the 125 GeV Higgs boson into pairs of these light particles, which leads to deviations from the SM expectation. Here we investigate the deviations from the SM for a specific mass combination in the grid of Fig. \ref{fig2} characterized by allowed decays into neutralinos (CASE Ia). In this case $m_{H1}=90~\mathrm{GeV},m_{H3}=2000~\mathrm{GeV}$ and $m_{A1}=200~\mathrm{GeV}$ ($m_0=m_{1/2}=1$ TeV). The signal strength $\mu_{\tau\tau}^{VBF/VH}$ is  required to deviate from the SM expectation by  fitting it to a value $\mu_{theo} \neq 1$. This is accomplished in the fit by increasing the invisible BR via the decrease of the neutralino mass, which  can be changed in the fit to a specific value by adjusting the free parameters $\mu_{eff}$, $\lambda$ and $\kappa$ ($m_{\tilde{\chi}_0^1}\sim 2\frac{\kappa}{\lambda}\mu_{eff}$), as can be observed from the NMSSMTools output for benchmark points in  Appendix \ref{output} in Tables \ref{t1}-\ref{t6} for two CASE I examples, called P1 and P2 for $\mu_{theo}=1$ and 0.7, respectively.
For CASE Ib ($m_{A_1}<60$ GeV) and CASE Ic ($m_{H_1}<60$ GeV) one cannot study the deviations for a fixed mass combination, since they require changes in $m_{A_1}$ and/or $m_{H_1}$, which contradicts the study for a fixed mass combination. These cases become apparent, if one samples over all mass combinations, as discussed in Sect. \ref{muscan}.

In the top left panel of Fig. \ref{fig4} the signal strengths are shown for all eight signal strengths as function of $\mu_{sel}=\mu_{\tau\tau}^{VBF/VH}$ between 0.3 and 1. They are overlapping implying the same (=correlated) deviation as the deviation imposed on $\mu_{sel}=\mu_{\tau\tau}^{VBF/VH}$. The fact that they {\it all} stay identical means they are nearly 100\% correlated, which is not unexpected from the correlations  in the couplings, defined in Eq. \ref{coupling3}. It also means that one does not have to fix all signal strengths to a certain value, but it is enough to require only one signal strength ($\mu_{sel}$) to be constrained. It also explains why one can choose another signal strength for $\mu_{sel}$. These correlated changes are independent of the production mode, as shown by the overlapping solid (dashed) lines for the production via VBF/VH (ggf), respectively.  Note that for the $b\bar{b}$ final states the ttH production mode   is selected instead of  ggf, as indicated in Eq. \ref{coupling5} before.   
Fig. \ref{fig4}(b) shows the reduced coupling $c$ as function of the selected signal strength $\mu_{\tau\tau}^{VBF/VH}$ for $c_u$, $c_d$ and $c_\gamma$, which are all close to one. The same is true for $c_{W/Z}$ and $c_{gluon}$ in Fig. \ref{fig4}(c). This implies that for the chosen values of $m_0=m_{1/2}=1$ TeV the contribution from stop loops is small, since the stop loops mainly affect $c_{gluon}$.  Since the signal strengths are the product of couplings squared and BRs relative to the SM values, but the couplings stay constant (panels (c)and (d)), the variation in signal strengths in panel (a) must originate in varying BRs. This is indeed the case, as shown in panel (d). One observes the correlated change of  BRs into visible final states (overlapping lines with a positive slope), which are anti-correlated with the invisible BRs (here into neutralinos) (line with negative slope). 
For $\mu_{\tau\tau}^{VBF/VH}=1$ the invisible BR is zero, while all other reduced BRs  are equal one, so no deviations from the SM expectations are observed. However, requiring $\mu_{\tau\tau}^{VBF/VH}<1$ is easiest obtained by reducing the neutralino mass, as discussed above. This partial width into neutralinos increases the total width, thus decreasing the BRs of the visible final states, since these are given by the ratios of partial and total widths. The partial widths of visible final states stay constant for constant reduced couplings. These results are independent of the production modes, as demonstrated by the overlap of the solid and dashed lines in (a) representing the signal strengths for the VBF/VH and ggf production mode, respectively.

\begin{figure}[]
\begin{center}
\includegraphics[width=0.45\textwidth]{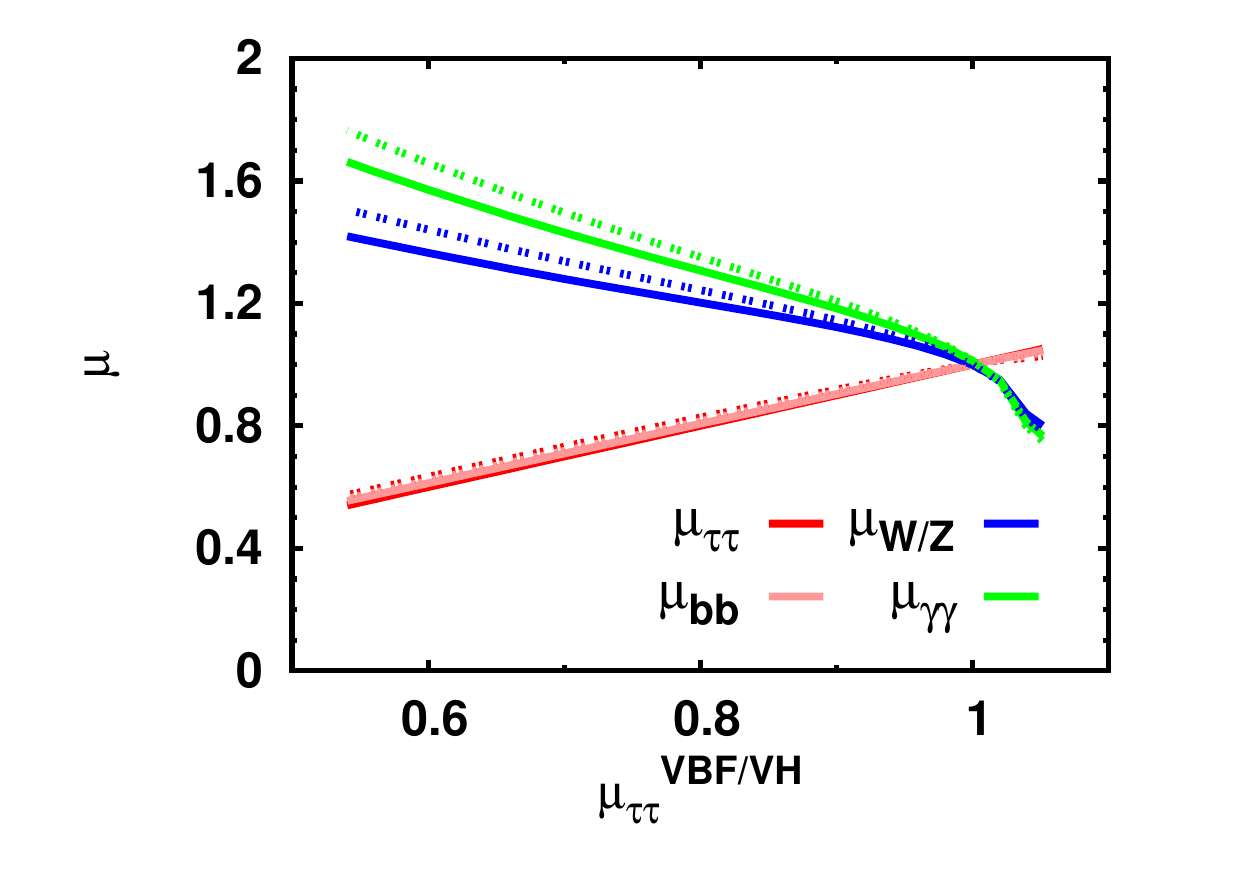}          
\includegraphics[width=0.45\textwidth]{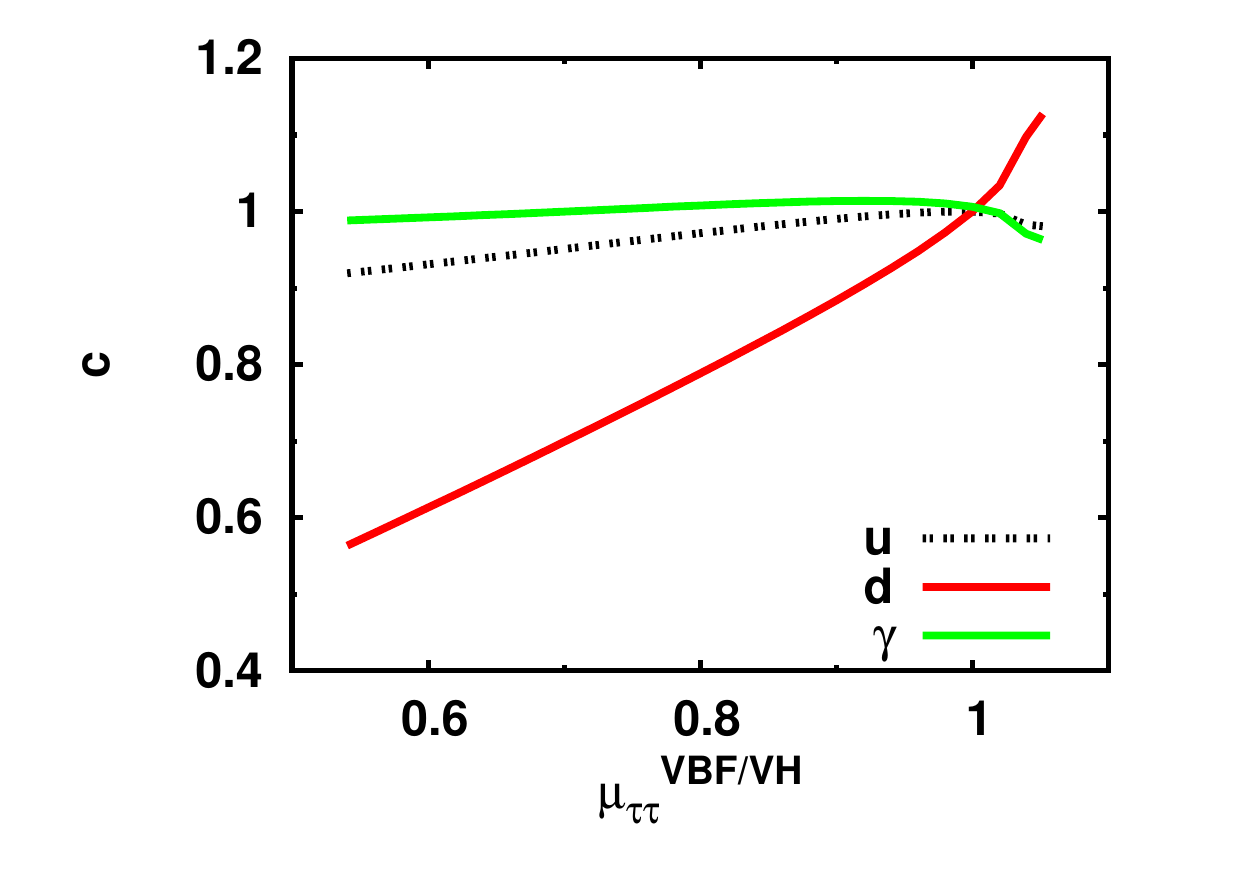} \\
\hspace{0.05\textwidth} (a) \hspace{0.4\textwidth} (b) \\
 \includegraphics[width=0.45\textwidth]{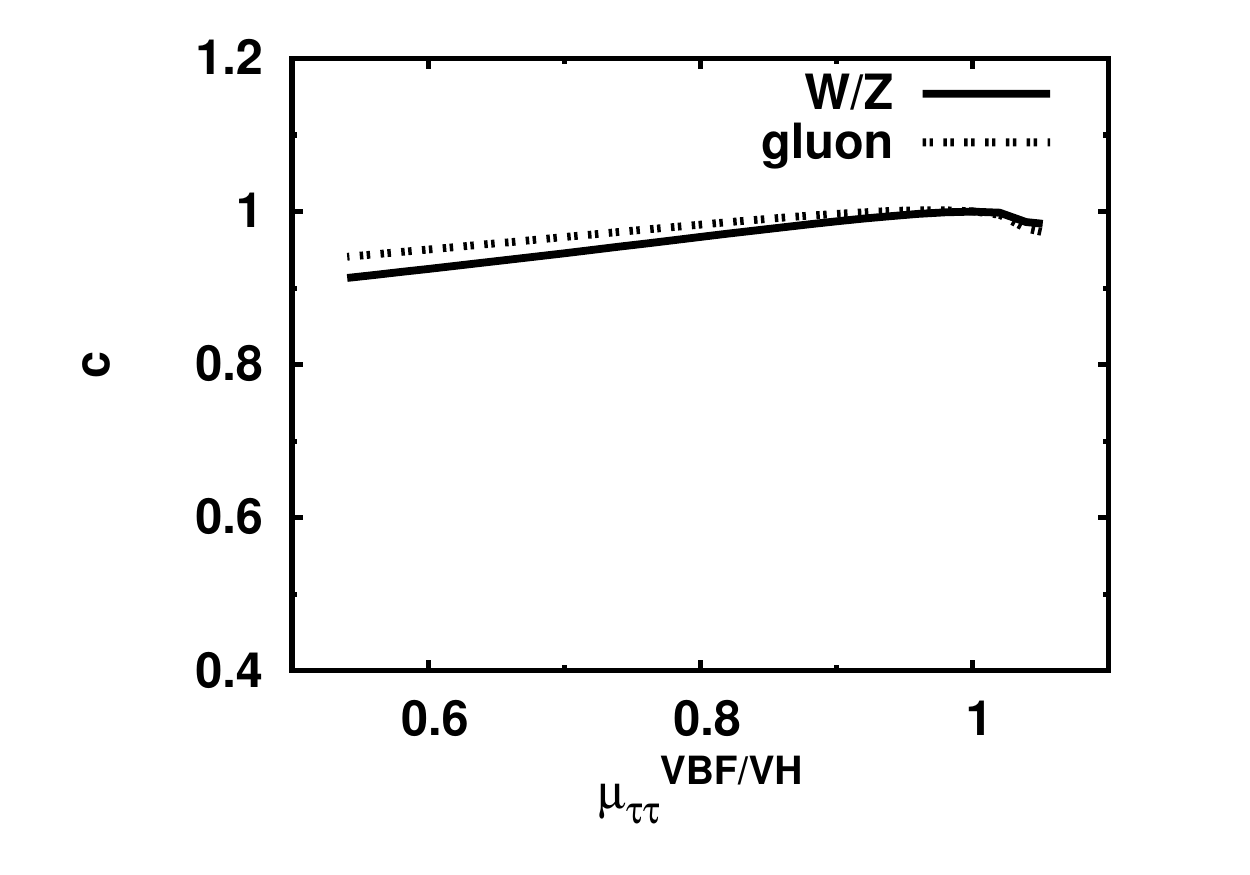}          
 \includegraphics[width=0.45\textwidth]{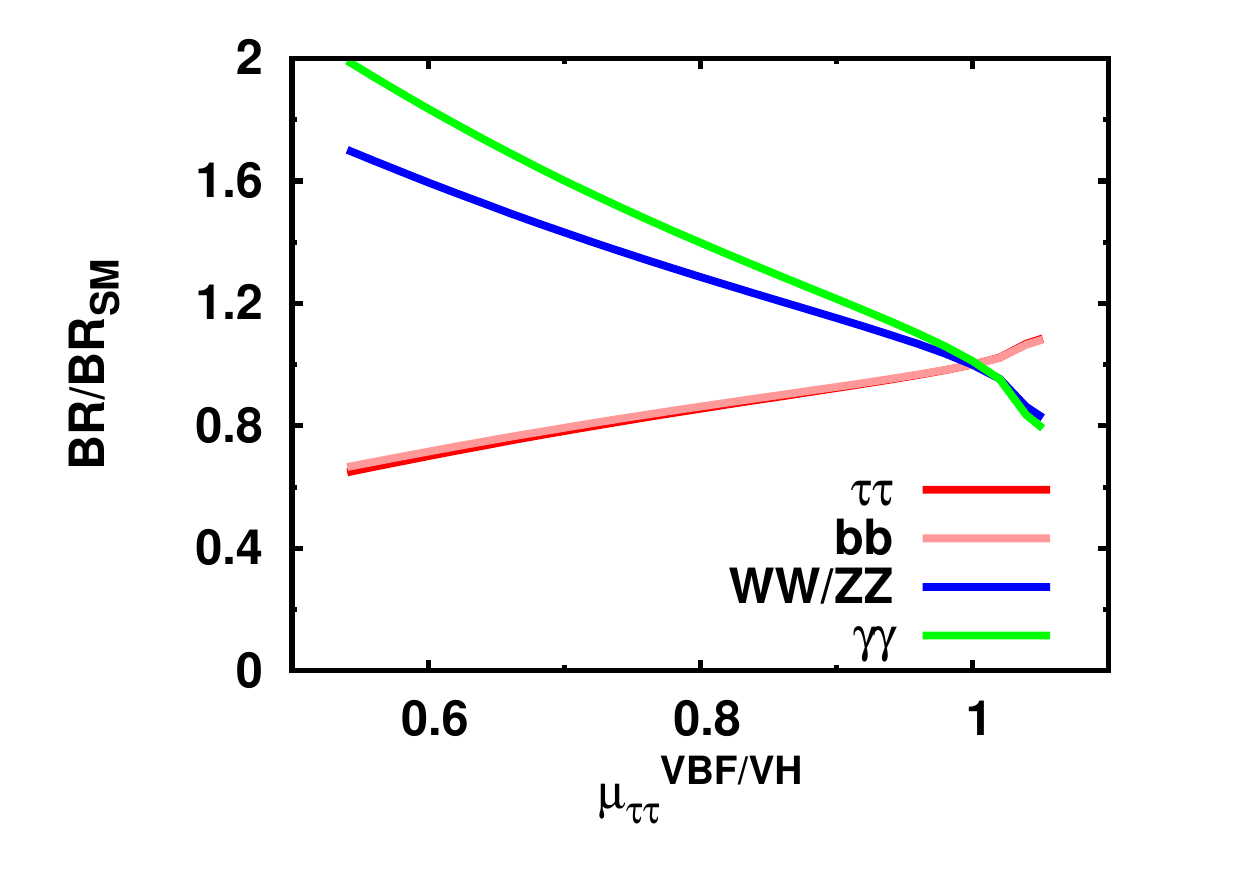} \\
\hspace{0.05\textwidth}(c)  \hspace{0.4\textwidth} (d) \\
\caption[]{ As Fig. \ref{fig4}, but for CASE II.  In this case the change in signal strengths in panel (a) originates from the change in reduced couplings,  shown in panels (b) and (c). The variation of the signal strengths is obtained in the fit by varying predominantly the coupling to down-type fermions $c_d$ (lowest line in panel (b)), which varies predominantly the corresponding BRs (lowest line in panel (d)). The decrease of the BR to down-type fermions can be compensated by an increase in the \textit{bosonic} BRs, since the sum of all BRs stays almost constant, if the total width stays almost constant. The anti-correlation in BRs leads to anti-correlations in the signal strengths, as shown in panel (a) by the lines with negative (positive) slopes for \textit{bosonic} (\textit{fermionic}) final states. These results are again nearly independent of the production modes, as demonstrated by the solid and dashed lines in (a) representing the signal strengths for the VBF/VH and ggf production mode, respectively.
}
\label{fig5}
\end{center}
\end{figure} 

\subsection{CASE II: Deviations by weak Higgs mixing}\label{II}

For CASE II  a mass combination with all particles above 60 GeV is selected, so decays of the 125 GeV boson into light particles are kinematically suppressed. For this CASE II the mass combination $m_{H1}=90~\mathrm{GeV}, m_{H3}=1000~\mathrm{GeV}$ and $m_{A1}=200~\mathrm{GeV}$ ($m_0=m_{1/2}=1$ TeV) was selected.
In comparison with the masses chosen in CASE I the mass $m_{H3}$ was reduced from 2 TeV to 1 TeV, while the other masses stayed the same.   By the decrease of the heavy Higgs boson mass the neutralino mass increases, since they both  depend partially on the same parameters,  as can be seen from Appendix \ref{higgsmixing}.

As before, regions with deviations from the SM expectations are searched for by minimizing the $\chi^2$ value under the constraint $\mu_{sel}=\mu_{\tau\tau}^{VBF/VH} =\mu_{theo} \neq 1$ and the additional constraints defined in the right column of Table \ref{t1-diff}.  The fit accomplishes this by modifying the NMSSM parameters leading to deviations in the Higgs mixing, since no additional decays are allowed for the selected Higgs mass combination. The change in mixing can be observed from a comparison of  $S_{2s}$ in Table \ref{t3} in  Appendix \ref{output} for  two CASE II benchmark points, P3 and P4, which were fitted to $\mu_{theo}=1$ and 0.7, respectively.  

By fitting the down-type signal strength $\mu_{\tau\tau}^{VBF/VH}$ to $\mu_{theo} \neq 1$ the reduced coupling $c_d$ follows closely, as shown in Fig. \ref{fig5}(b) (lowest line), while all other reduced couplings, shown in Fig. \ref{fig5}(b) and \ref{fig5}(c), vary less. The reduction in $c_d$ with much flatter dependencies for $c_u$ and $c_{W/Z}$ can be easily derived from Eq. \ref{coupling3} for larger values of $\tan\beta$. For the reduced couplings $c_{gluon}$ and $c_\gamma$ additional deviations from one can be caused by SUSY contributions in  loop diagrams. These are small in this case, where the lightest stop mass is about 1.2 TeV, as shown in Table \ref{t2} for the benchmark points in  Appendix \ref{output}. The small effect of stop loops is also apparent from the small difference between the couplings $c_{gluon}$ and  $c_{W/Z}$, shown in Fig. \ref{fig5}(c).
The decrease in $c_d$ leads to decreasing BRs into down-type fermion final states, which are displayed in Fig. \ref{fig5}(d)  by the overlapping lines with a positive slope. The total width stays almost constant, if no new decay modes open up, so the sum of the partial widths has to stay constant. Therefore, a decrease of the \textit{fermionic} BRs must be  compensated by a larger BR for \textit{bosonic} final states (lines with negative slopes in panel (d)), leading to an anti-correlation of the corresponding signal strengths, as demonstrated in Fig. \ref{fig5}(a): The signal strengths for \textit{fermionic} final states  (lines with  positive slopes) follow $\mu_{sel}$, while the signal strengths for \textit{bosonic} final states  (lines with  negative slopes) increase with decreasing $\mu_{\tau\tau}^{VBF/VH}$.
These anti-correlations  in signal strengths and BRs are obvious for the benchmark points, as can be seen  by comparing e.g. the signal strengths (BRs) for $H_2 \rightarrow bb$ and $H_2 \rightarrow ZZ/WW$ in Tables \ref{t5} and \ref{t6} of Appendix \ref{output}  for benchmark point P4. 
This anti-correlation contrasts the correlation in CASE I, shown in Fig. \ref{fig4}, where all signal strengths decrease simultaneously and follow approximately $\mu_{\tau\tau}^{VBF/VH}$. In addition, the reduced couplings, especially $c_d$, vary significantly, while for CASE I the reduced couplings equal one as function of $\mu_{\tau\tau}^{VBF/VH}$.
 
\begin{figure}[]
\begin{center}
\includegraphics[width=0.45\textwidth]{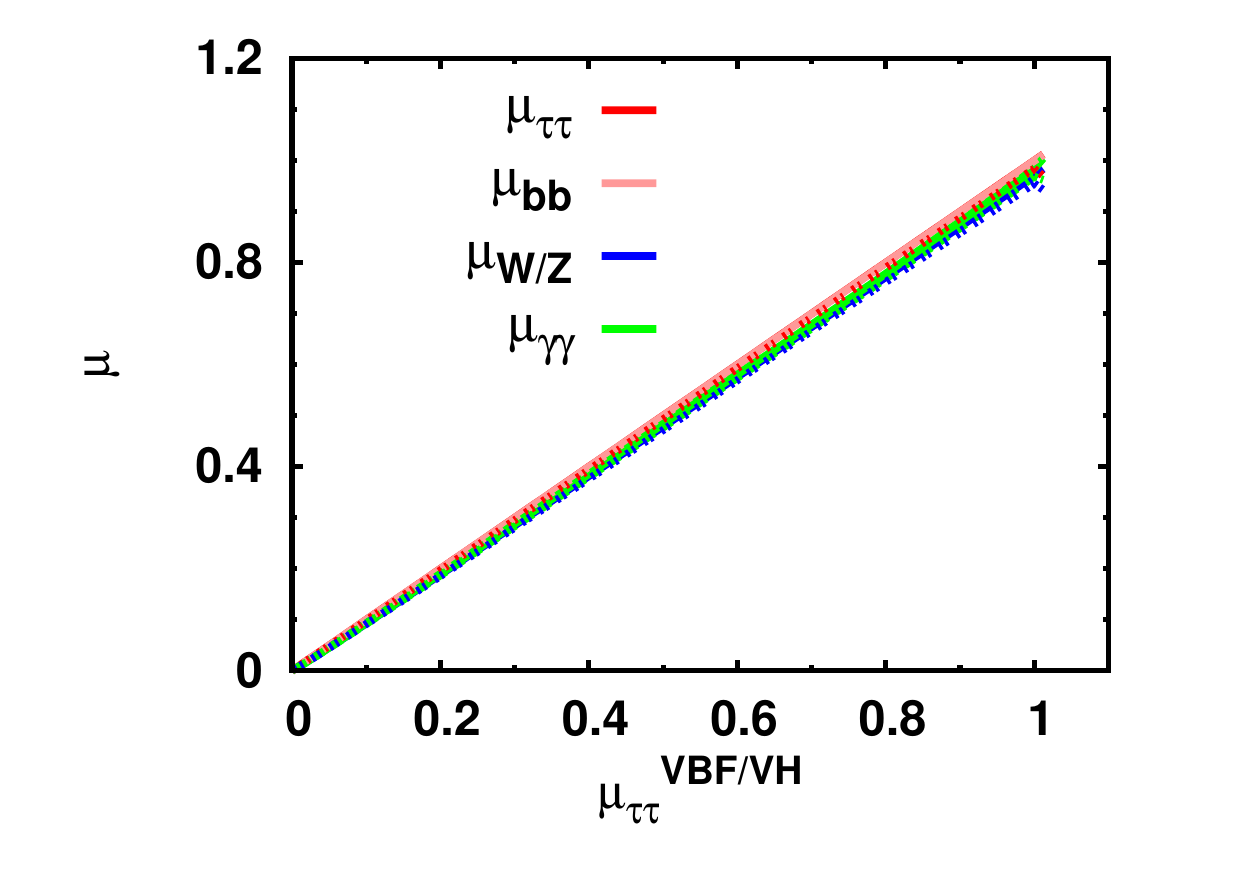}         
\includegraphics[width=0.45\textwidth]{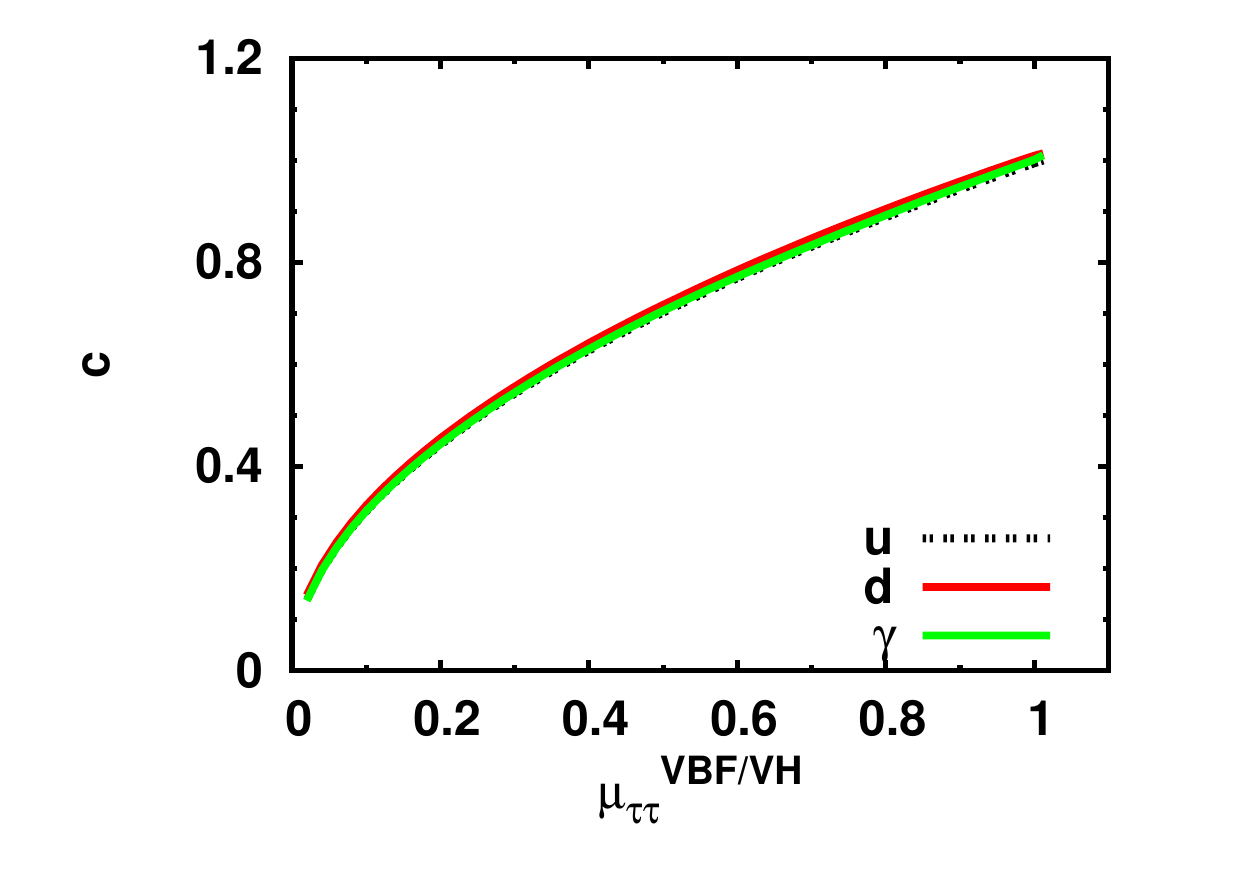}\\
\hspace{0.05\textwidth} (a) \hspace{0.4\textwidth} (b) \\
       \includegraphics[width=0.45\textwidth]{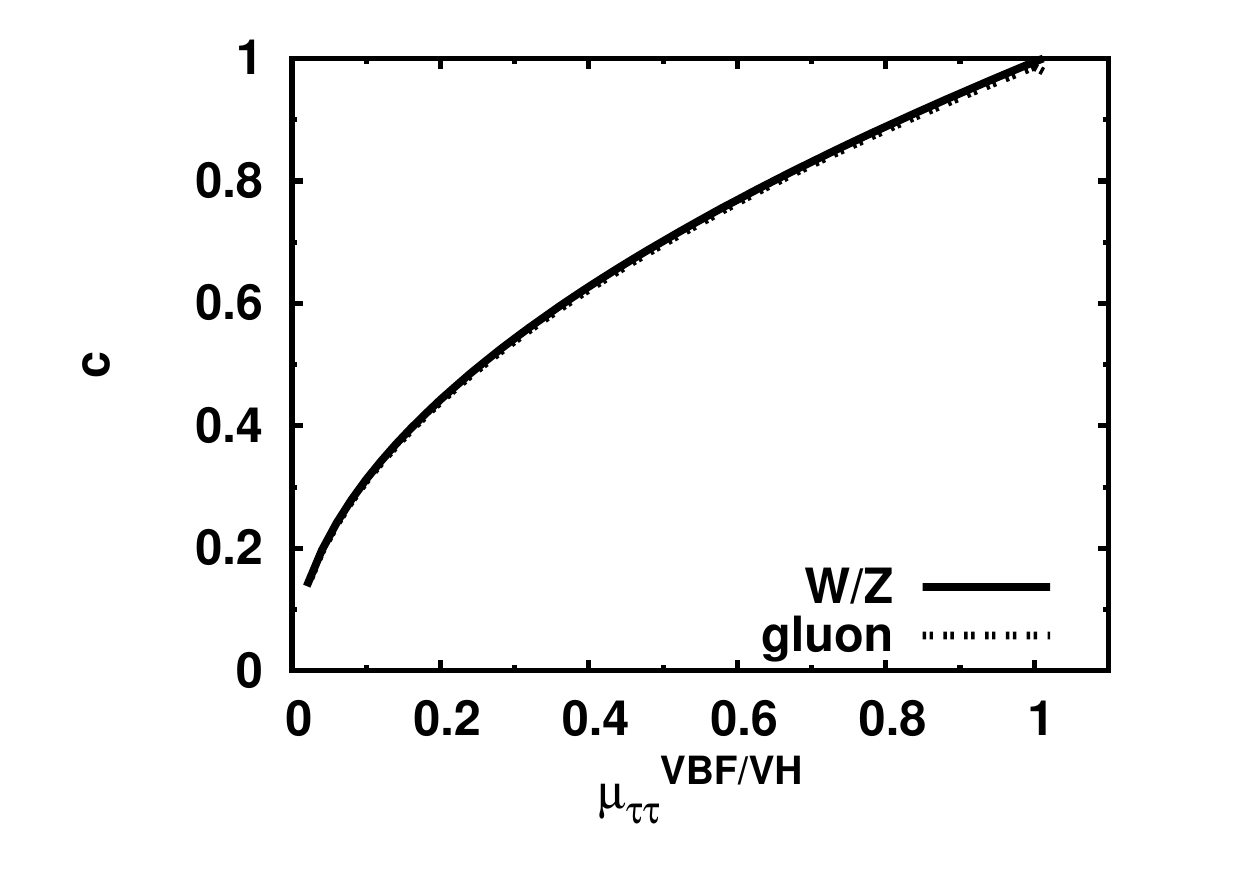}             
        \includegraphics[width=0.45\textwidth]{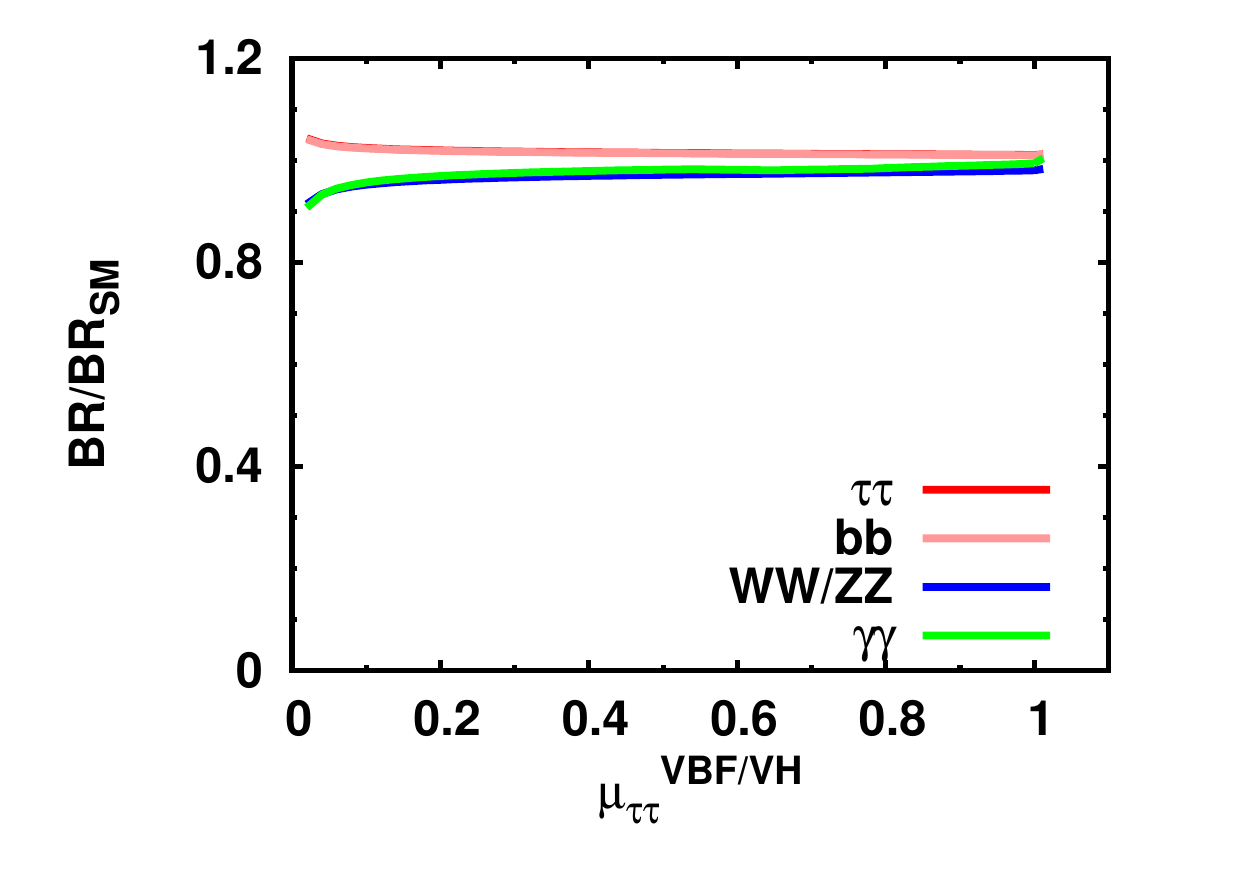}\\
\hspace{0.05\textwidth} (c) \hspace{0.4\textwidth} (d) \\
\caption[]{
As Fig. \ref{fig4}, but for CASE III. The variation of the signal strengths originates from a variation of  all reduced couplings simultaneously (see (b) and (c)), while the BRs stay close to 1 (see (d)). The simultaneous variation of all reduced couplings is caused by the increase of the singlet component $S_{2s}$ of the 125 GeV Higgs boson. In this case all signal strengths change in a correlated way (see overlapping lines in (a)). These results are again independent of the production modes, as demonstrated by the solid and dashed lines in (a) representing the signal strengths for the VBF/VH and ggf production mode, respectively.
}
\label{fig6}
\end{center}
\end{figure}  
\begin{figure}[t]
\begin{center}
\hspace{0.05\textwidth}\textbf{CASE Ia} \hspace{0.2\textwidth} \textbf{CASE Ib} \hspace{0.2\textwidth} \textbf{CASE Ic} \\
       \includegraphics[width=0.32\textwidth]{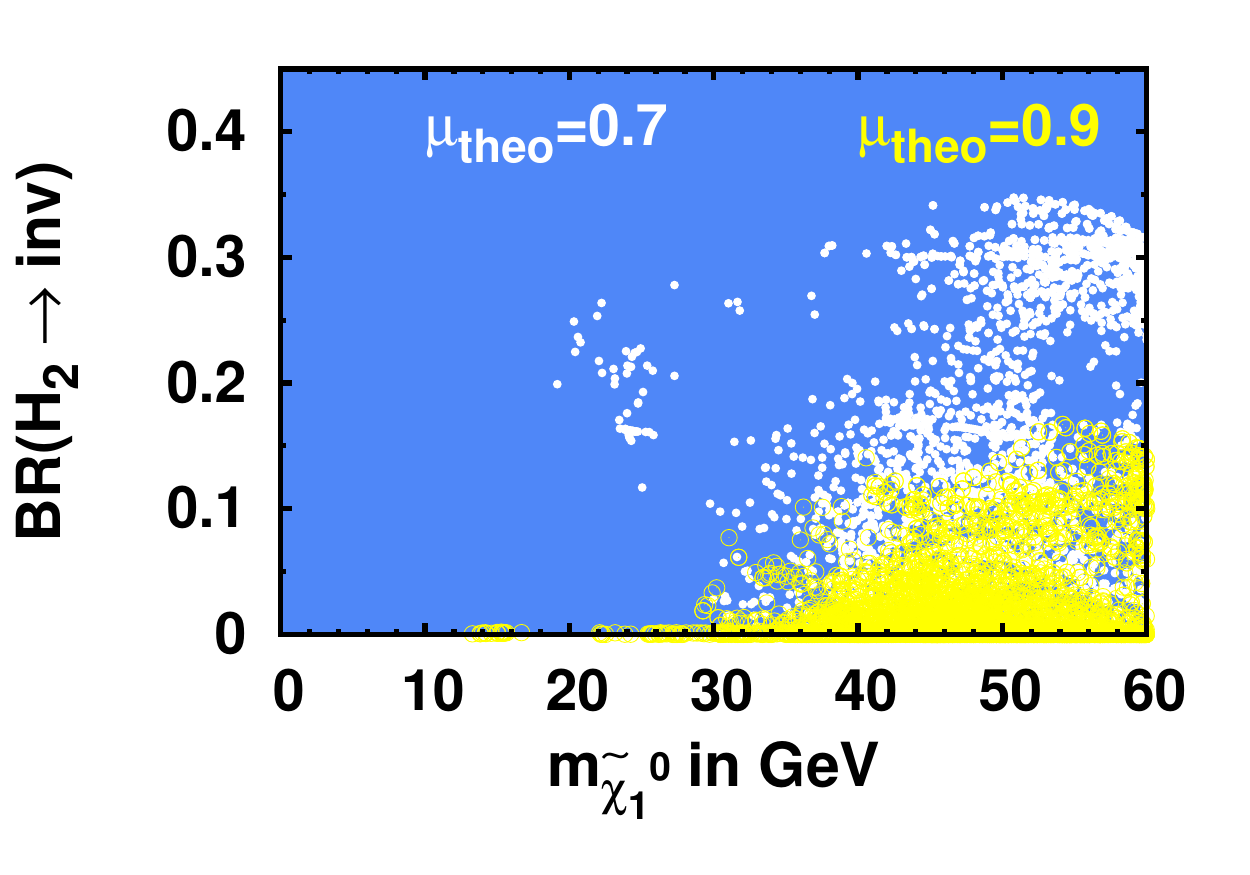}   
       \includegraphics[width=0.32\textwidth]{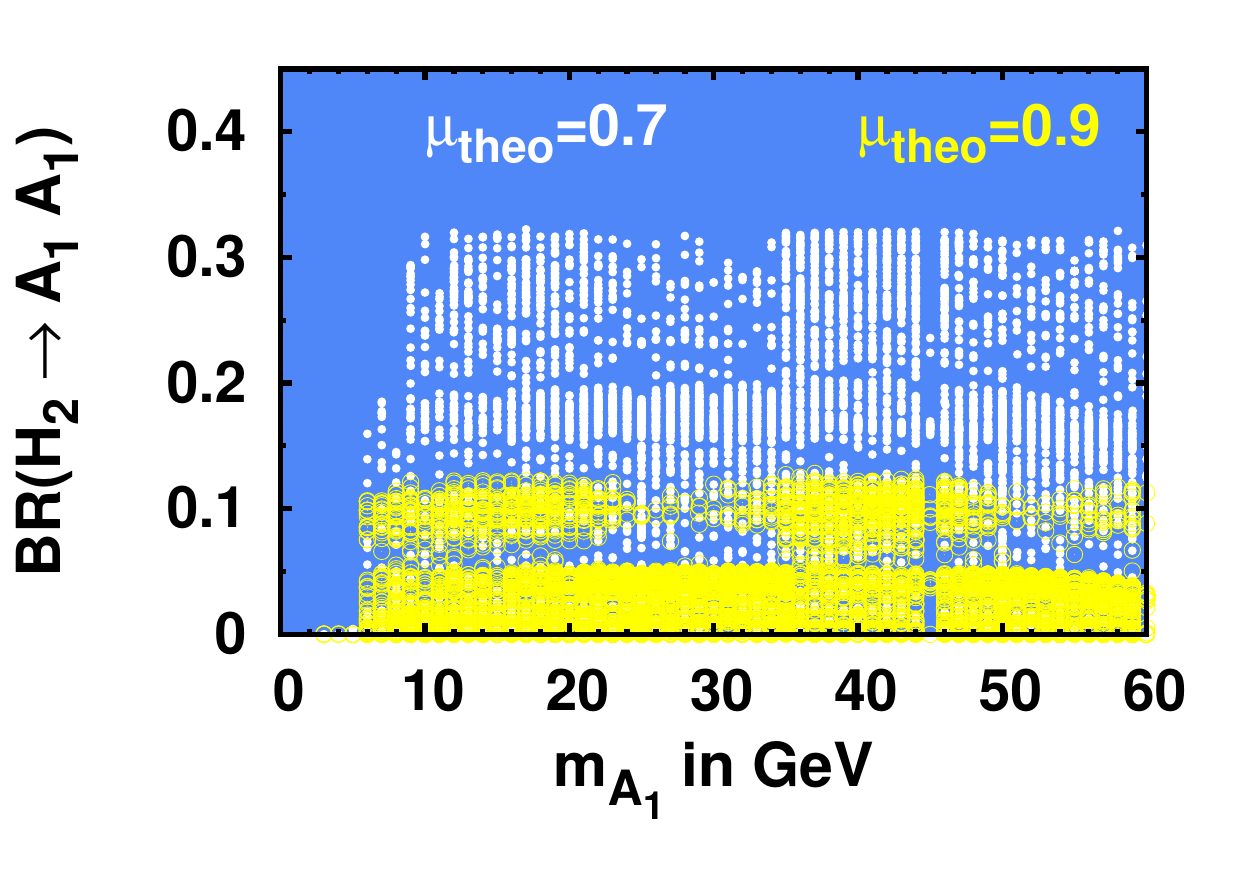} 
       \includegraphics[width=0.32\textwidth]{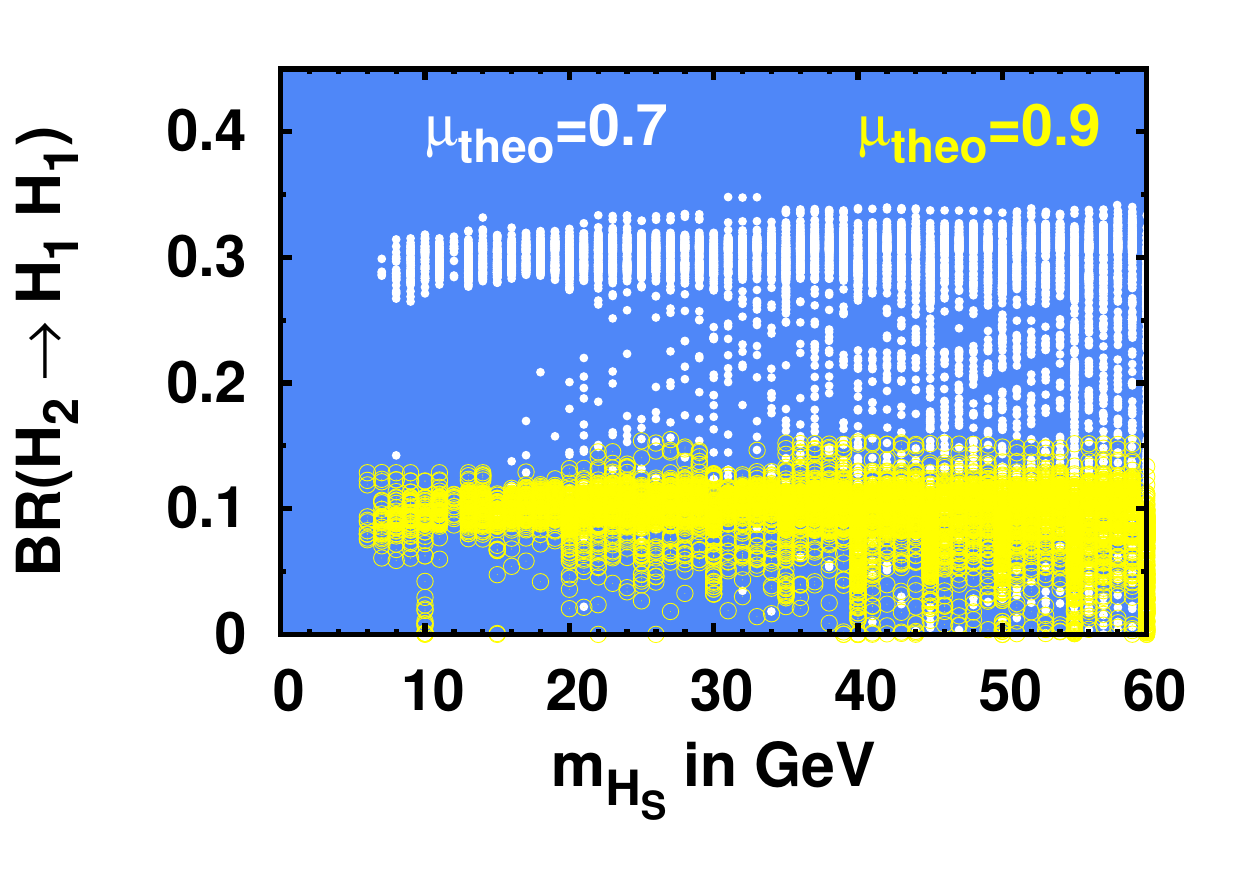} \\
\caption[]{BRs of the 125 GeV Higgs boson into lighter particles as function of the mass of the lighter particle, which can be neutralinos (CASE Ia), the light pseudo-scalar Higgs boson $A_1$ (CASE Ib) or the lightest Higgs boson $H_1$ (CASE  Ic). The  BRs are shown for $\mu_{sel}=\mu_{\tau\tau}^{VBF/VH}$ and $\mu_{theo}$=0.7 (white points) and 0.9 (yellow points), respectively.  The large spread in BRs is caused by the fact that one sums over all mass combinations, which lead to different values of the NMSSM parameters for a given mass value on the horizontal axis. One nevertheless still observes that the BRs  hint to the relation $BR\approx 1-\mu_{theo}$ (see text). 
}
\label{fig7}
\end{center}
\end{figure}

\begin{figure}[ht]
\begin{center}
\hspace{0.075\textwidth}
\footnotesize{\boldmath$\mu_{sel}=\mu_{\tau\tau}^{VBF/VH}\approx\mu_{theo}=0.9$} \hspace{0.15\textwidth} \footnotesize{\boldmath$\mu_{sel}=\mu_{\tau\tau}^{VBF/VH}\approx\mu_{theo}=0.7$}  \\
\includegraphics[width=0.45\textwidth]{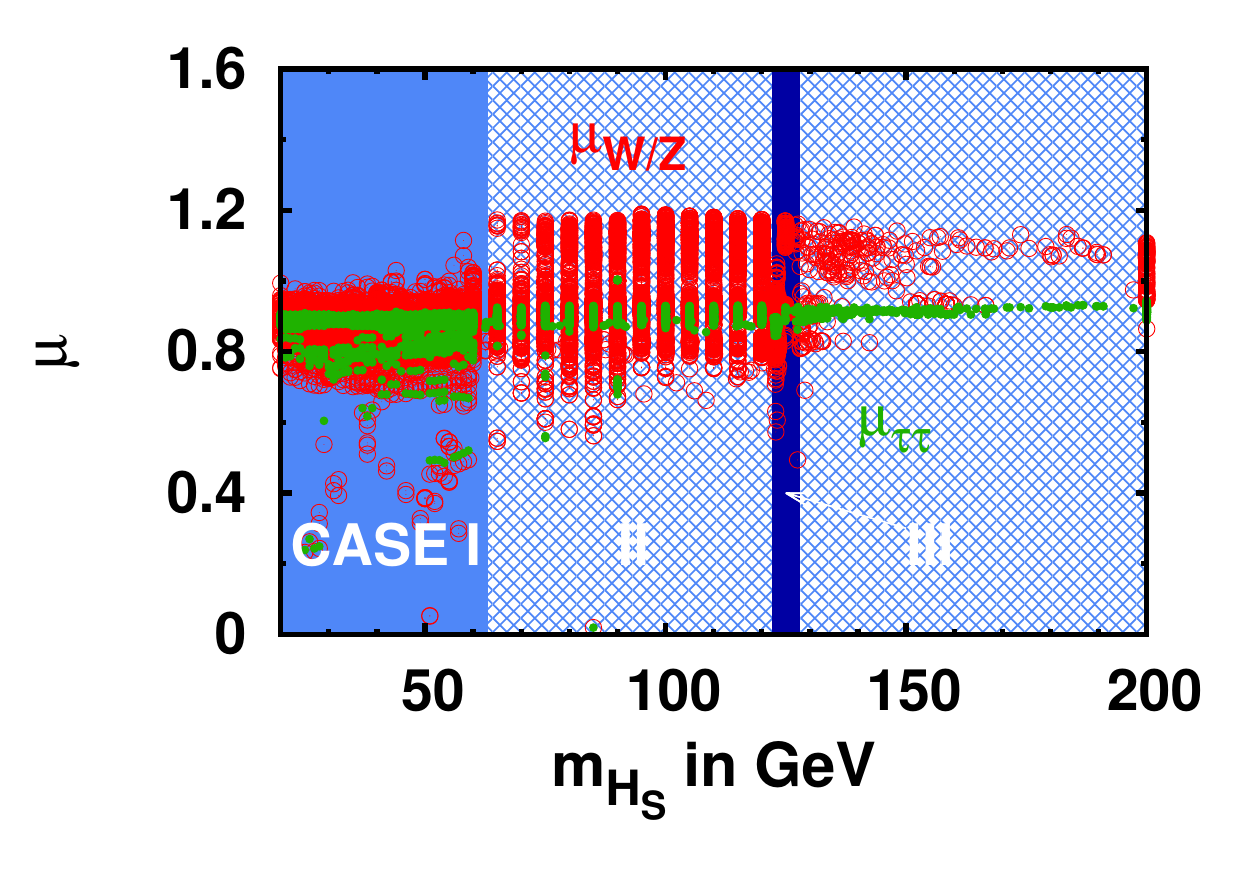}  
 \includegraphics[width=0.45\textwidth]{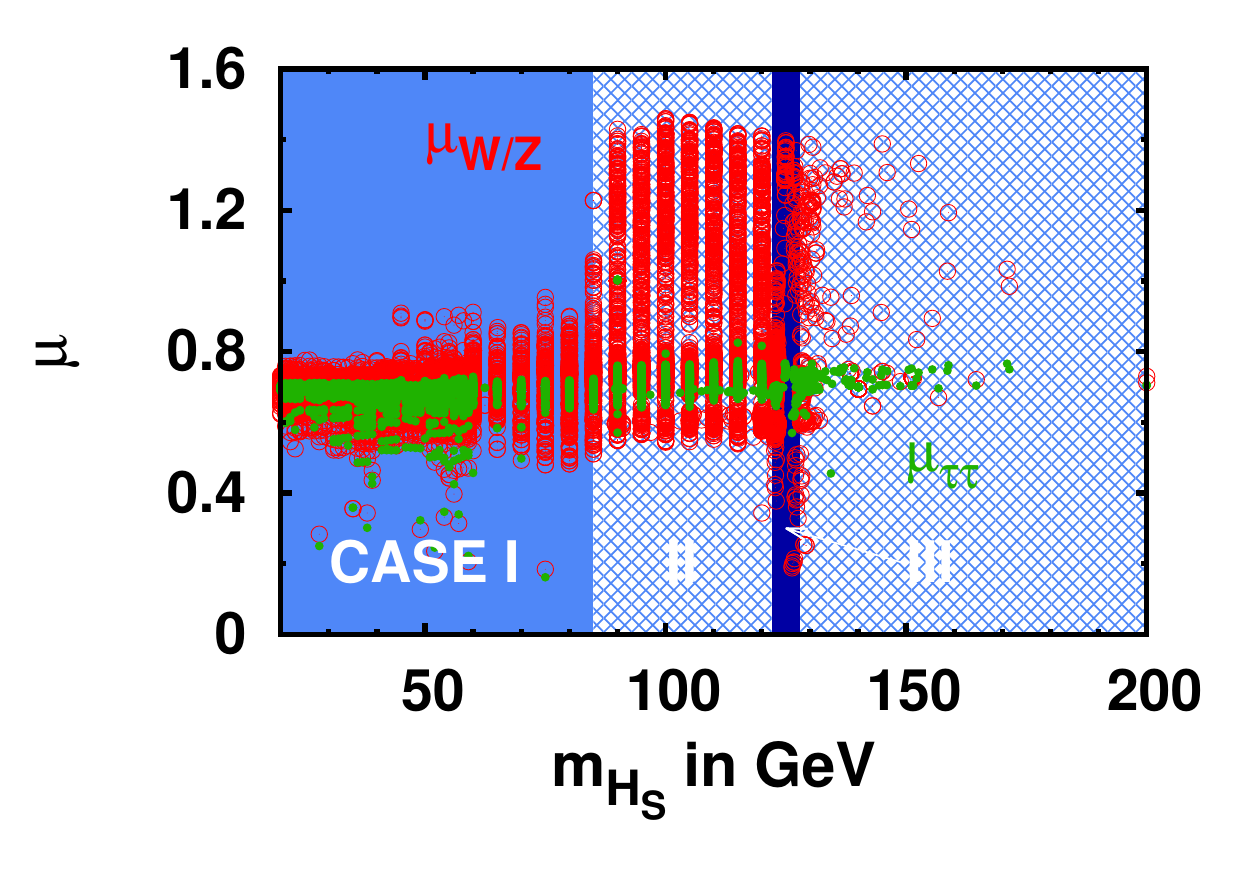} \\
\includegraphics[width=0.45\textwidth]{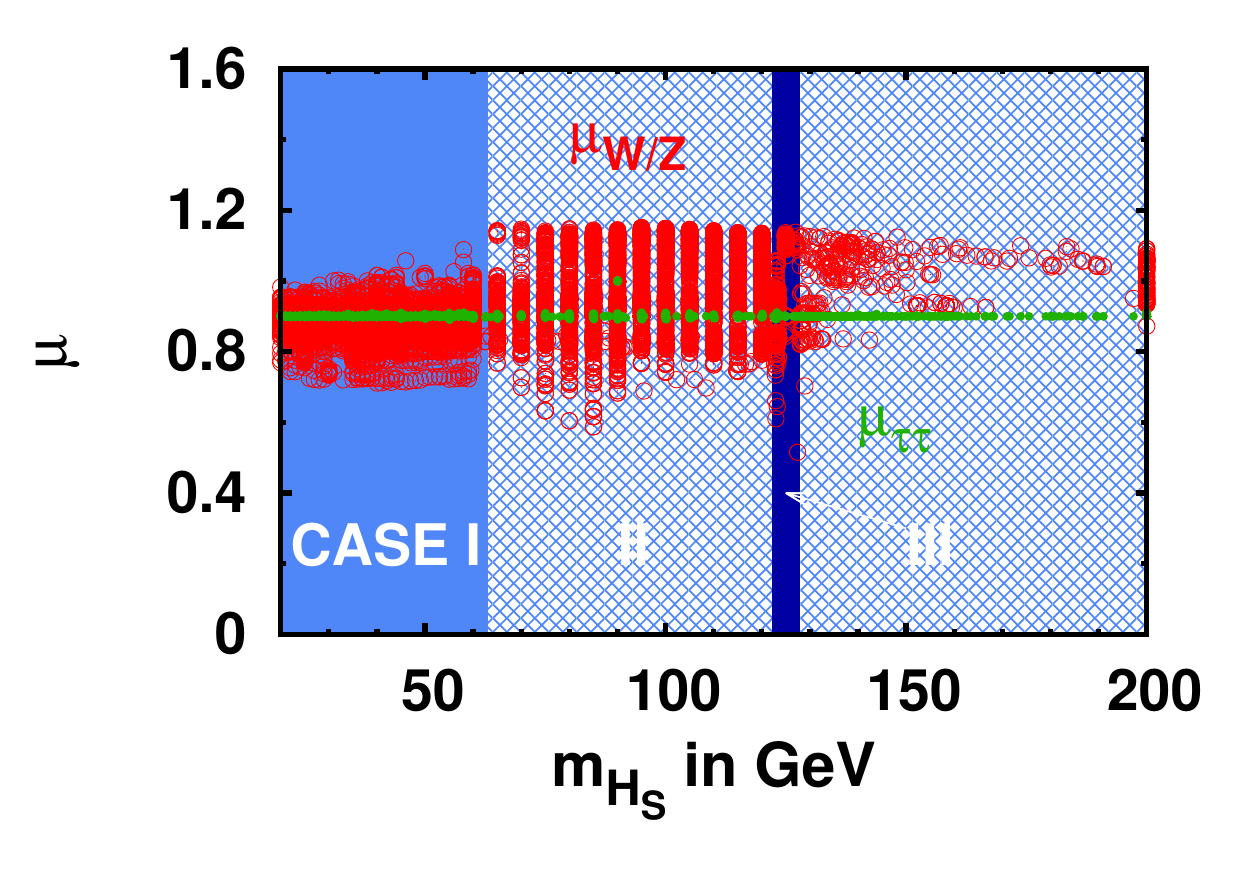}  
 \includegraphics[width=0.45\textwidth]{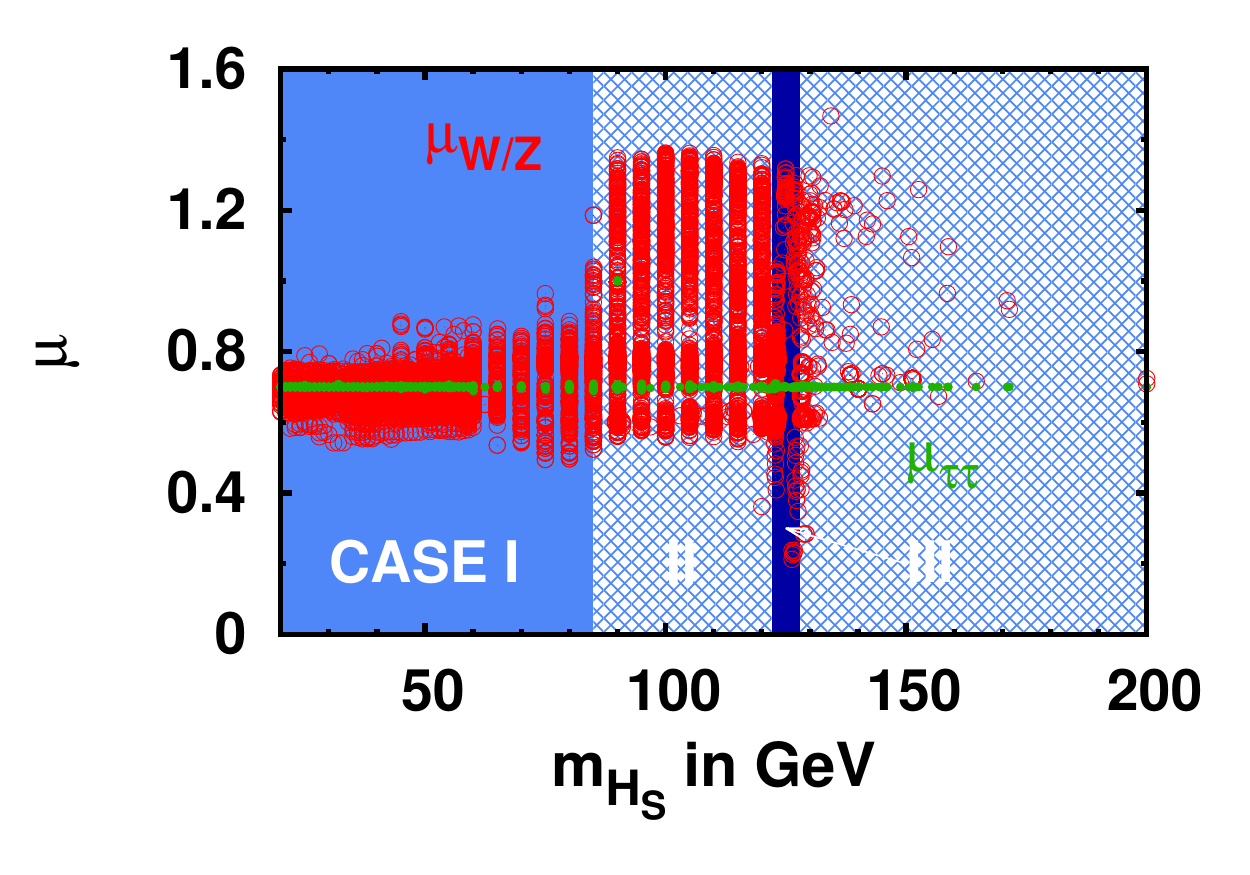} \\
\caption[]{Fitted signal strengths $\mu$ as function of the singlet-like Higgs mass - called $H_S$ on the horizontal axis - for  different values of $\mu_{sel}$, different production cross sections and different final states: 
the left(right) panels show results for $\mu_{theo}$ 0.9(0.7), while the top and bottom rows correspond to the ggf and VBF/VH production modes, respectively. The  regions with $m_{H_S}< 60$ GeV  reflect   CASEs Ia, Ib and Ic in Fig. \ref{fig7}. The  region with $m_{H_S}> 60$ GeV reflect CASE II, while the narrow stripe for $m_{H_S}\sim 125$ GeV reflect CASE III. The green (red) points correspond to $\mu_{\tau\tau}$ ($\mu_{W/Z}$), which are representative for fermionic and bosonic final states. The spread in $\mu_{\tau\tau}$ is small for the lower row (VBF/VH), because it was forced by the fit to be equal $\mu_{theo}$. In the upper row (ggf) the spread is larger, since the points are summed over values of $\mzero = \mhalf$ = 1, 1.5 and 2 TeV. The values of $\mu_{W/Z}$ (red points) show a much larger spread, since one sums over all Higgs mass combinations, which lead to different values of the NMSSM parameters for a given mass value on the horizontal axis.  One nevertheless observes that for $m_{H_S}< 60$ GeV the red and green points corresponding to bosonic and fermionic final states have deviations of the same sign and of the order of $\mu_{theo}$, as expected (see text), while for  $m_{H_S}> 60$ GeV deviations with an opposite sign occur, as expected for the anti-correlations observed for a fixed mass combination for CASE II in Fig. \ref{fig5}.
}
\label{fig8}
\end{center}
\end{figure}

\subsection{CASE III: Deviations by   strong Higgs mixing}\label{III}

In CASE III we select $m_{H1}=122~\mathrm{GeV},m_{H3}=1300~\mathrm{GeV},m_{A1}=200~\mathrm{GeV}$ ($m_0=m_{1/2}=1$ TeV), so $m_{H_1}$ is close to $m_{H_2}=$ 125 GeV), which leads to   a stronger mixing between $H_1$ and $H_2$ than in  CASE II.  As for the previous cases, we force $\mu_{\tau\tau}^{VBF/VH}$ to deviate from one by requiring $\mu_{sel}=\mu_{\tau\tau}^{VBF/VH}=\mu_{theo}\neq 1$. The required deviation in the fit is accomplished by increasing the mixing between $H_1$ and $H_2$, as demonstrated in Table \ref{t3} of Appendix \ref{output} for the benchmark points, P5 and P6, for $\mu_{theo}=1$ and 0.7, respectively. One observes that the singlet component   $H_{2s}$ of the 125 GeV Higgs boson becomes 0.56 for P6, while it is 0.008 for P5.
 The singlet component  does not couple to SM particles, so the  couplings to all \textit{fermions} and \textit{bosons} decrease from one for benchmark point P5  to about 0.83  for  benchmark point P6,  as shown in Table \ref{t4} of Appendix \ref{output} in the $H_2$ block. The dependencies of the reduced couplings on $\mu_{\tau\tau}^{VBF/VH}$ are displayed in  Figs. \ref{fig6}(b) and   \ref{fig6}(c).  In contrast to CASE I (Fig. \ref{fig4}) and CASE II (Fig. \ref{fig5}) the BRs of the 125 GeV Higgs boson stay in CASE III close to the SM expectation of one as function of $\mu_{\tau\tau}^{VBF/VH}$, as shown in Fig. \ref{fig6}(d). The constant BRs and decreasing couplings lead to correlated deviations of the signal strengths, as shown in Fig. \ref{fig6}(a) for \textit{fermionic} and \textit{bosonic} final states.

\section{Genuine NMSSM contributions in the whole parameter space}
\label{muscan}

In this section we scan over all mass combinations defined in the range in Eq. \ref{range}  for the  grid in Fig. \ref{fig2}. The corresponding fit was performed for two different values of hypothetical deviations from the SM expectations  ($\mu_{theo}=0.7$ and 0.9) and three different values of the common SUSY masses ($m_0=m_{1/2}=1,1.5,2$ TeV). 

For CASE I deviations are obtained by decays into light neutralinos (CASE Ia), light pseudo-scalar Higgs bosons (CASE Ib) and light singlet-like Higgs bosons (CASE Ic). The BRs, summed over the whole 3D mass space and the different values of $m_0,m_{1/2}$ pairs,  are shown in Fig. \ref{fig7}  as function of the mass of the final state particles for two values of $\mu_{theo}$, namely 0.7 and 0.9, respectively. Here $\mu_{sel}=\mu_{\tau\tau}^{VBF/VH}$ was chosen, but similar results are found for other choices of $\mu_{sel}$. The neutralino mass is calculated  from the fitted NMSSM parameters for each cell in the 3D Higgs mass space which leads to a non-equidistant grid in contrast to the Higgs masses in the two right panels in  Fig. \ref{fig7}. The fit usually converges with a $\chi^2$ value close to zero meaning that the mass combination  is theoretically allowed and fulfills all experimental constraints.
 
 One observes in the BRs a  correlation with the chosen value of $\mu_{theo}$. This can be understood as follows. Additional decays into light particles increase the total width $\Gamma_{tot}$ of the 125 GeV Higgs boson: $\Gamma_{tot}^*=\Gamma_{tot}+\Gamma_{light} > \Gamma_{tot}$. This leads to a reduction of the BRs of the 125 GeV SM-like Higgs boson, where it is convenient to normalize to the  BR of  the SM. Using the BR into $\tau$ leptons as an example one can write: ${BR_{\tau\tau}^*}/{BR_{\tau\tau,SM}}={\Gamma_{tot}}/{(\Gamma_{tot}+\Gamma_{light})}\approx 1-\Gamma_{light}/\Gamma_{tot}\approx 1-BR_{light}$.  
 Sometimes it happens that several particles are simultaneously light, since the masses are correlated, which can be seen already from the approximate expressions $m_{A_1} \sim \lambda \mu_{eff}$ and $m_{\tilde{\chi}_0^1} \sim 2 \kappa/\lambda \mu_{eff}$ in Appendix \ref{higgsmixing} or that the mixing between $H_1$ and $H_2$ (CASE II) changes simultaneously with the total width (CASE I). This leads to the broadening of the bands in Fig. \ref{fig7}, which are additional broadened by the fact that for a given value of the selected mass on the horizontal axis one sums over the masses of the other Higgs bosons, which leads to different fitted values of the NMSSM parameters. This broadening obscures  the relation  $\mu\approx 1- BR_{light}$, but the correlation between deviations of BRs and $\mu_{theo}$ is still apparent.

To study the CASEs II and III, where the deviations of the signal strengths are caused by the Higgs mixing with the singlet, we concentrate on the signal strength as function of the mass of the singlet-like Higgs boson, denoted by $m_{H_S}$, which can be either $m_{H_1}$ or $m_{H_2}$ and the points are summed over both possibilities.
 We select again $\mu_{sel}=\mu_{\tau\tau}^{VBF/VH}$ and $\mu_{theo}=0.9$ or $0.7$. Fig. \ref{fig8} shows the signal strengths for \textit{fermionic} and \textit{bosonic} final states  as function of $m_{H_S}$ for two production channels: ggf (top row in Fig. \ref{fig8}) and VBF/VH (bottom row in Fig. \ref{fig8}). 
One observes the following  features: The fitted  signal strengths for the fermionic final state  $\mu_{\tau\tau}$  are close to the chosen value of $\mu_{sel}=\mu_{theo}$ for the complete Higgs mass range, as shown by the green points in all panels, as expected, since this was imposed by the fit.  For the gluon fusion process (top row) the spread is larger, since here the stop masses contribute and the points are summed over the common SUSY masses $m_0=m_{1/2}=1,1.5,2$ TeV. The   signal strengths for the bosonic final state  $\mu_{W/Z}$, shown by the red points, tend to be similar to the ones for the fermionic final state (green points) in the region $m_{H_S} < 60$ GeV, which corresponds to CASE I, as discussed in Sect. \ref{I}. 

For  $m_{H_S}>60$ GeV the picture changes. In this case the signal strengths for the bosonic final state (red points) tend to be above the SM expectation of one in contrast to the  signal strengths for the fermionic final state (green points). These correspond to the anti-correlated signal strengths for fermions and bosons, as discussed for the single mass combination in Sect. \ref{II}.  The spread in the red points is large for the same reasons discussed above, but the anti-correlation is better seen,
 if one requires larger deviations, as shown on the right side of Fig. \ref{fig8} for  $\mu_{theo}=0.7$. 
In the narrow region for $m_{H_S}\approx 125$ GeV, the two lightest scalar masses are almost degenerate, which leads to a strong mixing with correlated signal strengths, as discussed for CASE III in Sect. \ref{III}. Unfortunately, the spread in the points is large, if one sums over all Higgs mass combinations in the 3D mass space and  SUSY masses for $m_0=m_{1/2}=1,1.5,2$ TeV, so the clear correlation from Fig. \ref{fig6} for signal strengths of fermionic and bosonic final states is not obvious anymore. 
 The difference between CASE II and CASE III  originates from the mixing between the observed Higgs boson and the singlet, which is weak in CASE II for a large mass splitting and strong in CASE III for a small mass splitting, i.e.  $m_{H_S}\approx 125$ GeV. In case of weak mixing the total width is  changed only mildly, so changes in BRs of fermions, as forced by the fit, have to be compensated by a change in BRs of bosons. In case of strong mixing the singlet component of the observed Higgs boson increases, so the couplings to SM particles decrease, which leads to a correlated change of the signal strengths. These results are exemplified in Figs. \ref{fig5} and \ref{fig6} for CASE II and III, respectively and by the benchmark points for the signal strengths of  $H_2$ in Table \ref{t5} of Appendix \ref{output}: for example between P3 and P4 (CASE II) the signal strengths for all decays into fermions  (top 4 rows in middle block) decrease from about 1 to about 0.7, while the signal strengths for decays into bosons (following 4 rows) increase to about 1.3.  For P5 and P6 (CASE III) the signal strengths {\it all} decrease from about 1 to about 0.7, if $\mu_{theo}$ is changed from 1 to 0.7, the values for P5 and P6, respectively.

\section{Summary}
In Supersymmetry one expects at least one Higgs boson with SM-like couplings. But SM-like means that  signal strengths may deviate from the SM expectations. 
In the MSSM they may occur through the contributions of SUSY particles to loop diagrams. In the NMSSM one can have additional effects, which we call ``genuine" NMSSM effects. They occur by additional decay channels of the 125 GeV Higgs boson into neutralinos or lighter Higgs bosons and/or the mixing between the 125 GeV Higgs boson and the singlet.  The effects from MSSM and NMSSM deviations can be disentangled by searching for correlations in the deviations, since in the MSSM the loop corrections lead to specific patterns in the signal strengths from loop-induced processes with only small effects in the no-loop signal strengths, as shown in Fig. \ref{fig3}.

We searched systematically for the ``genuine" NMSSM effects  by scanning the NMSSM parameter space in a efficient way by mapping the 7D NMSSM parameter space onto the 3D Higgs mass space and determining the NMSSM parameters for each mass combination of the 3D Higgs mass space, spanned by the $m_{A_1}$, $m_{H_1}$ and $m_{H_3}$ masses using the fit procedure outlined in Fig. \ref{fig2} and discussed in detail in Appendix \ref{method}. Using the deterministic scan guarantees a complete coverage of the Higgs mass space, in which we found three different regions with ``genuine" NMSSM deviations. In contrast to the deviations occuring in loop-induced processes they are largely independent of the production mode and  distinguish themselves  by correlations  between final states having either  fermions or bosons.  What is surprising: some regions have positive correlations between final states with fermions and bosons, while others show negative correlations. To understand these correlations we have studied them for three represenative examplse.In CASE I the deviations originate from additional decays, which lead to correlated deviations, as shown  in Fig. \ref{fig4}(a).  CASE II (III), corresponding to weak (strong) Higgs mixing, leads to anti-correlated (correlated) deviations between fermionic and bosonic final state, as shown  in Fig. \ref{fig5}(a) (\ref{fig6}(a)). The different correlations between weak and strong mixing originate from the fact, that in case of strong mixing the singlet fraction of the 125 GeV boson becomes so strong, that it reduces the couplings to all SM particles, which leads to correlated changes for all decay channels. In case of weak mixing the couplings to SM particles are only mildly changed, so the total width stays rather constant. The fit requires deviations for the selected tau-channel, which   changes the coupling for fermions. Given the almost constant total width then results in an anti-correlated change for bosons. The mixing is a strong function of the mass difference (see Fig. \ref{fig8}, where CASE III only corresponds to the narrow vertical stripe with $m_{H_1}\approx m_{H_2}$), so looking for correlations in possible deviations of the 125 GeV boson from SM expectations could give useful hints about the existence and mass of the singlet Higgs boson.   The features of these ``genuine" NMSSM effects  have been detailed for representative benchmark points in  Appendix \ref{output}.

Precision measurements at future colliders will allow to measure the 125 GeV Higgs boson signal strengths with much higher precision than presently available, e.g. a precision of  the order of 1\% can be expected at  future large circular electron-positron colliders, anticipated  in  tunnels for  large proton colliders \cite{CEPCStudyGroup:2018ghi,Abada:2019zxq}  or a future linear electron-positron collider. \cite{deBlas:2018mhx} This would allow to study correlations of possible deviations in much more detail. 



\providecommand{\href}[2]{#2}\begingroup\raggedright\endgroup

\newpage

\appendix

\section{Effectively Scanning the NMSSM  parameter space}
\label{method}

 The Higgs masses and their couplings to fermions and bosons can be calculated from the seven NMSSM parameters in Eq. \ref{parameter}, which form a 7D parameter space. Experimentally the Higgs masses span only a 3D parameter space, since one Higgs mass has to correspond to the observed 125 GeV Higgs boson and all heavy Higgs bosons are all related to the mass of the heavy pseudo-scalar Higgs mass $m_{H_3}$, so one is left with three unknown Higgs masses, e.g.  $m_{A_1}$, $m_{H_1}$ and $m_{H_3}$, which form a 3D mass space.   In the next section we discuss the advantages of sampling the 3D mass space instead of the 7D NMSSM parameter space. Mapping one space into the other can be done by fitting the 7 NMSSM parameters from the three Higgs masses in a cell of the 3D mass space together with one or more assumed values of the signal strengths, which can either be chosen as corresponding to the SM expectation or to deviate  from it. The relation between masses and parameters has been encoded in the publicly available software package NMSSMTools 5.2.0 \cite{Das:2011dg}, which can be obtained from the web site \cite{NMSSMToolsweb}.  The procedure has been illustrated in Fig. \ref{fig2}.


 Relevant questions for such a procedure are: What are the advantages of sampling the 3D mass space instead of the 7D parameter space? Which constraints are used in the fit? Are the solutions unique? Does a scan of the 3D mass space lead to 
  complete coverage of the 7D parameter space? 
  We discuss these questions in the following subsections.
    \subsection{What are the advantages of sampling the 3D mass space instead of the 7D parameter space?}\label{3dsampling}
 There are several observations in favor of SUSY, like the prediction of a Higgs boson with a mass below 125 GeV, the prediction of electroweak symmetry breaking for a top mass between 140 and 190 GeV, the prediction of a massive dark matter candidate with an extremely small cross section to interact with matter (upper limit from direct DM searches), but a 10 orders of magnitude larger cross section to interact with itself (annihilation cross section from Hubble expansion, if the DM candidate is a thermal relic) and SUSY may be part of a larger theory, given the observation of gauge coupling unification, Yukawa coupling unification and the natural connection with gravity. These results have been reviewed many times, see e.g. Refs. \cite{Haber:1984rc,Kane:1993td,deBoer:1994dg,Martin:1997ns,Kazakov:2010qn,Kazakov:2015ipa}.

One can sample  the 7D parameter space to search for regions, where all these observations are taken into account. However, scanning the large 7D parameter space is a daunting task, so one usually resorts to a stochastic sampling of the parameter space with a ``flat prior'' for the parameters. This sounds like a reasonable  ``unbiased''  approach by assuming equal probabilities for all values of all parameters. However, if the parameters are highly correlated not all  parameter combinations are allowed and the prior volume is too large, thus leading to too low probabilities. Examples where this happens vary from probability calculations for accidents with nuclear power plants or space shuttle flights \cite{lesswrong}. 
The most famous example for the importance of choosing the prior is certainly the search for lost submarines, which was used in reality, but is also used in a Web simulator  to assist in the teaching of Bayes` Theorem \cite{websim}. 
As an example, one can take as prior volume the distance a submarine  might have traveled from the latest known  position and assign equal probabilities to positions on a grid centered around this position (``flat prior''). However, if one assumes the captain might have tried to reach a harbor after some accident, then positions in directions of a harbor are assigned a higher probability, which reduces the probability of all other positions in a correlated way.  In contrast, new data on negative searches will enhance the probabilities of the remaining positions, so the posterior probability taking the new data into account can be used to reduce the prior volume, i.e. the posterior becomes the prior for  future observations.

The more efficient sampling in a reduced prior volume was the main motivation  for switching from the 7D NMSSM parameter space to the 3D Higgs mass space after we found a way to transform from one space to another. Sampling the 3D space has two main advantages: 
\begin{itemize} 
\item the reduced space allows for a deterministic scan in a short time, since ALL mass combinations in the grid on the 3D mass space can be fitted simultaneously in a cluster of parallel processors. A deterministic scan guarantees a complete coverage of the 3D mass space. The corresponding coverage of the 7D parameter space is discussed below in Sect. \ref{coverage}.
 \item The Higgs mass space turns out to be very smooth, since all mass combinations turn out to lead to acceptable fits, which is in strong contrast to the spiky likelihood distributions  of the 7D NMSSM parameters from correlated parameters. These correlations  lead to many  forbidden regions, as can be seen from Fig. \ref{fig9} showing some correlations of accepted points in projections of the 7D space. 
 \end{itemize}
 
 In the left panel of Fig. \ref{fig9} one of the so-called prior dependences, meaning the difference between  sampling in linear or logarithmic representations of the parameters, becomes obvious: the low  $\tan\beta$  values are allowed more often, so  an equidistant sampling of a logarithmic distribution will more often yield success thus leading to  different probabilities for  linear or logarithmic sampling of  $\tan\beta$. This prior dependence is considered a serious problem in sampling SUSY parameter spaces, see e.g. Refs. \cite{Trotta:2008bp,Athron:2017qdc}.   If  the sampling would be guided along these higher probability regions, very much like the search for submarines is guided along the regions favored by the priors correlating positions with harbor directions, the prior dependence would largely disappear. Alternatively, one can try a multistep fitting technique \cite{Beskidt:2012sk}, but transforming to a different space with less correlated parameters, like the 3D mass space,  with a smooth likelihood distribution and a much smaller prior volume, is much simpler. Furthermore,   a stochastic scan with flat priors may not find highly correlated regions in a 7D parameter space, just like it is difficult to find a submarine with the assumption of flat priors for all locations instead of correlating the positions with harbor directions. Unfortunately, SUSY has in this respect many ``harbors'', because the marginal data  allows  many parameter combinations, so it is hard work to find the correlations between the many parameters. Avoiding the need for using a correlation matrix during the sampling and simultaneously getting rid of the prior dependence  - both possible  by avoiding a stochastic sampling - are the main hallmarks for sampling the 3D Higgs mass space instead of the 7D NMSSM parameter space.

  \subsection{Which constraints are used in the fit?}\label{constraints}
As statistic for the fit determining the NMSSM parameters from the Higgs masses we choose the $\chi^2$ function, which  can be minimized  by Minuit \cite{James:1975dr}.  The following constraints are included in the $\chi^2$ function:
\beq\label{eq5}
\chi^2_{tot}=\chi^2_{H_S}+\chi^2_{H_3}+\chi^2_{A_1}+\chi^2_{H_{125}}+\chi^2_{\mu_{125}}+\chi^2_{LEP}+\chi^2_{LHC},
\eeq 
which are defined as:
\begin{itemize} 
\item $\chi^2_{H_S}=(m_{H_S} - m_{grid,H_S})^2/\sigma^2_{H_S}$: 
The term $\chi^2_{H_S}$ requires the NMSSM parameters to be adjusted such that the mass of the singlet-like light Higgs boson mass $m_{H_S}$ agrees with the chosen point in the 3D mass space $m_{grid,H_S}$. 
The value of $\sigma_{H_S}$ is set to $1\permil$ of $m_{grid,H_S}$.  A small error on the chosen Higgs mass   avoids a smearing in the 3D Higgs mass space and a corresponding smearing by the projection onto the 7D parameter space.
\item $\chi^2_{H_3}=(m_{H_3} - m_{grid,H_3})^2/\sigma^2_{H_3}$: as $\chi^2_{H_S}$, but for the heavy scalar Higgs boson $H_3$.
\item $\chi^2_{A_1}=(m_{A_1} - m_{grid,A_1})^2/\sigma^2_{A_1}$: as $\chi^2_{H_S}$, but for the light pseudo-scalar Higgs boson $A_1$.
\item $\chi^2_{H_{125}}=(m_{H_125} - m_{obs})^2/\sigma^2_{125}$: This term is analogous to the term for $m_{H_S}$, except that the  Higgs  mass $m_{H_125 }$ is required to agree with the observed Higgs boson mass, so $m_{obs}$ is set to $125.2$ GeV. The corresponding uncertainty $\sigma_{125}$ was set to $1\permil$ of $m_{obs}$. Note that the much larger error on the mass of the observed 125 GeV boson is not taken into account, since we want to obtain a precise mapping between the 3D Higgs mass space and the 7D NMSSM parameter space. Once the accepted region of NMSSM parameters has been determined one can look for the region, where the predicted Higgs mass is within the uncertainties of the observed Higgs mass (including theoretical uncertainties).
\item $\chi^2_{\mu_{125}}=\sum_i (\mu^i_{H_{125}} - \mu_{theo})^2/\sigma^2_{\mu}$: This term requires the Higgs boson $H_{125}$ to have specific couplings  $\mu_{theo}$ for selected signal strengths $i$. For the search for correlations between signal strengths with and without loop diagrams one sums over the four no-loop signal strengths in Eq. \ref{coupling6}. For the search for  ``genuine'' NMSSM effects only one signal strength $\mu_{sel}$  is used for this constraint. In all cases  $\sigma_{\mu}$ was chosen to be 0.005.   
\item $\chi^2_{LEP}$ includes the LEP constraints on the couplings of a light Higgs boson below 115 GeV and the limit on the chargino mass. These constraints include upper limits on the decay of light Higgs bosons into b-quark pairs, which are particular important for the singlet Higgs, if it is the lightest one. The LEP constraints are in principle implemented in NMSSMTools \cite{NMSSMToolsweb}, but small corrections were applied, as discussed in Ref. \cite{Beskidt:2014kon}.      
\item $\chi^2_{LHC}$ includes constraints from the LHC concerning light scalar and pseudo-scalar Higgs bosons, as implemented in NMSSMTools \cite{NMSSMToolsweb}. 
\end{itemize}
We assume that the constraints on SUSY mass limits from LEP and LHC as well as the Higgs masses are uncorrelated. These limits have been implemented in NMSSMTools, so one can check for each fitted point, if they are allowed by accelerator constraints. Cosmological constraints can be applied  in NMSSMTools as well, but as long as the nature of the dark matter is unknown, we refrain from doing this.  We also refrain from using data on the anomalous magnetic moment, which  has a large theoretical uncertainty. Its more than two standard deviations from the SM expectation prefers rather small SUSY masses, which have been excluded in  constrained models \cite{Beskidt:2012sk}.  Adding constraints for excluded regions only adds an offset to the test statistic, but it does not further constrain the parameter space.
Note that we either assume the lightest Higgs $H_1$ to be the singlet-like Higgs boson  and the second lightest Higgs boson $H_2$ to be the 125 GeV SM Higgs boson  or vice versa. In the first (second) case, the singlet-like Higgs boson has a mass below (above) 125 GeV. 
There are also solutions where $m_{H_1}=125$ GeV, $m_{H_2}>125$ GeV and $m_{H_3}$ is singlet-like, but we focus on  scenarios where the heavy Higgs bosons are MSSM-like. 

We restrict the range of the Higgs masses in the 3D mass space to the following values:
\begin{align}
5~\mathrm{GeV} &<  m_{H_S} < 500~\mathrm{GeV},\notag\\
125~\mathrm{GeV} & < m_{H_3} < 2000~\mathrm{GeV},\label{range}\\
5~\mathrm{GeV} & < m_{A_1} < 500~\mathrm{GeV}.\notag
\end{align}
Although heavier Higgs bosons outside the range in Eq. \ref{range} are not forbidden, they are not relevant for LHC physics, so we did not investigate them here. The LHC constraints require a minimal SUSY breaking scale, which avoids bino-like solutions for the lightest neutralino in most, but not all, regions of the parameter space, since in some regions  the lower limit  of  the bino mass element in Eq. \ref{neutralino} (M$_1\approx 0.4\mhalf  \approx 300$ GeV) can be smaller than  the diagonal matrix element  of  the Higgsinos  ($\mu_{eff}$). The latter  can reach values up to 600 GeV in small regions of parameter space, as can be seen from the left panel of Fig. \ref{fig9}.  All accepted mass combinations   were found to have $m_{H_3}$  in the alignment limit, where all heavy Higgs masses have  mass splittings below 3\%. But  the mass splittings are not used in the fit, so the results are independent of the splittings. The splittings have  simply been calculated from the  parameters found by the fit.

  \subsection{Does the fit lead to unique results?}\label{unique}
\begin{figure}
\begin{center}
\includegraphics[width=0.32\textwidth]{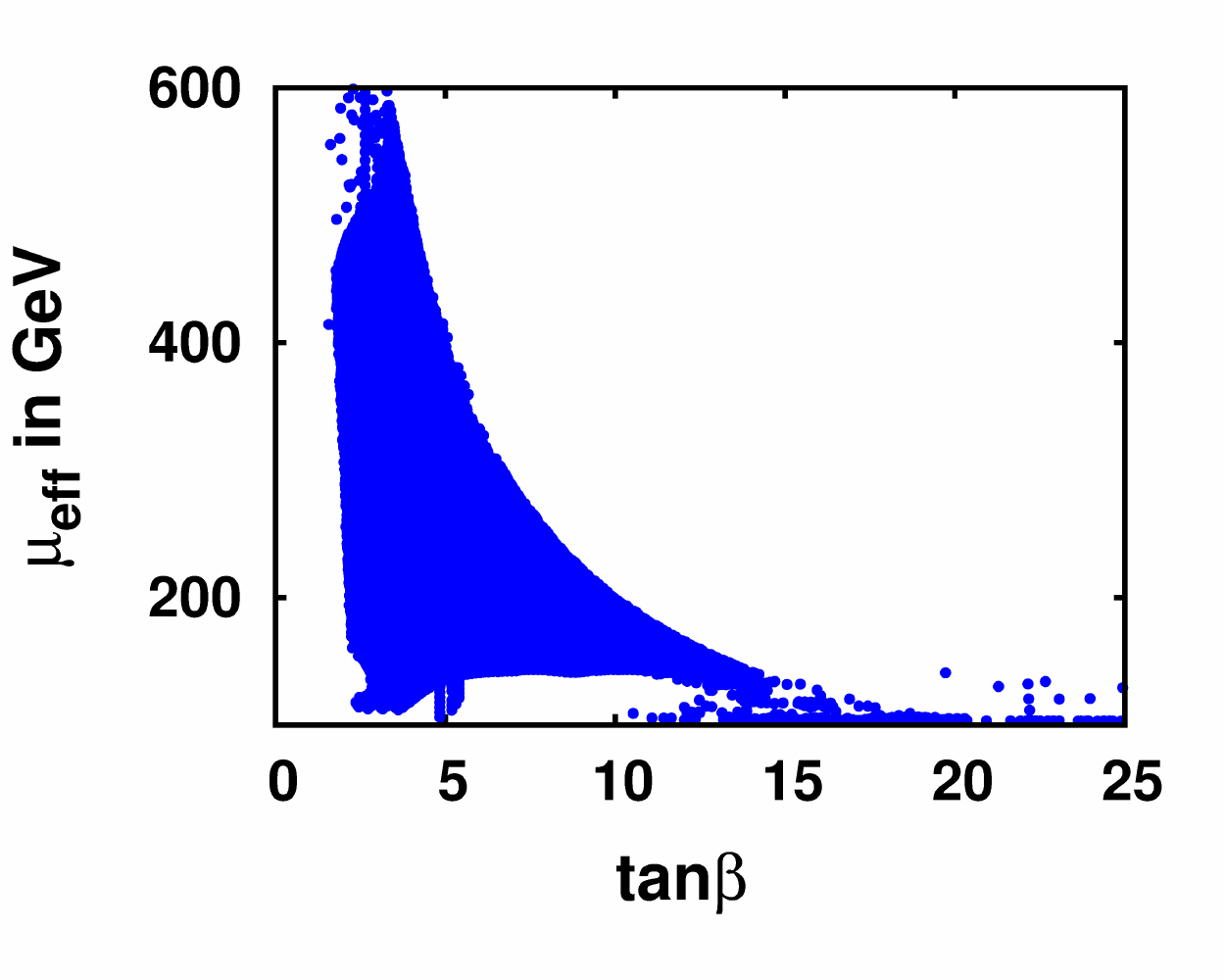}
\includegraphics[width=0.32\textwidth]{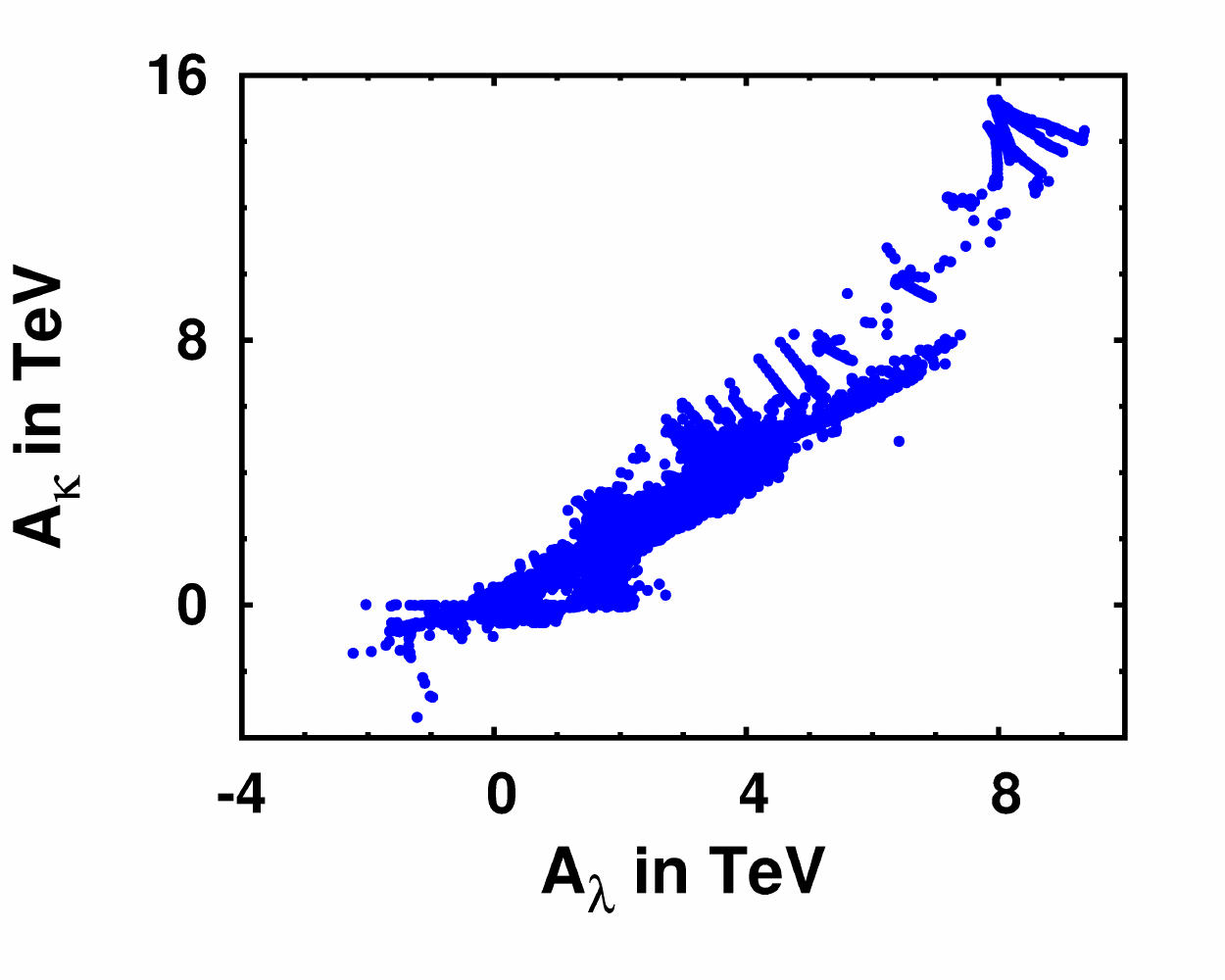}
\includegraphics[width=0.32\textwidth]{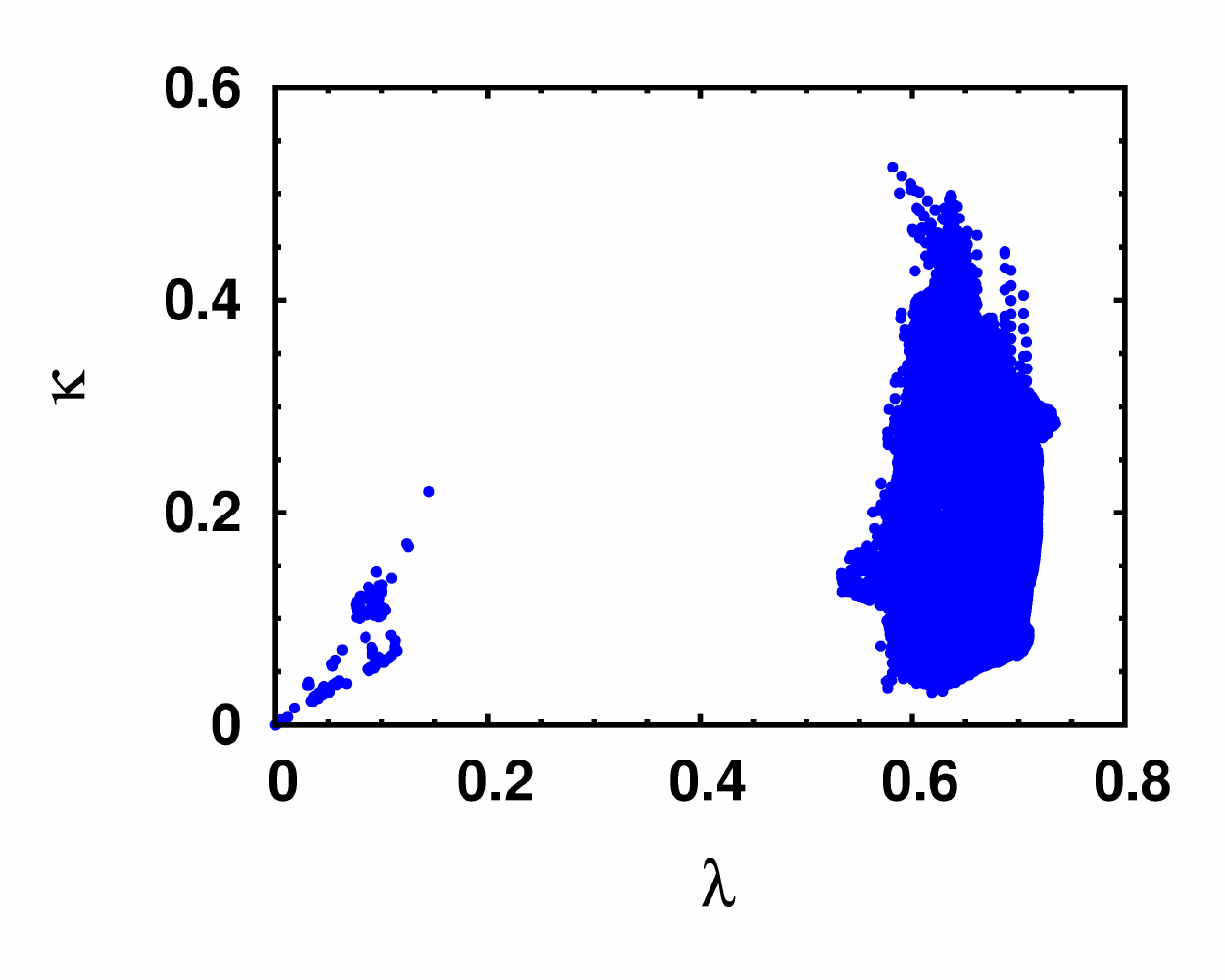}
\caption[]{ Accepted ranges  in a few  projections of the 7D NMSSM parameter space:  $\tan\beta-\mu_{eff}$, $A_\kappa-A_\lambda$ and $\lambda-\kappa$, respectively, for $m_0,m_{1/2}$=1 TeV.
}
\label{fig9}
\end{center}
\end{figure}

After fitting all cells in the 3D Higgs mass space  one obtains the  regions of the 7D parameter space allowed by the fit, which are shown for a few parameters in  two dimensional plots  in Fig. \ref{fig9}.
The parameters $\tan\beta$ and $\mu_{eff}$ show a strong negative correlation, while the trilinear couplings show a strong positive correlation for the GUT scale input parameters. 
The values of the trilinear couplings at the SUSY scale are much more restricted than their values at the GUT scale,  because of the fixed point solutions of the RGE for $A_\lambda$ and $A_\kappa$, which means that the SUSY scales are largely independent of the GUT scale values. However, the SUSY scale values depend on the chosen Higgs and SUSY masses, so the ranges of  $A_\lambda$ and $A_\kappa$ at the SUSY scale are still appreciable, as shown in Fig. 7 of Ref. \cite{Beskidt:2019mos}. 

In the  allowed range of $\lambda-\kappa$   one can identify two  preferred regions,  as shown in the right panel of Fig. \ref{fig9}: a region with NMSSM-like solutions with large couplings $\kappa$ and $\lambda$ and a region with small couplings, which is more MSSM-like, meaning a smaller mixing between the Higgs doublets and the Higgs singlet. The solutions in the different regions are quasi-degenerate in the  sense, that they describe the same mass combination on the grid and close to the same signal strengths of the 125 GeV bosons, since this was imposed by the fit and only points with a good fit, meaning a $\chi^2$ value close to zero, were accepted. 
How degenerate the solutions in a given region are depends on the $\chi^2$ cut and the flatness of the $\chi^2$ distribution, which  depends on the SUSY mass parameters as can be seen from the $\chi^2$ distributions in left  panels of Fig. \ref{fig3}. The solutions are only degenerate with respect to the SM-like Higgs boson, since  the signal strengths of the heavy scalar boson experience the $\tan\beta$ enhancement, which lifts the degeneracy,   as
discussed in one of our previous papers \cite{Beskidt:2016egy}. 

The origin of the splitting into two allowed regions can be understood if one considers the approximate expression  for the mass of the   125 GeV Higgs boson at tree level \cite{Ellwanger:2009dp}: 
\beq\label{eq4}
M_{H}^2\approx M_Z^2\cos^2 2\beta+ \Delta_{\tilde{t}} + \lambda^2 v^2 \sin^2 2\beta - \frac{\lambda^2}{\kappa^2}(\lambda-\kappa \sin 2 \beta)^2.  
\eeq 
The first two terms are present in the MSSM already, while the last two terms are unique to the NMSSM, which increase the 125 GeV Higgs mass already at tree level. 
The first tree level term can become at most $M_Z^2$ for large $\tan\beta$, so  the diffe\-rence between $M_Z$ and 125 GeV has to originate either from the lo\-garithmic stop mass corrections $\Delta_{\tilde{t}}$ or from the two remaining NMSSM terms. The two regions originate then from a trade-off between the various contributions: one solution for large $\tan\beta$ with large MSSM contributions  and one with small $\tan\beta$ leading to large NMSSM contributions.

In the high $\tan\beta$ regime the first term is maximal and the last two NMSSM terms can be smaller, meaning smaller values of the NMSSM couplings $\kappa$ and $\lambda$. In the small $\tan\beta$ regime the opposite happens and 125 GeV is only reached for large values of the couplings  $\kappa$ and $\lambda$. These two types of solutions are exemplified in the right panel of Fig. \ref{fig9}. Here we summed over all Higgs combinations, so the regions are not so small and their size furthermore depends on the mass of the SUSY particles, as shown in Fig. 5  of our previous paper \cite{Beskidt:2019mos}. But even  a single Higgs mass combination does not lead to a single point, but a range of $\tan\beta$  and a range of couplings in Fig. \ref{fig9}, meaning quasi-degenerate solutions for certain sets of parameters. Quasi-degenerate, since in all cases the final states correspond to SM signal strengths, as required by the fit. If no good fit is obtained the combination of parameters is not entered in the plot.

\subsection{Is the coverage complete? }
\label{coverage}

 If one scans over all mass combinations in the 3D mass space, does one cover all regions in the 7D parameter space?
This can only be verified by scanning over the 7D parameter space, check for allowed regions and compare these regions with the allowed regions from the 3D scan.
Unfortunately, it is prohibitive from available computing power to scan the large 7D space systematically, so one usually resorts to stochastic methods, which makes it difficult to check the completeness of the scan, especially if the parameters in 7D are highly correlated. This is the case, as shown exemplary in Fig. \ref{fig9} before. 
\begin{table}
\footnotesize
\centering
\caption{ \label{tf}
Differences of the fitting procedures for the standard fits to Higgs mass combinations $m_{A_1},m_{H_1},m_{H_3}$ in the 3D mass space (left column) and the fit parameters for the $\lambda-\kappa$ scan (right column), in which case the masses $m_{A_1},m_{H_1},m_{H_3}$ are free parameters, while the couplings $\lambda,\kappa$ are fixed. The listed input parameters are for the case that $m_{H_2}=$125 GeV, but the fits have been repeated for the case $m_{H_1}=$125 GeV. The $\chi^2$ constraints are detailed in Eq. \ref{eq5}. 
}
\begin{tabular}{l|c|c}
	\hline\noalign{\smallskip}
	Procedure & standard & $\lambda-\kappa$ scan \\
	\noalign{\smallskip}\hline\noalign{\smallskip}
	Input & $m_{A_1},m_{H_1},m_{H_3}$ & $\lambda,\kappa$\\
	Constraints & $m_{H_2}=$125 GeV, $\mu_{no-loop}=$1 & $m_{H_2}=$125 GeV, $\mu_{no-loop}=$1\\
	Output &  $\tan\beta,A_0,A_\kappa,A_\lambda,\mu_{eff},\lambda,\kappa$ & $\tan\beta,A_0,A_\kappa,A_\lambda,\mu_{eff},m_{A_1},m_{H_1},m_{H_3}$ \\
	$\chi^2$ contribution & $\chi^2_{H_1},\chi^2_{H_3},\chi^2_{A_1},\chi^2_{H_{2}},\chi^2_{\mu_{}},\chi^2_{LEP},\chi^2_{LHC}$ & $\chi^2_{H_{2}},\chi^2_{\mu_{H_2}},\chi^2_{LEP},\chi^2_{LHC}$ \\
	\noalign{\smallskip}\hline
\end{tabular}
\end{table}
We perform  the test in an alternative way: instead of scanning randomly over all seven parameters, we scan in a plane of two variables only, but for each point  in this plane the test statistic is optimized with respect to all other parameters. If the 7D parameter space would have been covered completely by the fit to the 3D mass scan, then the resulting distributions in the projected plane of two variables would be the same in both scans. In principle, any two variables can be chosen for the plane on which the 7D parameter space is projected. Given the ambiguities in the $\lambda,\kappa$ plane discussed above, we choose these variables to define the plane.

\begin{figure}[h]
\begin{center}
       \includegraphics[width=0.47\textwidth]{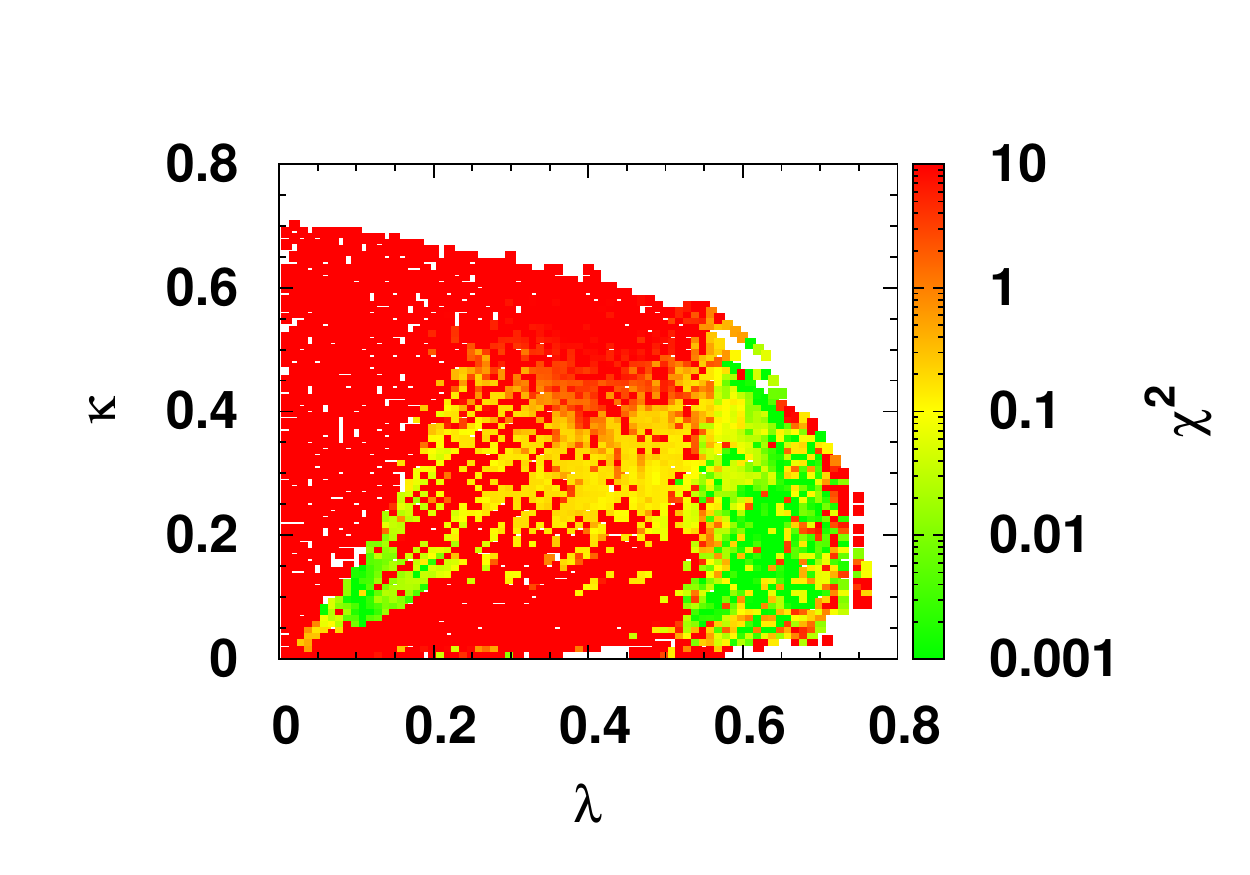} 
       \includegraphics[width=0.47\textwidth]{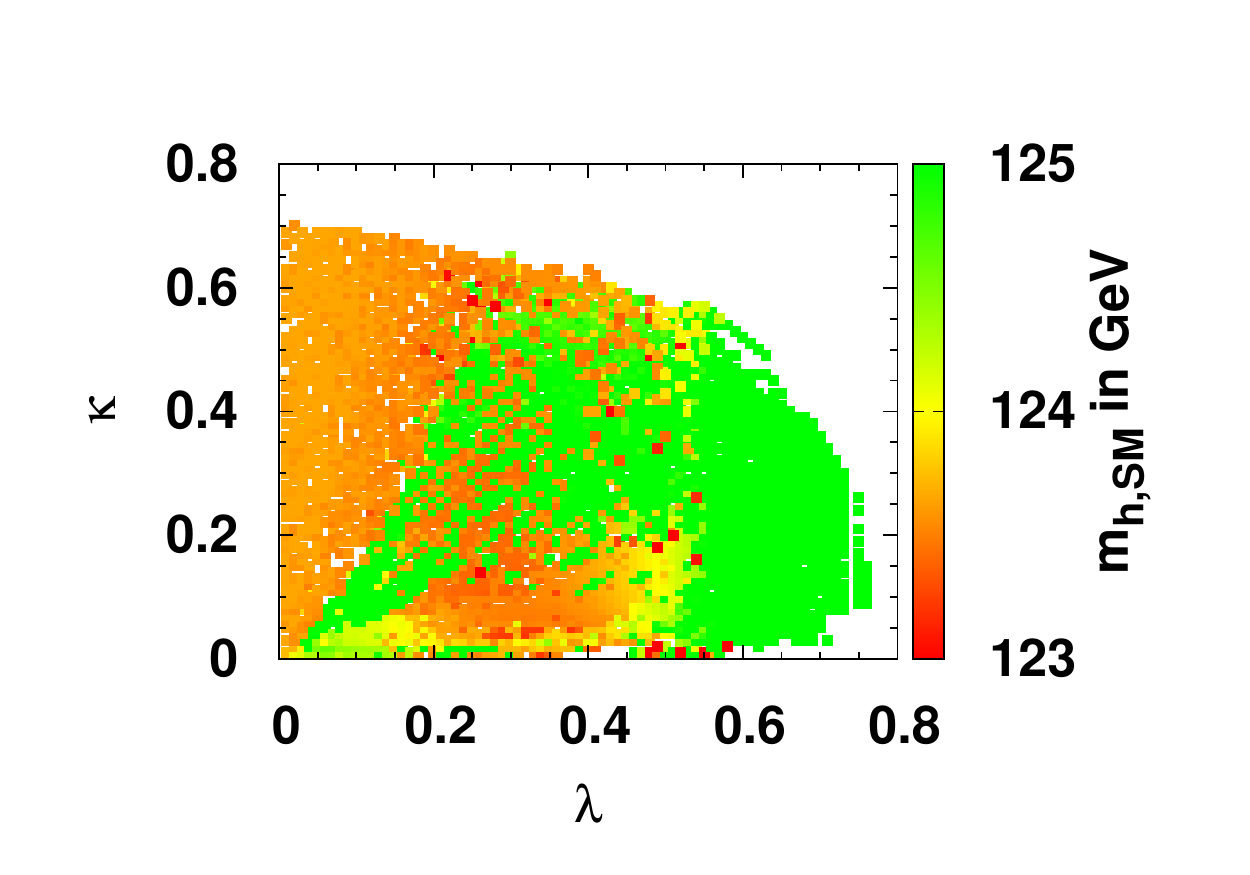}\vspace*{-3mm}
       \hspace*{0.02\textwidth}(a)\hspace*{0.46\textwidth} (b)\\[-4mm]
       \includegraphics[width=0.47\textwidth]{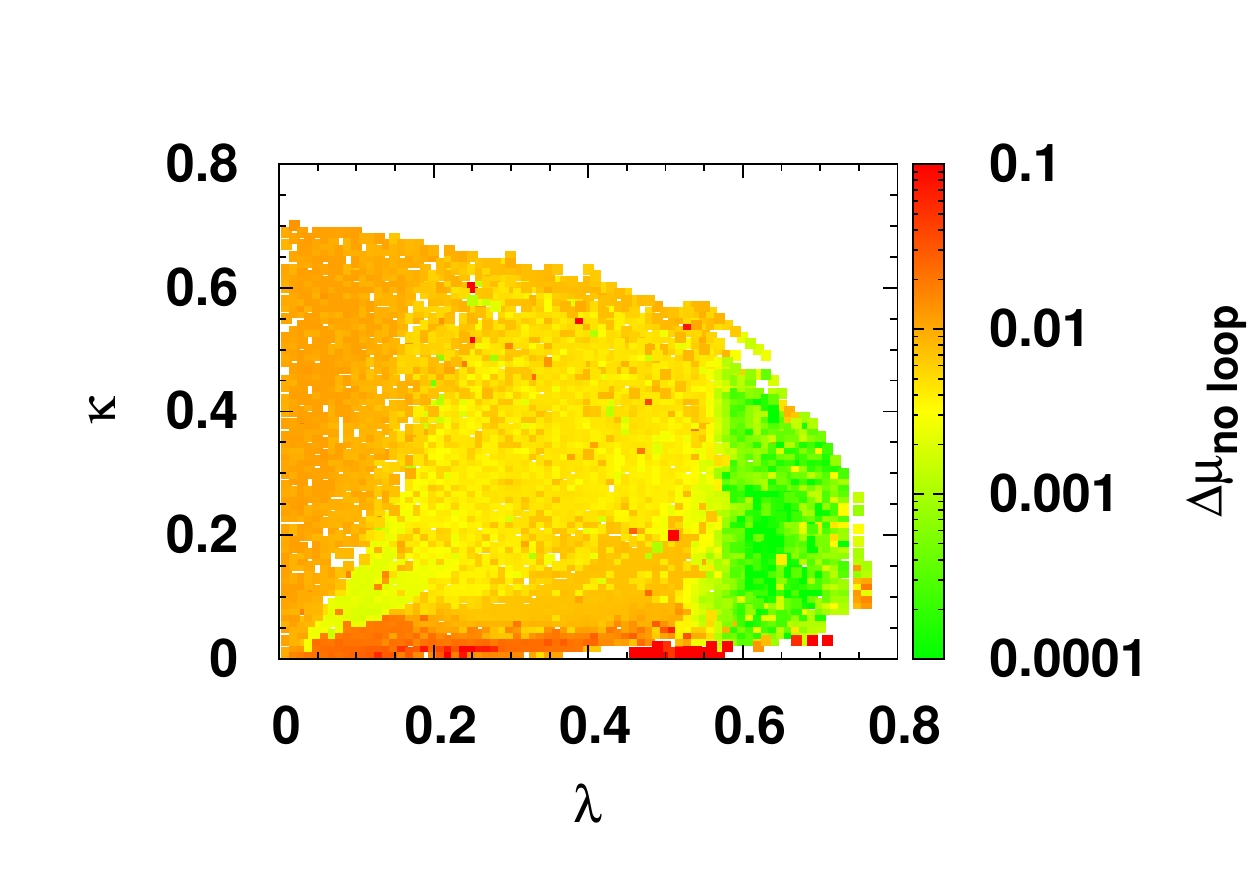}
       \includegraphics[width=0.47\textwidth]{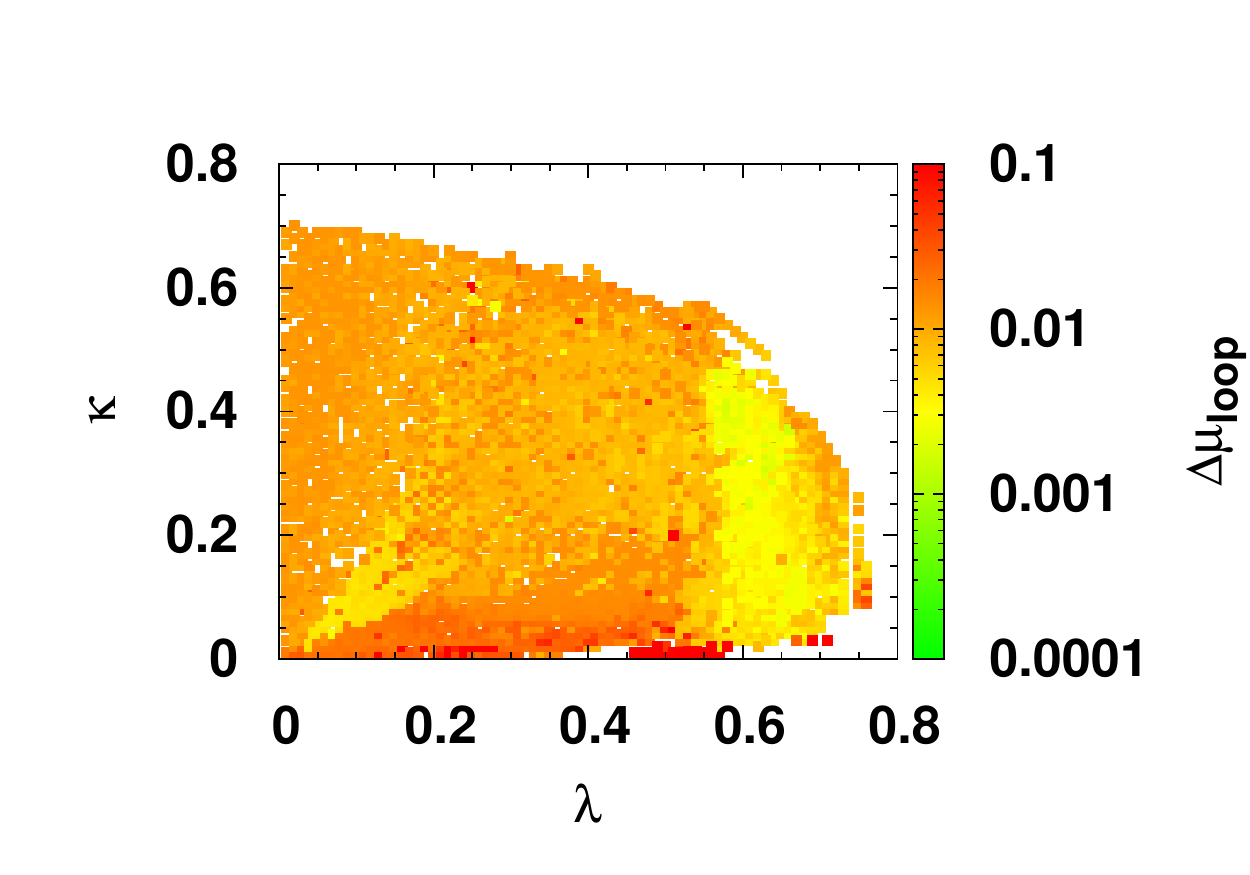}\vspace*{-3mm}
        \hspace*{0.02\textwidth}(c)\hspace*{0.46\textwidth} (d)\\[-1mm]
     \caption{ 
The  total $\chi^2$ function in the  this plane is shown by the color coding in (a). The two green regions with the smallest  $\chi^2$ values   are called Region I (II) for the region on the right (left). Here all constraints are fulfilled. Note that the absolute $\chi^2$ value depends on the choice of the error for the selected  Higgs mass combination, but the minimum of the $\chi^2$ distribution is always in the green regions. Different errors  only change the color coding.   For small values of $\lambda$ and large values of $\kappa$  the Higgs mass is too low, as shown by the color coding in (b).  For intermediate values of $\lambda$ and $\kappa$ the main contribution to the total $\chi^2$ function is coming from the fermion signal strength.
The remaining signal strength $\mu_{loop}$, shown by the averaged difference from the SM value (d), deviates from one in almost the whole $\lambda-\kappa$ plane, except for Region I. 
}
     \label{fig10}
\end{center}
   \end{figure}

 In order to do a scan in a projected plane in the 7D parameter space
the standard fit procedure was slightly modified by requiring the NMSSM parameters $\lambda$ and $\kappa$ to be fixed in the fit and leave the parameters in the 3D mass space free, including the masses.
 The differences with the standard fit procedure using a scan in the 3D mass space have been summarized in Table \ref{tf}. 

The two  regions with the smallest  $\chi^2$ values in the $\lambda$-$\kappa$ plane are  the green regions  in Fig. \ref{fig10}(a). 
These regions are in agreement with the two regions shown on the  right panel of Fig. \ref{fig9}, which is expected if the sampling in the 3D mass scans covers the same regions from sampling in the 7D NMSSM parameter space. Since no other regions were found, this is a significant consistency check, which gives us confidence that we  found all solutions.
The remaining regions - called intermediate regions - have a larger $\chi^2$-value for various reasons: for small values of $\lambda$ and large values of $\kappa$ the Higgs mass of the observed Higgs boson is too low, as can be seen from Fig. \ref{fig10}(b), where the color coding corresponds to the Higgs boson mass. The low  Higgs mass  (around 123 GeV) originates from the fact, that $m_0=m_{1/2}=$ 1 TeV was chosen.  Larger Higgs masses are obtained for larger values of $m_0$ and $m_{1/2}$  because this would lead to larger stop corrections. In addition, the fitted signal strength $\mu_{no-loop}$ deviates from 1 as can be seen from Fig. \ref{fig10}(c), where the color coding corresponds to the averaged difference from the SM value, i.e. $\Delta\mu_{no-loop}=\frac{1}{4}\sum_{i=1}^4 (\mu^i - \mu_{theo})^2/\sigma^2_{\mu}$ with $\mu^i=\mu_{\tau\tau}^{VBF/VH},\mu_{bb}^{ttH},\mu_{bb}^{VBF/VH}$ and $\mu_{ZZ/WW}^{VBF/VH}$. 
The averaged difference of the remaining signal strengths $\mu_{loop}$, i.e. $\Delta\mu_{loop}=\frac{1}{4}\sum_{i=1}^4 (\mu^i - \mu_{theo})^2/\sigma^2_{\mu}$ where $\mu_i$ includes $\mu_{\tau\tau}^{ggf},\mu_{ZZ/WW}^{ggf},\mu_{\gamma\gamma}^{VBF/VH}$ and $\mu_{\gamma\gamma}^{ggf}$ is shown in Fig. \ref{fig10}(d). 
Here, the deviations of the signal strengths $\mu_{loop}$ of the order of a few percent  correspond to the orange region.
\section{Experimental data on the signal strengths of the 125 GeV Higgs boson}
\label{error}
\begin{figure}
\begin{center}
\hspace{-1cm}
\includegraphics[width=0.6\textwidth]{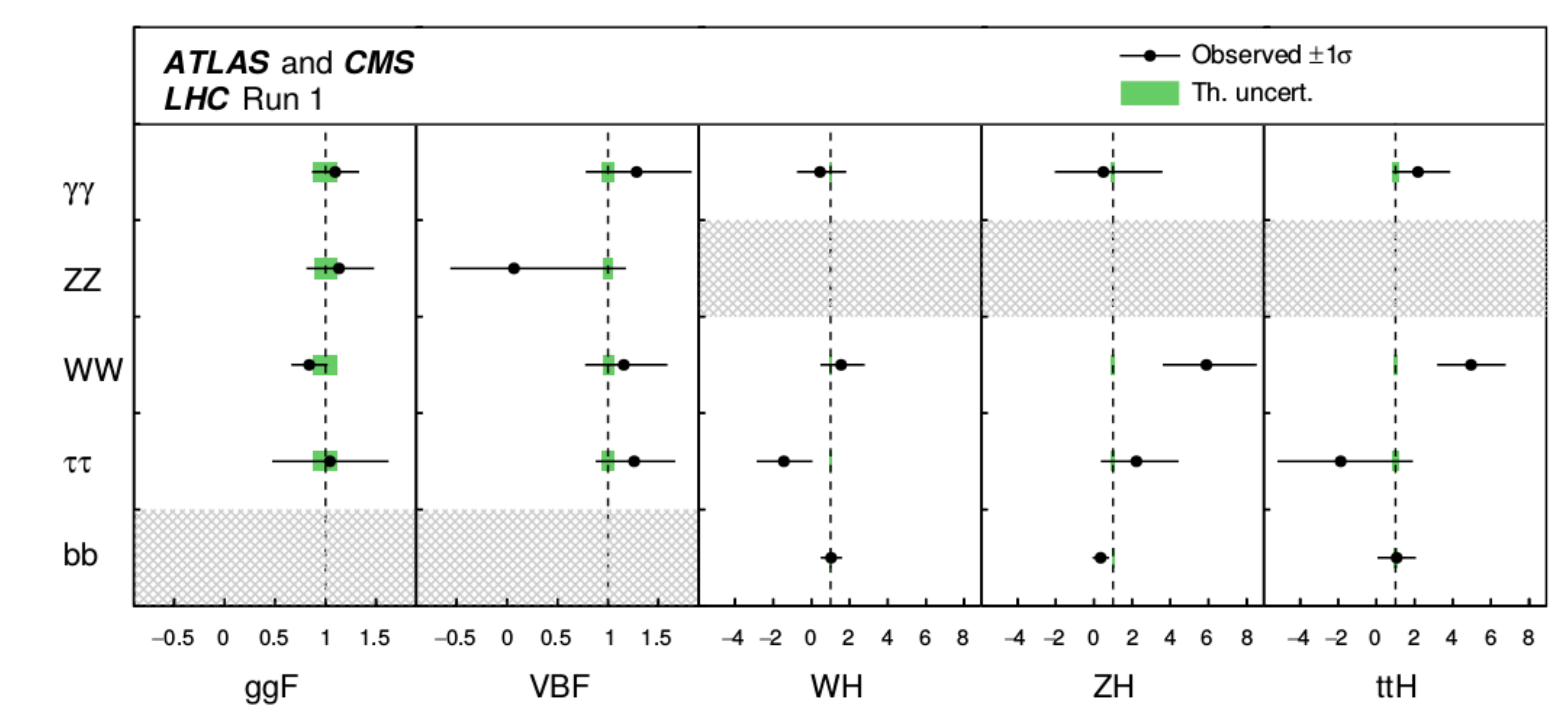}
\caption[]{ Combined measurements of the products $\sigma \cdot$ BR for the five main production  and five main decay modes on the horizontal and  vertical axis, respectively. The hatched combinations require more data for a meaningful confidence interval to be provided. The figure is taken from Ref. \cite{Tanabashi:2018oca}. 
}
\label{fig11}
\end{center}
\end{figure}

All LHC measurements have been combined by the Particle Data Group to obtain the most precise results \cite{Tanabashi:2018oca}. A Summary of the combinations is shown in Fig. \ref{fig11}.

\section{Higgs and neutralino mixing matrix in the NMSSM}
\label{higgsmixing}
The neutral components from the two Higgs doublets and singlet mix to form three physical CP-even scalar ($S$) bosons and two physical CP-odd pseudo-scalar ($P$) bosons, so the scalar Higgs bosons $H_i$, where the index $i$ increases with increasing mass, are mixtures of the CP-even weak eigenstates $H_d, H_u$ and $S$.  The elements of the corresponding mass matrices at tree level read \cite{Miller:2003ay}:
\begin{eqnarray}\label{smixing}
{\cal M}^2_{S,11}&=&M_A^2+(M_Z^2-\lambda^2v^2)\sin^22\beta,\nonumber\\
{\cal M}^2_{S,12}&=&-\frac{1}{2}(M_Z^2-\lambda^2v^2)\sin4\beta,\nonumber\\
{\cal M}^2_{S,13}&=&-\frac{1}{2}(M_A^2\sin2\beta+\frac{2\kappa\mu^2}{\lambda})\frac{\lambda v}{\mu}\cos2\beta,\nonumber\\
{\cal M}^2_{S,22}&=&M_Z^2\cos^22\beta+\lambda^2v^2\sin^22\beta,\\
{\cal M}^2_{S,23}&=& 2 \lambda \mu v \left[1 - (\frac{M_A \sin 2\beta}{2 \mu} )^2
-\frac{\kappa}{2 \lambda}\sin2\beta\right],\nonumber\\
{\cal M}^2_{S,33}&=& \frac{1}{4} \lambda^2 v^2 (\frac{M_A \sin 2\beta}{\mu})^2
+ \frac{\kappa\mu}{\lambda} (A_\kappa +  \frac{4\kappa\mu}{\lambda} )
 - \frac{1}{2} \lambda \kappa v^2 \sin 2 \beta,\nonumber 
\end{eqnarray}

\begin{eqnarray}\label{pmixing}
{\cal M}^2_{P,11}&=&\frac{\mu (\sqrt{2}A_\lambda+\kappa \frac{\mu}{\lambda})}{\sin 2 \beta}=M^2_A,\nonumber\\
{\cal M}^2_{P,12}&=&\frac{1}{\sqrt{2}}\left( M^2_A \sin 2 \beta - 3 \frac{\kappa}{\lambda}\mu^2 \right)\frac{v\lambda}{\mu},\\
{\cal M}^2_{P,22}&=&\frac{1}{2}\left( M^2_A \sin 2 \beta + 3 \frac{\kappa}{\lambda} \mu^2 \right ) \frac{v^2}{\mu^2}\lambda^2 \sin 2 \beta - \frac{3}{\sqrt{2} \frac{\kappa}{\lambda}\mu A_\kappa}.\nonumber
\end{eqnarray}

One observes that the element ${\cal M}^2_{S,22}$, which corresponds to the tree-level term of the lightest MSSM Higgs boson, can be above $M^2_Z$ because of the $\lambda^2v^2\sin^2 2 \beta$ term.
The diagonal element ${\cal M}^2_{P,11}$ at tree level corresponds to the pseudo-scalar Higgs boson in the MSSM limit of small $\lambda$, so it is called $M_A$.

Within the NMSSM the singlino, the superpartner of the Higgs singlet, mixes with the gauginos and Higgsinos, leading to an additional fifth neutralino. The resulting mixing matrix reads: \cite{Ellwanger:2009dp,Staub:2010ty}

\beq\label{neutralino}
{\cal M}_0 =
\left( \ba{ccccc}
M_1 & 0 & -\frac{g_1 v_d}{\sqrt{2}} & \frac{g_1 v_u}{\sqrt{2}} & 0 \\
0 & M_2 & \frac{g_2 v_d}{\sqrt{2}} & -\frac{g_2 v_u}{\sqrt{2}} & 0 \\
-\frac{g_1 v_d}{\sqrt{2}} & \frac{g_2 v_d}{\sqrt{2}} & 0 & -\mu_\mathrm{eff} & -\lambda v_u \\
\frac{g_1 v_u}{\sqrt{2}} & -\frac{g_2 v_u}{\sqrt{2}}& -\mu_\mathrm{eff}& 0 & -\lambda v_d \\
0& 0& -\lambda v_u&  -\lambda v_d & 2 \kappa s 
\ea \right)
\eeq
with the gaugino masses $M_1$, $M_2$, the gauge couplings $g_1$, $g_2$ and the Higgs mixing parameter $\mu_{eff}$ as parameters.
Furthermore, the vacuum expectation values of the two Higgs doublets $v_d$,$v_u$, the singlet $s$ and the Higgs couplings $\lambda$ and $\kappa$ enter the neutralino mass matrix. 
 
The upper left $4 \times 4$ submatrix of the neutralino mixing matrix corresponds to the MSSM neutralino mass matrix, see e.g. Ref. \cite{Martin:1997ns}. 

The neutralino mass eigenstates are obtained from the diagonalization of ${\cal M}_0$ in Eq. \ref{neutralino} and are linear combinations of the gaugino and Higgsino states: 
\beq\label{neutralino1}
\footnotesize{
\tilde{\chi}^0_i = {\cal N}(i,1) \left | \tilde{B} \right\rangle +{\cal N}(i,2) \left | \tilde{W}^0 \right\rangle 
  +{\cal N}(i,3) \left | \tilde{H}^0_u \right\rangle +{\cal N}(i,4) \left | \tilde{H}^0_d \right\rangle+{\cal N}(i,5) \left | \tilde{S} \right\rangle.}
\eeq

Typically, the diagonal elements in Eq. \ref{neutralino} dominate over the off-diagonal terms, so the neutralino masses are of the order of $M_1$, $M_2$, while  the heavier Higgsinos are of the order of the  mixing parameter $\mu_{eff}$  and   the lightest (singlino-like)  neutralino is of the order of $2 \kappa s \ = \ 2 (\kappa/ \lambda) \mu_{eff}$.

\section{Output of NMSSMTools for  six representative  benchmark points  P1 to P6}
\label{output}
\hspace{1.5cm}

In Sect. \ref{single} examples of mass combinations for the three possible cases for ``genuine'' NMSSM deviations of  the signal strengths of the 125 GeV Higgs boson from the SM-expectation were discussed. To see which parameters the fit changes to impose the  deviations from the SM-expectation (regulated by $\mu_{theo}$)  we present for each example the parameters without (with) deviation, i.e. $\mu_{theo}$, which can be used as benchmark points to search for  ``genuine'' NMSSM effects. For CASE I we select the mass combinations P1 and P2,  where P2 corresponds to the deviations presented  in Fig. \ref{fig4}. Similarly, we select P3 and P4 for CASE II, where P4 corresponds to the deviations presented in Fig. \ref{fig5} and  P5 and P6  for CASE III with the deviations presented in Fig. \ref{fig6}.   
The fitted NMSSM parameters for the representative mass combinations  are listed in Table \ref{t1}. Here one compare the difference in NMSSM parameters required by the fit for a deviation of zero and 30\%, respectively (corresponding to  $\mu_{theo}$=1 or 0.7). From these  parameters  NMSSMTools calculates all masses  (shown in Table \ref{t2}) and the Higgs and neutralino mixing matrices (shown in Table \ref{t3}). 
The calculated reduced couplings are listed in Table \ref{t4}.  The calculated signal strengths  are shown in Table \ref{t5}. One observes that the value of $\mu_{sel}= \mu_{\tau\tau}^{VBF/VH}$ of one or 0.7 is reached for $H_2$ for all six examples. The other fermionic final states follow closely, but for the bosonic final states the deviations are either the same or of an opposite sign, as can be seen from the second block in  Table \ref{t5} for the rows corresponding to P2, P4 and P6. For P2 and P6 one finds a positive correlation between the fermionic and bosonic signal strengths, while for P4 one finds a negative correlation, as discussed in Sec. \ref{single}. The BRs are shown in  Table \ref{t6}.  If the deviation in the signal strengths originate from changes in couplings or BRs can be seen from a comparison of Table  \ref{t5}  with Tables \ref{t4} and \ref{t6}. All values in the tables are obtained from the output of NMSSMTools. All points use $m_0=m_{1/2}=1$ TeV. 
     
\begin{table}
\centering
\caption{List of fitted NMSSM parameters for  representative  mass combinations  P1 to P6, which used $\mu_{sel}=\mu_{\tau\tau}^{VBF/VH}$ and either $\mu_{theo}=1$ or 0.7, as indicated in the second row. \label{t1}}
\begin{tabular}{l|c|c||c|c||c|c|}
	\hline\noalign{\smallskip}
P & 1 & 2 & 3 & 4 & 5 & 6 \\
	\noalign{\smallskip}\hline\noalign{\smallskip}
$\mu_{\tau\tau}^{VBF/VH}$ & 1 & 0.7 & 1 & 0.7 & 1 & 0.7 \\
	\noalign{\smallskip}\hline\noalign{\smallskip}
	$\tan\beta$ &   4.12  & 3.96 & 5.36 & 6.57 & 12.87 &  19.47   \\ 
	$A_0$ in GeV &  -654.85 &  -553.88 & -617.09 & 520.12 & -2467.63 & -2295.48   \\
	$A_\kappa$ in GeV& 3779.52 &  3724.97 &  4673.16 &  4444.88  &  -156.54 & -156.55    \\
	$A_\lambda$ in GeV &  4325.18 & 4415.52 & 3300.37 & 3553.43 & 319.80 & -126.42  \\ 
	$\lambda \cdot 10^{-1}$&  6.31 &  6.31 & 6.97 &  6.50 & 0.04 &   0.04 \\ 
	$\kappa \cdot 10^{-1}$ &  0.37 & 0.28 & 2.51 &  3.20 & 0.03 &  0.04 \\
	$\mu_{eff}$ in GeV &   459.38 &  475.24 & 184.48 &  146.19 & 103.64 & 104.11   \\
	\noalign{\smallskip}\hline
\end{tabular}
\end{table}

\begin{table}
\centering
\caption{Masses of Higgs bosons, sparticles and gauginos in GeV for the representative  mass combinations P1 to P6 for $m_0=m_{1/2}=1$ TeV. The values of   $m_{H_1}, m_{H_3}$ and $m_{A_1}$ represent the chosen mass combination in the grid of Fig. \ref{fig2}. One observes the approximate mass degeneracy of the heavy Higgs masses ($m_{H_3}, m_{A_2}$ and  $m_{H^\pm}$). $m_{H_2}$ is the observed 125 GeV Higgs mass and $m_{H_1}$ the singlet-like boson mass (see large value of $S_{1s}$ in Table \ref{t3}).  \label{t2}}
\begin{tabular}{l|c|c||c|c||c|c|}
	\hline\noalign{\smallskip}
P & 1 & 2 & 3 & 4 & 5 & 6 \\
	\noalign{\smallskip}\hline\noalign{\smallskip}
$\mu_{\tau\tau}^{VBF/VH}$ & 1 & 0.7 & 1 & 0.7 & 1 & 0.7 \\
	\noalign{\smallskip}\hline\noalign{\smallskip}
	$m_{H_1}$ & 90.0 & 90.0 & 90.0 & 90.0 &  122.9  &  122.9 \\
	$m_{H_2}$ &  125.2 & 125.2 &  125.2 & 125.2 &  125.2  &  125.3 \\ 
	$m_{H_3}$ &  2000.0 & 2000.0 & 1000.0 & 1000.0 & 1300.8   &  1300.0 \\ 
	$m_{A_1}$ & 200.0 & 200.0 & 200.0 & 200.0 & 200.0  & 200.0  \\
	$m_{A_2}$ &  2000.5 & 2000.6 & 998.1 & 997.7 & 1300.7  & 1300.0  \\
	$m_{H^\pm}$ & 1996.5 & 1996.6 &  990.1 & 990.7 &  1303.4 & 1302.7    \\
	$m_{\tilde{d}_L}$ &   2214.6 & 2214.3 & 2211.4 & 2209.9 & 2217.3 & 2218.3    \\ 
	$m_{\tilde{d}_R}$& 2137.3 & 2137.0 & 2128.1 & 2128.6 &  2134.1 &   2136.6 \\ 
	$m_{\tilde{u}_L}$ &   2213.4 & 2213.1 & 2210.2 & 2208.6 & 2216.0 &  2217.0   \\
	$m_{\tilde{u}_R}$ &  2165.3 & 2165.0 & 2184.8 & 2173.3 & 2187.0 &  2180.8    \\
	$m_{\tilde{s}_L}$ &   2214.6 & 2214.3 & 2211.4 & 2209.9 & 2217.3 & 2218.3    \\ 
	$m_{\tilde{s}_R}$&  2137.3 & 2137.0 & 2128.1 & 2128.6 & 2134.1 &  2136.6 \\ 
	$m_{\tilde{c}_L}$ &  2213.4 & 2213.1 & 2210.2 & 2208.6 & 2216.0 & 2217.0  \\
	$m_{\tilde{c}_R}$ & 2165.3 & 2165.0 &  2184.8 & 2173.3 & 2187.0 &  2180.7    \\
	$m_{\tilde{b}_1}$ &  1773.2 & 1777.3 &  1796.2 & 1884.5 & 1706.0 &  1701.9   \\ 
	$m_{\tilde{b}_2}$ & 2131.4 &  2131.6 & 2121.0 & 2122.2 &  2083.7 & 2030.4  \\ 
	$m_{\tilde{t}_1}$ & 1064.3 & 1078.1 & 1189.0 & 1427.4 &  950.8 & 1034.0   \\
	$m_{\tilde{t}_2}$ &   1792.4 & 1796.3 & 1814.5 & 1899.8 & 1731.7 & 1728.3    \\
	$m_{\tilde{e}_L}$ &   1206.3 & 1206.2 & 1233.7 & 1222.9 & 1232.0 & 1224.2   \\ 
	$m_{\tilde{e}_R}$&    1021.0 & 1021.2 & 959.2 & 986.2 & 968.0 &  987.5 \\ 
	$m_{\tilde{\nu}^e_L}$ &  1204.1 & 1204.0 & 1231.5 & 1220.6 & 1229.6 &   1221.7  \\
	$m_{\tilde{\mu}_L}$ &  1206.3 & 1206.2 &  1233.7 & 1222.9 & 1232.0 &  1224.2  \\ 
	$m_{\tilde{\mu}_R}$&  1021.0 &  1021.2 & 959.2 & 986.2 & 968.0 &  987.5  \\ 
	$m_{\tilde{\nu}^\mu_L}$&1204.1 & 1204.0 & 1231.5 & 1220.6 &  1229.6 &1221.7   \\
	$m_{\tilde{\tau}_1}$ & 1016.1 & 1016.7 &  953.5 & 981.0 & 910.0  & 860.9   \\ 
	$m_{\tilde{\tau}_2}$& 1204.3 & 1204.3 &  1231.5 & 1220.8 & 1209.9  &   1176.0  \\ 
	$m_{\tilde{\nu}^\tau_L}$&1202.1& 1202.1 & 1229.2 & 1218.5 & 1207.4 &  1173.3   \\
	$m_{\tilde{g}}$ &  2237.4 & 2237.3 & 2239.2 & 2240.0 & 2240.3 &  2239.3   \\
	$m_{\tilde{\chi}^0_1}$&  62.5 &  52.1 & 103.2 & 88.8 & 98.1 &  99.0      \\
	$m_{\tilde{\chi}^0_2} $& 407.2 & 410.7 & -220.3 &  -179.5 &  -110.9 & -111.6   \\
	$m_{\tilde{\chi}^0_3} $&  483.9 & 494.7 & 238.2 & 224.7 & 174.8 & 174.8   \\
	$m_{\tilde{\chi}^0_4} $&  -483.9 & -499.8 & 433.6 & 431.2 & 431.8 &  431.7  \\
	$m_{\tilde{\chi}^0_5} $&  831.1 & 831.8 & 823.1 & 820.0 & 824.2 & 824.3   \\
	$m_{\tilde{\chi}^\pm_1} $& 454.0 & 469.2 &  183.8 & 146.2 & 104.1 &  105.0   \\
	$m_{\tilde{\chi}^\pm_2} $&  831.0 & 831.6 & 823.1 & 820.0 & 824.2 &  824.2  \\
	\noalign{\smallskip}\hline
\end{tabular}
\end{table}

\begin{table}
\centering
\caption{Higgs mixing matrix elements (multiplied by 100 to exhibit small values) for the scalar Higgs bosons $S_{ij}$ with $i=1,2,3$ and $j=d,u,s$. The Higgs mixing matrix elements for the pseudo-scalar Higgs bosons are labeled as $P_{ij}$ with $i=1,2$ and $j=d,u,s$ while the neutralino mixing matrix elements are denoted by $N_{ij}$ with $i=1,...,5$ and $j=1,...,5$. The components are defined in Eq. \ref{neutralino1}.
One observes that for the mass combinations P1, P3 and P5 the 125 GeV Higgs boson has a large $S_{2u}$ component and a small singlet $S_{2s}$ component. The  singlet  component $S_{2s}$  increases in CASE II and CASE III, as shown for  P4 and P6 with $\mu_{\tau\tau}^{VBF/VH}$=0.7. For CASE I  the lightest neutralino (the dark matter candidate) is singlino-like with $N_{15}$=0.97, while for the other cases  with larger mixing (P3 to P6) the singlino component $N_{15}$  decreases and the Higgsino components $N_{13}$ and $N_{14}$ increase. \label{t3}}
\begin{tabular}{l|c|c||c|c||c|c|}
	\hline\noalign{\smallskip}
P & 1 & 2 & 3 & 4 & 5 & 6 \\
	\noalign{\smallskip}\hline\noalign{\smallskip}
$\mu_{\tau\tau}^{VBF/VH}$ & 1 & 0.7 & 1 & 0.7 & 1 & 0.7 \\
	\noalign{\smallskip}\hline\noalign{\smallskip}
	$S_{1d}$ &   4.64 &  4.83 &  11.41 & 15.83 & 0.12 & -2.86   \\
	$S_{1u}$ &  -1.60 & -0.59 & -2.54 & 30.50 &  0.83 & -55.51  \\
	$S_{1s}$ &  99.88 & 99.88  & 99.31 & 93.91 & 99.99 & 83.13 \\
	$S_{2d}$ & 23.62 &  24.48 & 18.42 & 10.52 &  7.86  & 4.34  \\
	$S_{2u}$ &  97.17 &  96.96 & 98.29 &  94.05 & 99.69  & 83.02 \\
	$S_{2s}$ &  0.46 & -0.62 & 0.40 & -32.32 & -0.83  &  55.58  \\
	$S_{3d}$ &   97.06 & 96.84 &  97.62 & 98.18 & 99.69  &  99.86  \\
	$S_{3u}$ & -23.57 &  -24.48 & -18.24 &  -15.00 &  -7.86 &  -5.20\\
	$S_{3s}$ &  -4.88 & -4.83 & -11.69 &  -11.68 & -0.05  &  -0.04 \\
	\noalign{\smallskip}\hline\noalign{\smallskip}
	$P_{1d}$ &   -5.08 & -5.13 & -9.43 & -9.13 & -0.05  & -0.03  \\
	$P_{1u}$ &  -1.23 & -1.29 & -1.76 & -1.39 &  $<$ -0.01  & $<$ -0.01   \\
	$P_{1s}$ &  99.86 & 99.86 &  99.54 & 99.57 &  99.99 & 99.99   \\
	$P_{2d}$ &   97.04 & 96.81 & 97.85 & 98.44 & 99.70  &  99.87  \\
	$P_{2u}$ &   23.56 &  24.46  & 18.26 & 14.98 & 7.75  & 5.13 \\
	$P_{2s}$ &    5.23 & 5.29 & 9.59 &  9.24 & 0.05  &  0.03     \\
	\noalign{\smallskip}\hline\noalign{\smallskip}
	$N_{11}$ & 2.48 &  2.30 &  8.90 & 9.48 & 9.55  &  9.34  \\
	$N_{12}$ & -2.26 & -2.12 & -7.61 & -8.27 &  -8.19 & -8.00 \\
	$N_{13}$ &   -1.83 & -2.59 & 31.09 & 41.22 & 72.51  &  72.53 \\
	$N_{14}$ &  -22.24 & -21.45 &  -62.58 & -68.91 & -67.70  & -67.73   \\
	$N_{15}$ &   97.42 & 97.59  & 70.57 & 58.26 & 0.66  & 0.68  \\
	$N_{21}$ &   83.62 & 87.64 &  -3.39 & -3.78 & -5.32 & -5.47    \\
	$N_{22}$ &   -8.55 & -7.61 &  3.98 & 4.32 & 5.78 &  5.94   \\
	$N_{23}$ & 40.14 & 35.43  & 70.42 & 70.86 &  68.76  & 68.74  \\
	$N_{24}$ &  -35.08 & -30.67 & 65.19 & 64.68 & 72.18  &  72.17     \\
	$N_{25}$ &  -9.58 & -8.03 & 27.64 & 27.60 & 0.19  &  0.19  \\
	$N_{31}$ &   -54.69 & 47.99 & 11.46 & -7.93 & -0.06 &  -0.06  \\
	$N_{32}$ &  -16.71 & 18.17 &  -7.06 & 5.09 &  0.04 &  0.05 \\
	$N_{33}$ & 57.86 &  -60.72 & 63.23 & -56.87 &  -0.61  &  -0.63   \\
	$N_{34}$ &  -57.11 & 59.69 &   -39.62 & 28.89 & 0.30  & 0.32   \\
	$N_{35}$ &   -10.95 & 10.77  & -65.20 & 76.43 & 99.99  &  99.99 \\
	$N_{41}$ &  -2.38 & -2.31 & 98.88 & 99.15 & 99.39 & 99.41 \\
	$N_{42}$ &   3.13 & 3.05  &  2.86 & 2.53 &  2.19  & 2.17 \\
	$N_{43}$ &  70.10 & 70.09 & -7.66 & -5.74 &  -3.26 &  -3.02   \\
	$N_{44}$ &  69.09 & 69.12  &  12.34 & 11.24 &  10.26 & 10.23 \\
	$N_{45}$ &   17.23 & 17.18 & 2.15 & 1.59 & $<$ 0.01  & $<$ 0.01  \\
	$N_{51}$ &  -1.89 & -1.97 & -1.22 & -1.19 & -1.10 & -1.09    \\
	$N_{52}$ &   98.15 & 97.97 & 99.34 & 99.40 & 99.47 & 99.48 \\
	$N_{53}$ & 11.07 &  11.78  &  4.27 & 3.41 & 2.05 & 1.79  \\
	$N_{54}$ &  -15.50 &  -16.07 & -10.57 & -10.32 & -9.99 & -9.98   \\
	$N_{55}$ &  -1.01 &  -1.04 & -0.40 & -0.31 &  $<$ -0.01 &  $<$ 0.01  \\
	\noalign{\smallskip}\hline
\end{tabular}
\end{table}

\begin{table}
\centering
\caption{The reduced couplings for all Higgs bosons to up-type fermions $c_u$, down-type fermions $c_d$, b-quarks $c_b$, W/Z-bosons $c_{W/Z}$, gluons $c_{gluon}$ and photons $c_\gamma$. One observes that for CASE I, represented by P1 and P2, the reduced couplings of $H_2$ are close to 1, even if the signal strength $\mu_{\tau\tau}^{VBF/VH}$=0.7.  The deviations from the SM-like signal strength in P2 is caused by the change in BRs to neutralinos, as can be seen from the last row in Table \ref{t6}. For the other cases the signal strength changes by a change of the reduced couplings,  see columns for P4 and P6 in the $H_2$ block. \label{t4}}
\begin{tabular}{ll|c|c||c|c||c|c|}
	\hline\noalign{\smallskip}
\multicolumn{2}{c|}{ P }& 1 & 2 & 3 & 4 & 5 & 6 \\
	\noalign{\smallskip}\hline\noalign{\smallskip}
\multicolumn{2}{c|}{$\mu_{\tau\tau}^{VBF/VH}$} & 1 & 0.7 & 1 & 0.7 & 1 & 0.7 \\
	\noalign{\smallskip}\hline\noalign{\smallskip}
\multirow{6}{*}{$H_1$} &	$c_u$ &  -0.017 &   -0.006 & -0.026 &  0.308 &0.008 & -0.556    \\
	& $c_d$ &   0.196 &  0.197 & 0.622 & 1.052 & 0.016 & -0.557    \\ 
	& $c_b$ &  0.196 &   0.196 &  0.622 & 1.050 & 0.016 &  -0.557  \\ 
	& $c_{W,Z}$ & -0.004 &  0.325 & -0.005 & 0.006 &   0.008 & -0.556    \\
	& $c_{gluon}$ &  0.046 & 0.038  & 0.126 & 0.291 & 0.008 & 0.553 \\
	& $c_\gamma$ &  0.053 & 0.041 &  0.139 & 0.177 & 0.007 &  0.560 \\ 
	\noalign{\smallskip}\hline\noalign{\smallskip}
\multirow{6}{*}{$H_2$} &	$c_u$ &  1.000 & 1.000 & 1.000 &  0.951 & 1.000 & 0.831  \\
	& $c_d$ & 1.001 & 0.999 & 1.004 &  0.699 &  1.014 &  0.847   \\ 
	& $c_b$ & 1.001 & 0.999 &  1.004 &  0.7000 & 1.014 &  0.847   \\ 
	& $c_{W,Z}$ &   1.000 & 1.000 & 1.000 & 0.946 & 1.000 &   0.831   \\
	& $c_{gluon}$ &  0.999 & 0.999  & 1.000 & 0.967 & 0.993 & 0.827   \\
	& $c_\gamma$ &   1.003 & 1.003  & 1.006 & 1.000 &  1.007 & 0.834   \\ 
	\noalign{\smallskip}\hline\noalign{\smallskip}
\multirow{6}{*}{$H_3$} &	$c_u$ & -0.243 & -0.252 &-0.186 & -0.152 &  -0.079 & -0.052  \\
	& $c_d$ & 4.114 & 3.953 & 5.322 &  6.524 &  12.868 &  19.471  \\ 
	& $c_b$ & 4.107 & 3.945 &  5.319 & 6.512 & 12.899 & 19.511   \\ 
	& $c_{W,Z}$ & $<$ -0.001 & $<$ -0.001 &  $<$ -0.001 & $<$ - 0.001 & -0.001 &  $<$ -0.001    \\
	& $c_{gluon}$ &  0.230 & 0.240 & 0.176 &  0.141 & 0.064 &  0.063  \\
	& $c_\gamma$ &   0.112 & 0.115 & 0.352 & 0.300 & 0.080 & 0.060   \\ 
	\noalign{\smallskip}\hline\noalign{\smallskip}
\multirow{6}{*}{$A_1$} &	$c_u$ & -0.013 & -0.013 & -0.018 & -0.014 & $<$ 0.001 & $<$ 0.001   \\
	& $c_d$ &-0.215 & -0.209 &  -0.514 & -0.607 &  -0.006  & -0.005  \\ 
	& $c_b$ &  -0.215 & -0.209  & -0.514 & -0.606 & -0.006  &  -0.005 \\ 
	& $c_{gluon}$ & 0.015  &   0.016 & 0.021 & 0.021 & $<$ 0.001  &  $<$ 0.001  \\
	& $c_\gamma$ &   0.048  & 0.046  & 0.150 & 0.195 &   0.003 &  0.003    \\ 
	\noalign{\smallskip}\hline\noalign{\smallskip}
\multirow{6}{*}{$A_2$} &	$c_u$ & 0.242  & 0.252  &  0.186 & 0.152 & 0.078 & 0.051 \\
	& $c_d$ & 4.114 &  3.952 &  5.335 & 6.542 & 12.870 & 19.472  \\ 
	& $c_b$ & 4.106 &  3.944 &  5.331 & 6.530 & 12.900  &  19.512 \\ 
	& $c_{gluon}$ & 0.272 &  0.283 & 0.226 &  0.191 &  0.128 &  0.125 \\
	& $c_\gamma$ &   0.131  & 0.135 & 0.416 & 0.353 & 0.113 & 0.100 \\ 
	\noalign{\smallskip}\hline
\end{tabular}
\end{table}

\begin{table}
\centering
\caption{Signal strengths from Eq. \ref{coupling5} for the scalar Higgs bosons $H_1, H_2$ and $H_3$ in the blocks  separated by horizontal lines.  One observes the correlated deviations of the signal strengths into \textit{fermionic} and \textit{bosonic} final states  for P2 and P6 and the anti-correlated changes in case of P4 by  comparing e.g. $H_2 \rightarrow bb$ with $H_2 \rightarrow ZZ/WW$. For P2 the deviation from the SM-like signal strength is caused by the increase into invisible states $H_2 \rightarrow invisible$ (last 2 lines of the middle block). For P4 the deviation from the SM-like signal strength is caused by a decrease of the BRs to down-type fermions, which is compensated by an increase of the BRs into bosons (compare  e.g. BR($H_2 \rightarrow bb$ and BR($H_2 \rightarrow WW$) in Table \ref{t6}).  For P6 the  deviation from the SM-like signal strength is caused by a correlated decrease of all reduced couplings, as shown in the last column of Table \ref{t4}  in the $H_2$ block. The explanations are given in  Sect. \ref{single}. \label{t5}}
\begin{tabular}{l|c|c||c|c||c|c|}
	\hline\noalign{\smallskip}
P & 1 & 2 & 3 & 4 & 5 & 6 \\
	\noalign{\smallskip}\hline\noalign{\smallskip}
$\mu_{\tau\tau}^{VBF/VH}$ & 1 & 0.7 & 1 & 0.7 & 1 & 0.7 \\
	\noalign{\smallskip}\hline\noalign{\smallskip}
	$VBF/VH \rightarrow H_1 \rightarrow \tau\tau$ &  $<$ 0.0001 & $<$ 0.0001 & $<$ 0.0001  & 0.1151 & $<$ 0.0001 & 0.3095 \\
	$ggf \rightarrow H_1 \rightarrow \tau\tau$ &  0.0023 & 0.0016 & 0.0173 & 0.0922 & $<$ 0.0001 &  0.3066  \\
	$VBF/VH \rightarrow H_1 \rightarrow bb$ & $<$ 0.0001 & $<$ 0.0001 &  $<$ 0.0001 & 0.1135 & $<$ 0.0001  & 0.3095  \\
	$ggf \rightarrow H_1 \rightarrow bb$ & 0.0003 & $<$ 0.0001 & 0.0007 & 0.1021 &  $<$ 0.0001  & 0.3095 \\
	$VBF/VH \rightarrow H_1 \rightarrow ZZ/WW$ & 0.0000 & 0.0000 &  0.0000 &  0.0000 & $<$ 0.0001 & 0.3081  \\
	$ggf \rightarrow H_1 \rightarrow ZZ/WW$ & 0.0000 & 0.0000 &  0.0000 &  0.0000  & $<$ 0.0001 & 0.3052 \\
	$VBF/VH \rightarrow H_1 \rightarrow \gamma\gamma$ &  $<$ 0.0001 & $<$ 0.0001 &  $<$ 0.0001 &  0.0032 & $<$ 0.0001 &  0.3124  \\
	$ggf \rightarrow H_1 \rightarrow \gamma\gamma$ &   0.0002 & $<$ 0.0001 & 0.0009 & 0.0026 & $<$ 0.0001 &  0.3095    \\
	\noalign{\smallskip}\hline\noalign{\smallskip}
	\noalign{\smallskip}\hline\noalign{\smallskip}
	$VBF/VH \rightarrow H_2 \rightarrow \tau\tau$ &  1.0009  & 0.7000 & 1.0026 & 0.7002 &  1.0100  & 0.7002     \\
	$ggf \rightarrow H_2 \rightarrow \tau\tau$ & 0.9984 &  0.6989 &  1.0026 & 0.7319 & 0.9951  &  0.6923  \\
	$VBF/VH \rightarrow H_2 \rightarrow bb$ &  1.0008  & 0.7000 & 1.0024 & 0.7100 & 1.0095  &  0.6998  \\
	$ggf \rightarrow H_2 \rightarrow bb$ & 1.0007 & 0.7001 & 1.0021 & 0.7186 & 1.0093  &   0.6997  \\
	$VBF/VH \rightarrow H_2 \rightarrow ZZ/WW$ &  0.9981 &  0.7009 & 0.9946 & 1.2802 & 0.9821 &  0.6739  \\
	$ggf \rightarrow H_2 \rightarrow ZZ/WW$ &    0.9955  & 0.6999  & 0.9946 & 1.3381 & 0.9675 & 0.6663 \\
	$VBF/VH \rightarrow H_2 \rightarrow \gamma\gamma$ & 1.0041 & 0.7057  &  1.0057 & 1.4317 & 0.9959 &  0.6788 \\
	$ggf \rightarrow H_2 \rightarrow \gamma\gamma$ &   1.0015 &  0.7047 & 1.0057 & 1.4964 & 0.9812 & 0.712  \\
	$VBF/VH \rightarrow H_2 \rightarrow invisible$ &   0.0002  &  0.2997 & 0.0000 & 0.0000 & 0.0000 & 0.0000  \\
	$ggf \rightarrow H_2 \rightarrow invisible$ &  0.0002 &   0.2993 & 0.0000 & 0.0000 & 0.0000 & 0.0000 \\
	\noalign{\smallskip}\hline\noalign{\smallskip}
	\noalign{\smallskip}\hline\noalign{\smallskip}
	$VBF/VH \rightarrow H_3 \rightarrow \tau\tau$ &$<$ 0.0001 & $<$ 0.0001 &  $<$ 0.0001 & 0.0004 &  0.0206 & 0.0149  \\
	$ggf \rightarrow H_3 \rightarrow \tau\tau$ &  69.2039 &  69.3713 & 24.7435 & 24.6346 & 68.8573 &  113.7163  \\
	$VBF/VH \rightarrow H_3 \rightarrow bb$ & $<$ 0.0001  & $<$ 0.0001 & $<$ 0.0001 & 0.0002 &  0.0142  & 0.0103 \\
	$ggf \rightarrow H_3 \rightarrow bb$ &  57.2843 & 57.0001  & 18.0317 & 18.6574 & 72.7133  & 53.4240 \\
	$VBF/VH \rightarrow H_3 \rightarrow ZZ/WW$ & $<$ 0.0001 & $<$ 0.0001  & $<$ 0.0001 & $<$ 0.0001  & $<$ 0.0001  & $<$ 0.0001 \\
	$ggf \rightarrow H_3 \rightarrow ZZ/WW$ & $<$ 0.0001  & $<$ 0.0001  &  $<$ 0.0001  & $<$ 0.0001  & $<$ 0.0001  & $<$ 0.0001    \\
	$VBF/VH \rightarrow H_3 \rightarrow \gamma\gamma$ & $<$ 0.0001  & $<$ 0.0001 & $<$ 0.0001  & $<$ 0.0001 & $<$ 0.0001   & $<$ 0.0001   \\
	$ggf \rightarrow H_3 \rightarrow \gamma\gamma$ & 0.0512 & 0.0584 & 0.1080 &  0.0520 &  0.0027  & 0.0011   \\
	$VBF/VH \rightarrow H_3 \rightarrow invisible$ & $<$ 0.0001  & $<$ 0.0001  & $<$ 0.0001  & $<$ 0.0001  & $<$ 0.0001  & $<$ 0.0001   \\
	$ggf \rightarrow H_3 \rightarrow invisible$ & 0.0013 & 0.0013 & 0.0051 & 0.0022 &  $<$ 0.0001  & $<$ 0.0001  \\
	\noalign{\smallskip}\hline
\end{tabular}
\end{table}

\begin{table}
\centering
\caption{ The BRs for the two light scalar Higgs bosons in \%. The lightest Higgs boson is predominantly singlet-like, while the second lightest Higgs boson corresponds to the 125 GeV Higgs boson, see Tables \ref{t2} and \ref{t3}.  $H_1$ decays predominantly into fermions.  Note the anti-correlation of BR($H_2 \rightarrow bb$) and 	BR($H_2 \rightarrow WW$)  for P3 and P4: the first one decreases by decreasing $\mu_{\tau\tau}^{VBF/VH}$, while the second one increases. For the other mass combinations one finds correlated changes. The explanations are given in Sect. \ref{single}. \label{t6}}
\begin{tabular}{l|c|c||c|c||c|c|}
	\hline\noalign{\smallskip}
P & 1 & 2 & 3 & 4 & 5 & 6 \\
	\noalign{\smallskip}\hline\noalign{\smallskip}
$\mu_{\tau\tau}^{VBF/VH}$ & 1 & 0.7 & 1 & 0.7 & 1 & 0.7 \\
	\noalign{\smallskip}\hline\noalign{\smallskip}
	BR($H_1 \rightarrow hadrons$) & 0.308 & 0.234 & 0.246 & 0.401 &  1.842 & 5.839   \\
	BR($H_1 \rightarrow ee$) & $<$ 0.001 & $<$ 0.001  & $<$ 0.001  & $<$ 0.001  & $<$ 0.001  & $<$ 0.001   \\
	BR($H_1 \rightarrow \mu\mu$) &  0.033 &  0.033 & 0.033 & 0.033 & 0.031  &  0.025   \\
	BR($H_1 \rightarrow \tau\tau$) & 9.315 & 9.325 &  9.301 & 9.244 & 8.872  & 7.010 \\
	BR($H_1 \rightarrow cc$) & 0.057 & 0.028 & 0.032 & 0.309 & 0.993  & 3.01 \\
	BR($H_1 \rightarrow bb$) &  90.277 & 90.375 & 90.382 & 90.007 &  81.848  &   65.278 \\
	BR($H_1 \rightarrow WW$) &  $<$ 0.001  & $<$ 0.001 & $<$ 0.001  & 0.002  & 5.729 & 16.739 \\
	BR($H_1 \rightarrow ZZ$) & -  & -  &  -  & -  & 0.588 & 1.720  \\
	BR($H_1 \rightarrow \gamma\gamma$) &  0.010 & 0.006 & 0.007 & 0.004 &  0.056 & 0.241  \\
	BR($H_1 \rightarrow Z\gamma$) & -  & -  &  - & -  & 0.041 &  0.142  \\
	\noalign{\smallskip}\hline\noalign{\smallskip}
	\noalign{\smallskip}\hline\noalign{\smallskip}
	BR($H_2 \rightarrow hadrons$) &  5.785 & 4.066 & 5.779 & 8.662 & 5.623 & 5.600 \\
	BR($H_2 \rightarrow ee$) & $<$ 0.001  & $<$ 0.001  & $<$ 0.001  & $<$ 0.001  & $<$ 0.001  & $<$ 0.001  \\
	BR($H_2 \rightarrow \mu\mu$) &  0.024 &  0.016  &  0.024 & 0.018 & 0.024  &  0.024    \\
	BR($H_2 \rightarrow \tau\tau$) & 6.651 & 4.650  & 6.660 & 5.202 &  6.708 & 6.718 \\
	BR($H_2 \rightarrow cc$) &  2.885 & 2.027  & 2.876 & 4.233 & 2.827 & 2.806 \\
	BR($H_2 \rightarrow bb$) &   61.752 & 43.177  & 61.827 & 48.973 & 62.252 & 62.334 \\
	BR($H_2 \rightarrow WW$) &  20.259 & 14.243 & 20.212 & 29.091 & 19.973 & 19.931 \\
	BR($H_2 \rightarrow ZZ$) &  2.223 & 1.563 & 2.219 & 3.193 & 2.193 & 2.192  \\
	BR($H_2 \rightarrow \gamma\gamma$) &  0.241 & 0.169 & 0.241 & 0.384 & 0.239 &  0.236   \\
	BR($H_2 \rightarrow Z\gamma$) & 0.162 & 0.114 & 0.162 & 0.244 &  0.160  & 0.159  \\
	BR($H_2 \rightarrow \tilde{\chi}^0_1\tilde{\chi}^0_1$) &    0.018 &  29.974 & - & -  & -  & - \\
	\noalign{\smallskip}\hline
\end{tabular}
\end{table}

\end{document}